\documentclass{aa}  

\usepackage{graphicx}
\usepackage{multirow}
\usepackage{lineno}
\usepackage{txfonts}
\begin{document}

   \title{Directly imaged exoplanets in reflected starlight. The importance of knowing the planet radius}

   %\subtitle{I. Overviewing the $\kappa$-mechanism}

   \author{\'O. Carri\'on-Gonz\'alez\inst{1}\thanks{ \email{o.carriongonzalez@astro.physik.tu-berlin.de, oscarcggs@gmail.com}}
          \and
          A. Garc\'ia Mu\~noz\inst{1}
          \and
          J. Cabrera\inst{2}
          \and
          Sz. Csizmadia\inst{2}
          \and
          N. C. Santos\inst{3,4}
          \and
          H. Rauer\inst{1,2,5}
          }

   \institute{Zentrum f\"ur Astronomie und Astrophysik, Technische Universit\"at Berlin, Hardenbergstrasse 36, D-10623 Berlin, Germany
         \and
             Deutsches Zentrum f\"ur Luft- und Raumfahrt, Rutherfordstraße 2, D-12489 Berlin, Germany
         \and
             Instituto de Astrof\'isica e Ci\^encias do Espa\c{c}o, Universidade do Porto, CAUP, Rua das Estrelas, 4150-762 Porto, Portugal
         \and
             Departamento de F\'isica e Astronomia, Faculdade de Ci\^encias, Universidade do Porto, Rua do Campo Alegre, 4169-007 Porto, Portugal
         \and
             Institute of Geological Sciences, Freie Universit\"at Berlin, Malteserstrasse 74-100, D-12249 Berlin, Germany
             }

   %\date{Received September 15, 1996; accepted March 16, 1997}

% \abstract{}{}{}{}{} 
% 5 {} token are mandatory
 
  \abstract
   {The direct imaging of exoplanets in reflected starlight will represent a major advance in the study of cold and temperate exoplanet atmospheres.
Understanding how basic planet and atmospheric properties may affect the measured spectra is key to their interpretation.}
   {We have investigated the information content in reflected-starlight spectra of exoplanets.
   We specify our analysis to Barnard's Star b candidate super-Earth, for which we assume a radius 0.6 times that of Neptune, an atmosphere dominated by H$_2$-He, and a CH$_4$ volume mixing ratio of 5$\cdot$10$^{-3}$. 
   The main conclusions of our study are however planet-independent.}
   {We set up a model of the exoplanet described by seven parameters including its radius, atmospheric methane abundance and basic properties of a cloud layer. 
    We generate synthetic spectra at zero phase (full disk illumination) from 500 to 900 nm and spectral resolution R$\sim$125-225.
    We simulate a measured spectrum with a simplified, wavelength-independent noise model at Signal-to-Noise ratio S/N=10. 
    With an MCMC-based retrieval methodology, we analyse which planet/atmosphere parameters can be inferred from the measured spectrum and the theoretical correlations amongst them.
    We consider limiting cases in which the planet radius is either known or completely unknown, and intermediate cases in which the planet radius is partly constrained.}
   {If the planet radius is known, we can generally discriminate between cloud-free and cloudy atmospheres, and constrain the methane abundance to within two orders of magnitude. 
    If the planet radius is unknown, new correlations between model parameters occur and the accuracy of the retrievals decreases.
    Without a radius determination, it is challenging to discern whether the planet has clouds, and the estimates on methane abundance degrade.
    However, we find the planet radius is constrained to within a factor of two for all the cases explored.
    Having \textit{a priori} information on the planet radius, even if approximate, helps improve the retrievals.}
   {Reflected-starlight measurements will open a new avenue for characterising long-period exoplanets, a population that remains poorly studied.
    For this task to be complete, direct-imaging observations should be accompanied by other techniques.
    We urge exoplanet detection efforts to extend the population of long-period planets with mass and radius determinations.
    }


   \maketitle
%
%-------------------------------------------------------------------

\section{Introduction}
\label{sec:introduction}

Upcoming space missions such as WFIRST\footnote{Recently renamed Nancy Grace Roman Space Telescope (NGRST).} \citep{spergeletal2013} and mission concepts such as LUVOIR \citep{bolcaretal2016} or HabEx \citep{mennessonetal2016} aim to measure the starlight reflected from cold and temperate exoplanets by direct imaging.
This offers opportunities to detect and characterise long-period exoplanets, a population that remains substantially unexplored due to current technological biases. 
Due to their lower equilibirum temperatures, these planets are expected to have very different atmospheric properties from those currently observed in transit.

Characterising long-period exoplanets will help complete the picture of extrasolar planetary systems and shape the formation and evolution theories that explain their architectures \citep{dangelo-bodenheimer2013, fabryckyetal2014}. 
The probability that these long-period planets are observed in transit is small, since it decreases with the orbital distance.
Furthermore, even if they transit, atmospheric refraction will limit the insight obtained from transmission spectroscopy \citep{garciamunozetal2012,misraetal2014}.
Hence, direct-imaging  will be key to analyse these planets.

In preparation for these missions, it is crucial to investigate what information can be extracted from measurements of reflected starlight from spatially-unresolved planets. 
The underlying question is certainly not new, and there is a vast literature examining the diagnostic possibilities of reflected light in general remote sensing applications. 
For instance, in the framework of Earth observations, \citet{stephens-heidinger2000} and \citet{heidinger-stephens2000} examined the potential of satellite observations of possibly cloudy views above our planet's surface. 
They concluded that six main parameters affect a reflected-light spectrum, namely: the optical thickness of the cloud, its geometrical thickness, the pressure level of the cloud top, the scattering phase function and single-scattering albedo of the cloud particles and the surface reflectivity.

The interpretation of direct-imaging observations of exoplanets in reflected starlight will rely on essentially the same physical principles. 
Although the above space missions are still years away, there is already significant interest to understand their prospective scientific output \citep[see e.g.][]{cahoyetal2010,greco-burrows2015,lupuetal2016,robinsonetal2016,nayaketal2017, batalhaetal2018,fengetal2018,guimond-cowan2018,damiano-hu2019,hu2019,lacyetal2019}.
Such investigations will have an impact on the design of future telescopes and their observing strategies.
They are also making the case for long-period planets as a target population that should be followed up by other techniques such as radial velocity and transit photometry.

\citet{cahoyetal2010} generated synthetic spectra of cold giant exoplanets
for a range of atmospheric compositions, orbital distances and 
star-planet-observer phase angles. 
They analysed how clouds modify the reflected-starlight spectra by altering the depth of gas absorption bands.
\citet{greco-burrows2015} studied the importance of the assumed scattering model for the atmospheres of giant exoplanets.
For instance, they showed that choosing a Lambertian or a Rayleigh-like scattering varies the fraction of the orbit predicted to be observable by up to 20\%.

\citet{robinsonetal2016} developed a comprehensive noise model and computed the integration times required to detect various atmospheric spectral features with WFIRST-like telescopes for a set of planets including Solar System analogs.
They concluded that detecting methane at a planet within 5 pc would be easy for cool Jupiter twins, feasible for cool Neptunes and challenging for super-Earths. 
They also argue that methane could be detected with reasonable integration times for super-Earths within 3 pc.

\citet{lupuetal2016} studied the characterisation prospects for giant exoplanets observed at zero phase, i.e. when the planet is fully illuminated. 
Their atmospheric models considered a known value of the planetary radius and H$_2$-He atmospheres with methane as the major absorber, and either one or two cloud layers. 
A similar study was carried out by \citet{damiano-hu2019}, who also accounted for non-uniform volume mixing ratio profiles of gaseous species (NH$_3$, H$_2$O) due to cloud formation.
However, in their analysis they do not consider as free parameters some of the cloud properties such as the optical depth, the single-scattering albedo or the scattering asymmetry factor of the aerosols.
Instead, they compute these values with an equilibrium-cloud and radiative-transfer model \citep{hu2019}.
This introduces new physics into the inversion problem, but it potentially makes the conclusions more dependent on the specifics of the equilibrium-cloud model.
Based on the models reported by \citet{lupuetal2016}, \citet{nayaketal2017} assumed the phase angle to be unknown at the time of the observations and studied the impact of this on the inferred planet radius R$_p$ (also unknown \textit{a priori}) and the atmospheric properties.

\citet{guimond-cowan2018} examined the Earth-twin yield from future direct-imaging searches.
They found that uncertainties in the planetary radius R$_p$ will lead to false-positive detections of Earth analogues.
In such cases, low-albedo planets with large R$_p$ are mistaken for highly-reflecting, Earth-sized planets.
Although they do not address the atmospheric characterisation of these planets, their work shows that not knowing
R$_p$ is a source of additional uncertainties in the interpretation of direct-imaging observations.
\citet{batalhaetal2018} considered the problem of classifying giant exoplanets with colour photometry, concluding that at least three colour filters are necessary to differentiate them on the basis of e.g. metallicity, cloud properties, and pressure-temperature profiles.

\citet{fengetal2018} studied the prospective science return of future LUVOIR- and HabEx-like space missions observing reflected starlight from cloudy Earth analogs at visible wavelengths. They concluded that weak detections of H$_2$O, O$_3$ and O$_2$ could be achieved if the spectra are measured with at least a spectral resolving power R=$\lambda$/$\Delta \lambda$=70 and signal-to-noise ratio S/N=15, or R=140 and S/N=10.
With the most up-to-date WFIRST design parameters available at the time, \citet{lacyetal2019} studied the capabilities of this space mission to constrain the planetary radius and methane abundance of exoplanets.
Since they focus on the impact of instrument design and noise levels, their atmospheric models contain a number of simplifications, in particular affecting the description of the clouds.
For instance, they assume that cloud properties and aerosol scattering functions are known.

In this work, we focus on cold gas exoplanets as they will likely be the first to be imaged in reflected starlight.
In particular, we investigate how well the size and atmospheric properties of such planets can be constrained.
We proceed through an exercise of retrievals that attempts to interpret a measured spectrum on the basis of atmospheric models, inferring best-fitting configurations and intervals of confidence. 
It is foreseeable that a variety of planet sizes and atmospheric configurations may result in spectra that are essentially indistinguishable.
We quantify such possibilities as part of our retrievals.

Our first goal is to understand how model parameters modify the reflected-light spectra and how  degeneracies between parameters might affect our conclusions from the retrieval.
Second, we aim to test the effects of an unknown planet radius on the characterisation.
If this parameter is unconstrained, we show below that new correlations between model parameters will play a role in the retrieval, changing the conclusions.
This comparison between retrievals with both a fixed value of R$_p$ and with R$_p$ as a free parameter, describing the importance of having a constraint on the planet size, has not been addressed in previous literature.

Although our work is essentially theoretical, we specify it to the Barnard's Star planetary system,  as the representative of a growing population of long-period planets discovered by radial velocity measurements around nearby stars.
Barnard's Star (Gl 699) has long been an object of interest due to its proximity to the Earth and its proper motion relative to the Sun, 10.3 arcsec/year, the highest among all known stars \citep{barnard1916}. 
At a distance of 1.8 pc from the Sun, this is the second closest stellar system after $\alpha$ Centauri \citep{giampapaetal1996}.
In a late stage of evolution (age of 7-10 Gyr) and with an effective temperature T$_{\rm{eff}}$=3100-3200K \citep{dawson-derobertis2004,ribasetal2018}, Barnard's Star is one of the known M dwarfs with lower magnetic activity and X-ray luminosity \citep{schmitt-liefke2004}.
Searches for planetary companions around this M dwarf during the 20th century resulted in the detection of a Jovian candidate by \citet{vandekamp1963} that was later refuted \citep{gatewood-eichhorn1973}. 
Recently, \citet{ribasetal2018} announced a super-Earth candidate (Barnard b) of minimum mass 3.23M$_{\oplus}$ around this star on an orbit of semimajor axis $a$=0.4 AU, period P=232 days and maximum angular separation $\Delta \theta$=220 milliarcseconds ($mas$).
This planet has not been observed in transit \citep{tal-oretal2019}, so its atmosphere can only be characterised by measuring the light either reflected or emitted by the planet.

For generality, we avoid predictions for specific telescopes.
Nevertheless, we take the designs of the WFIRST and LUVOIR missions for reference, as both are planned to be equipped with optical coronagraphs.
WFIRST and LUVOIR will detect exoplanets orbiting their host stars with angular separations greater than $\sim$150 and $\sim$50 $mas$, respectively \citep{traugeretal2016, starketal2015}.
Another critical specification for direct imaging is the minimum planet-to-star contrast ($C_{min}$) that these instruments can achieve. 
$C_{min}$ is expected to be $10^{-9}$ in the case of WFIRST \citep{traugeretal2016} and $10^{-10}$ for LUVOIR \citep{luvoirteam2018}.
As shown below, the reflected-light spectra of Barnard b are predicted to exceed these limits.
This makes Barnard b a prime target to be characterised with future direct-imaging missions.

The paper is structured as follows.
In Section \ref{sec:model} we describe the atmospheric model used to generate synthetic spectra and motivate its application to Barnard b.
In Section \ref{sec:retrieval}, we describe the retrieval procedure.
Results are presented in Section \ref{sec:results} and Section \ref{sec:conclusions} contains the final summary and conclusions.

\section{Model} \label{sec:model}

\subsection{Setting} \label{subsec:model_theory}
The planet-to-star contrast in reflected starlight \citep{horak1950} is given by:
\begin{equation}
\label{eq:contrast}
\frac{F_p}{F_{\star}}(\alpha, \lambda)= \left(\frac{R_N}{r} \right)^2 
\left(\frac{R_p}{R_N} \right)^2 A_g(\lambda; \;\boldsymbol{p}) 
\Phi(\alpha, \lambda; \;\boldsymbol{p})
\end{equation}

Here, $R_p/R_N$ is the planet radius normalized to that of Neptune (convenient because we focus on a sub-Neptune planet);
$r$ the planet-to-star distance;
$A_g$ the geometric albedo and $\Phi$ the normalized scattering phase function of the planet. 
$A_g$ depends on the observation wavelength $\lambda$ and on the specifics of the atmosphere, described here by the vector of atmospheric parameters $\boldsymbol{p}$ (see Section \ref{subsec:model_atmos}). 
$\Phi$ depends additionally on the phase angle $\alpha$. 
$\alpha$ and $r$ will generally vary as the planet moves on its orbit. 
In this study we will focus on phase angle $\alpha$=0$^\circ$ and $r$=0.4 AU, an orbital distance consistent with that of Barnard b.
We take $\alpha$=0$^\circ$ as an idealization, commonly adopted in previous works, that we take as an initial step.
The consideration of multiple phases, either separately or jointly, will be addressed in follow-up work.

\subsection{Bulk atmospheric composition of Barnard b}  \label{subsec:model_Barnard bcomposition} 
The lack of a radius measurement for Barnard b means that its bulk and atmospheric compositions are essentially unconstrained \citep{zengetal2019}.
The focus of our investigation is to elucidate the diagnostic possibilities of reflected starlight measurements, rather than inquire about the real nature of the planet.
We can nevertheless make some reasonable guesses for the size and composition of Barnard b that will allow us to define the basic properties of our model.

We will assume that the planet is enshrouded by an optically thick atmosphere mainly composed of H$_2$ and He.
Whether this assumption corresponds to the actual bulk composition and history of Barnard b cannot be stated with current data.
However, there are reasons to take this scenario as physically plausible based on the mass of the exoplanet.
It is expected that protoplanets with the size of the Moon (0.27 $R_{\oplus}$) will form a significant atmosphere surrounding the core \citep{inaba-ikoma2003}.
Cores with masses larger than 2--3 $M_{\oplus}$ might indeed undergo runaway gas accretion depending on the properties of the protoplanetary disk \citep{ikomaetal2001,inaba-ikoma2003}. 
Whether Barnard b kept its primordial envelope or not depends on its history of stellar activity and planet migration. 
Barnard b, being close to the stellar \textit{snow line}, might have experienced some atmospheric escape driven by the extreme ultraviolet (EUV) irradiation from its host star.
If it did not undergo migration, the minimum mass of the planet suggests that a fraction of its primordial atmosphere might have been lost \citep{pierrehumbert-gaidos2011}. 
However, migration towards the host star is thought to be common for super-Earths in low-mass disks after the end of the highly-active T Tauri phase \citep{ida-lin2008}.
Barnard b's present orbital distance is therefore plausibly the result of inward migration at some point of the 7-10 Gyr age of this stellar system. If the planet formed farther from its host star, the atmospheric escape would have been much less \citep{pierrehumbert-gaidos2011} and the planet could have retained a larger fraction of its primordial atmosphere.
Ultimately, our investigation offers atmospheric predictions that should be testable with direct imaging spectroscopy.

   \begin{figure}
   \centering
   \includegraphics[width=9.cm]{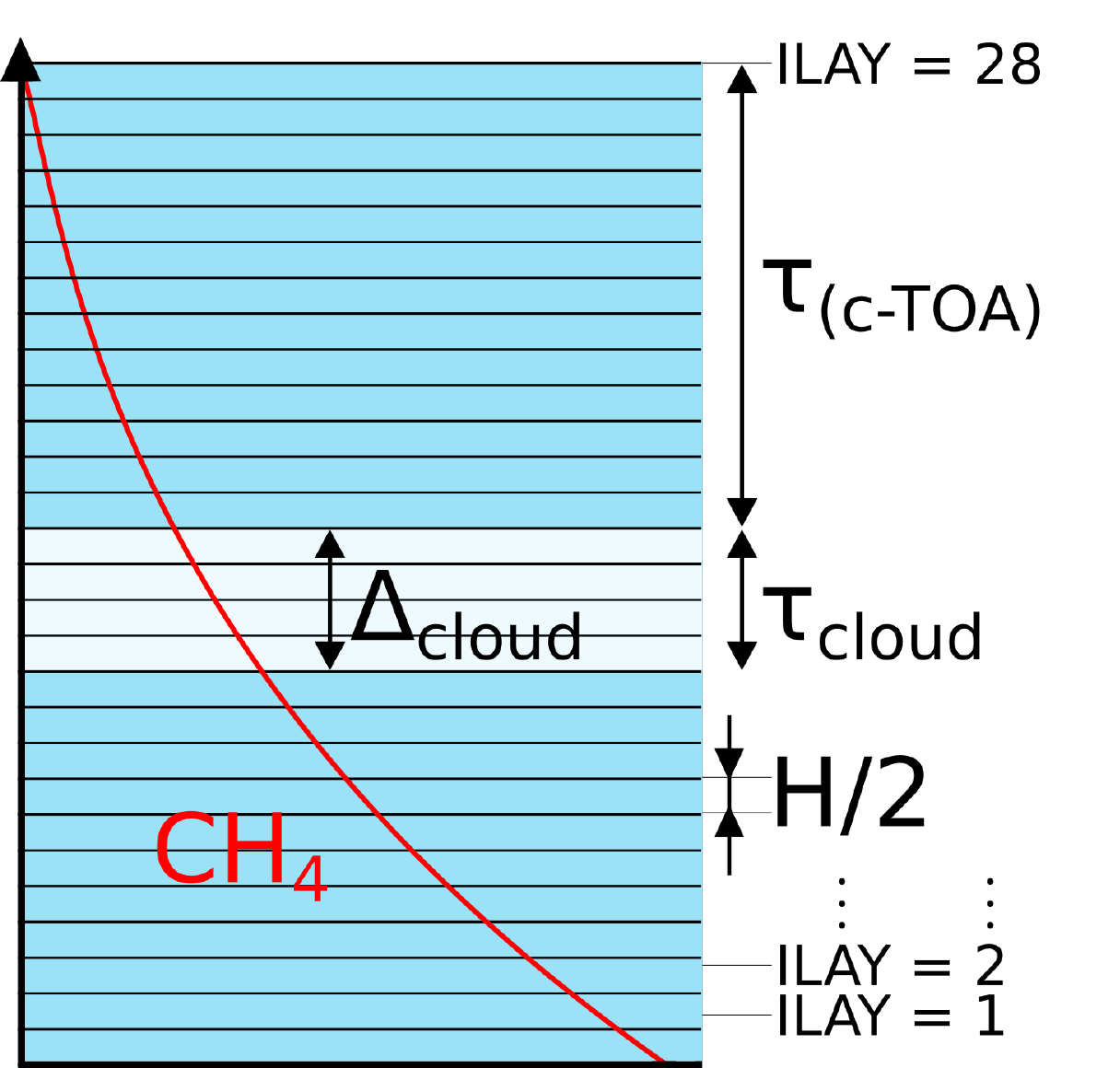}
      \caption{\label{fig:model_sketch} Sketch of the vertical structure in our atmospheric model.
      \textit{ILAY} denotes each of the slabs in which the atmosphere is divided, each with thickness $H_g/2$.
      Gas density decreases exponentially with height.
      The atmosphere is assumed to contain CH$_4$, with a certain fraction $f_{\rm{CH_4}}$.
      We include a cloud with optical thickness $\tau_{c}$ and a geometrical extension $\Delta_{c}$.
      The position of the cloud top is specified by $\tau_{c\rightarrow TOA}$.
      In addition, the aerosols of the cloud are described by their single-scattering albedo $\omega_0$ and effective-radius $r_{eff}$, as explained in the text.
     }
   \end{figure}

\subsection{Forward atmospheric model} \label{subsec:model_atmos}
We model the atmosphere of Barnard b to be composed of H$_2$-He gas in a Jovian ratio $x_{He}/x_{H_2}$=0.157 \citep{sanchezlavega2010}, an absorbing gas present in trace amounts (CH$_4$ in this work), and a cloud layer. 
We use CH$_4$ as this is the most abundant gaseous absorbing species in the atmospheres of all cold H$_2$-dominated atmospheres in the Solar System.
Assuming a constant temperature and gravity representative of the atmospheric layers reached by the stellar photons, the gas density and pressure decay exponentially in the vertical with a scale height H$_g$. 
The value of H$_g$ is not important in our treatment, as discussed below. 
We set the base of the atmosphere at a depth such that the Rayleigh optical thickness of the gas is  $\tau_{*}$=10 at the reference wavelength  $\lambda_*$=800 nm. The Rayleigh optical thickness varies with wavelength in the usual way $\propto$ $\lambda^{-4}$. 
The choice of $\tau_{*}$ avoids a prohibitive computational cost while ensuring that the boundary condition at the base of the atmospheric model is not critical \citep{buenzli-schmid2009}, where we set a zero surface reflectance. 
With the base of the atmosphere defined this way, the planetary radius corresponds to the distance between this level and the center of the planet.
For relatively massive, cold exoplanets, as we focus on here, the scale height of the atmosphere will be much smaller than e.g. that of hot Jupiters.
Thus, the planet radius relevant to direct imaging is practically equivalent to the radius that would be measured in transit.

Our adoption of a one-cloud model is motivated by simplicity, and is supported by recent works \citep[e.g.][]{nayaketal2017, damiano-hu2019}.
They show that one-cloud and multi-cloud models can fit comparably well a measured spectrum in disk-integrated observations for the expected S/N values.
Further, \citet{nayaketal2017} conclude that, in a two-cloud model, generally it is only possible to retrieve information about the lowermost cloud if the upper one is practically non-existent.
This, in fact, recovers a one-cloud atmospheric model.

\begin{figure}
   \centering
   \includegraphics[width=9.cm]{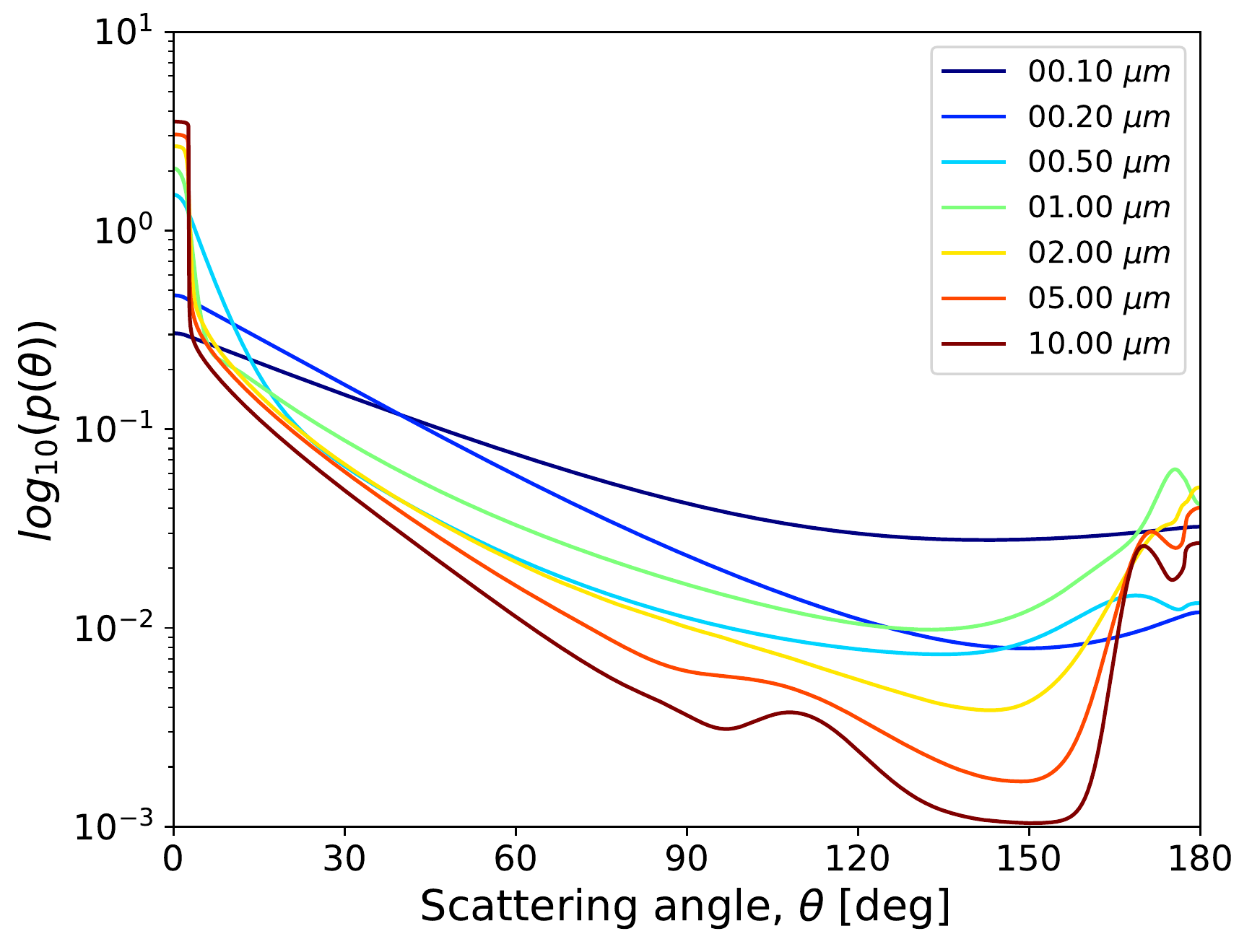}
      \caption{\label{fig:Miescattering} Scattering phase functions for the considered values of $r_{\rm{eff}}$. 
     }
\end{figure}

Our atmospheric model is sketched in Fig. \ref{fig:model_sketch} and described by the six parameters in the atmospheric vector $\boldsymbol{p}=$\{$\tau_{c}$, $\Delta_{c}$, $\tau_{c\rightarrow TOA}$, $r_{\rm{eff}}$, $\omega_0$, $f_{\rm{CH_4}}$ \}. 
Here, {$\tau_{c}$} is the optical thickness of the cloud and {$\Delta_{c}$} its geometrical vertical extension.
The cloud layer is assumed to be horizontally homogeneous over the whole planet. 
{$\tau_{c\rightarrow TOA}$} is the optical thickness of gas from the cloud top to the top of the atmosphere (TOA) at the reference wavelength $\lambda_*$. 
This is a measure of the altitude/pressure level at which the cloud top is located. 
{$\omega_0$} is the single scattering albedo of the cloud aerosols, 
that we leave as a free parameter to account for different aerosol compositions. 
Both {$\tau_{c}$} and {$\omega_0$} are assumed to be wavelength-independent.
{$r_{\rm{eff}}$} is the effective radius of the aerosols that, in our treatment, defines their scattering phase functions.
We have calculated the scattering phase function $p$($\theta$) for each {$r_{\rm{eff}}$} through Mie theory assuming a constant refractive index of 1.42 specific to NH$_3$. 
Our treatment preserves the expected trend that 
small effective radii {$r_{\rm{eff}}$} will result in nearly isotropic scattering phase functions, whereas large values will result in enhanced backward and forward scattering.
The scattering phase functions are shown in Fig. \ref{fig:Miescattering}.
By varying {$\omega_0$} and {$r_{\rm{eff}}$} independently, our exploration is generalised to clouds with arbitrary composition.
Last, {$f_{\rm{CH_4}}$} stands for the relative methane abundance or
volume mixing ratio relative to H$_2$-He, assumed constant over all layers. 
The CH$_4$ opacities are taken from \citet{karkoschka1994}.

Each of the six parameters is given the set of values summarized in Table \ref{table:grid}  
to obtain a dense grid of atmospheric configurations ($\sim$300,000 configurations after omitting impossible situations in which the cloud's vertical extension reaches below the model's deepest layer).
For each of these configurations we will obtain a synthetic spectrum, as described below.
This dense grid of spectra will be used during the retrieval process to obtain new spectra from interpolation within the grid at $\boldsymbol{p}$ configurations not represented in Table \ref{table:grid} (see Section \ref{subsec:retrieval_exploringparamspace}).

We discretize the atmosphere with a total of $N_{\rm{slabs}}$=28 vertical slabs, each of them of geometrical thickness H$_g$/2. 
The scattering and absorption coefficients are assumed to be constant within each slab, and we follow standard rules when averaging over the gas and aerosols. 
The optical thickness of gas above each interface $k$ (see Fig. \ref{fig:model_sketch}) is $\tau_k=\tau_{k=0} \exp{(-k/2)}$. 
Clouds are placed over an integer number of scale heights ($\Delta_{c}$/H$_g$=1, 2, 3, ..., 8). 
Once the atmospheric model is set up and $\boldsymbol{p}$ prescribed, we produce the reflected-starlight spectra by solving the radiative transfer equation for multiple scattering with a Backward Monte Carlo (BMC) model  \citep{garciamunoz-mills2015}.
The BMC model has been thoroughly tested against phase curve calculations  \citep{dlugach-yanovitskij1974, buenzli-schmid2009} with excellent agreement, and against whole-disk polarization and brightness phase curves of Venus and Titan \citep{garciamunozetal2014, garciamunoz-mills2015,garciamunoz-isaak2015, garciamunozetal2017, ilicetal2018}.
Two reasons for using the BMC model instead of other, probably faster, radiative transfer solvers are:
securing the accuracy of eventual calculations at all possible phase angles; being able to extend in the future the analysis to polarization.

\begin{table}[t]
\caption{Values for the parameters in the atmospheric vector $\boldsymbol{p}$
used to construct the grid of synthetic reflected-starlight spectra. }
\label{table:grid}
\begin{center}
 \begin{tabular}{ c  c } 
 \hline \hline
  Parameter    &   Values \\
 \hline
  $\tau_{c}$ &   0.05, 0.20, 0.50, 1.0, 2.0, 5.0, 10.0, 20.0, 50.0 \\
  $\Delta_{c}$/H$_g$   &   1, 2, 3, 4, 5, 6, 7, 8 \\
  $\tau_{c\rightarrow TOA}$  &   1.35, 0.50, 0.18, 6.7$\cdot$10$^{-2}$,\\
  & 2.5$\cdot$10$^{-2}$, 9.1$\cdot$10$^{-3}$, 4.5$\cdot$10$^{-4}$ \\
  {$r_{\rm{eff}}$} [$\mu$m] &   0.10, 0.20, 0.50, 1.0, 2.0, 5.0, 10.0 \\
  $\omega_0$  &   0.50, 0.60, 0.70, 0.75, 0.80,\\
  & 0.85, 0.90, 0.95, 0.98, 0.99, 1.0 \\
  $f_{\rm{CH_4}}$    &   1$\cdot$10$^{-5}$, 5$\cdot$10$^{-5}$, 1$\cdot$10$^{-4}$, 2$\cdot$10$^{-4}$, 5$\cdot$10$^{-4}$, \\
   & 1$\cdot$10$^{-3}$, 2$\cdot$10$^{-3}$, 5$\cdot$10$^{-3}$, 1$\cdot$10$^{-2}$, 2$\cdot$10$^{-2}$, 5$\cdot$10$^{-2}$ \\
  $\lambda$ &   500 nm $-$ 900 nm ($\Delta \lambda$=4 nm) \\
 \hline
\end{tabular}
\end{center}
\end{table}

\begin{figure*}[t]
       \centering
    \includegraphics[width=9.cm]{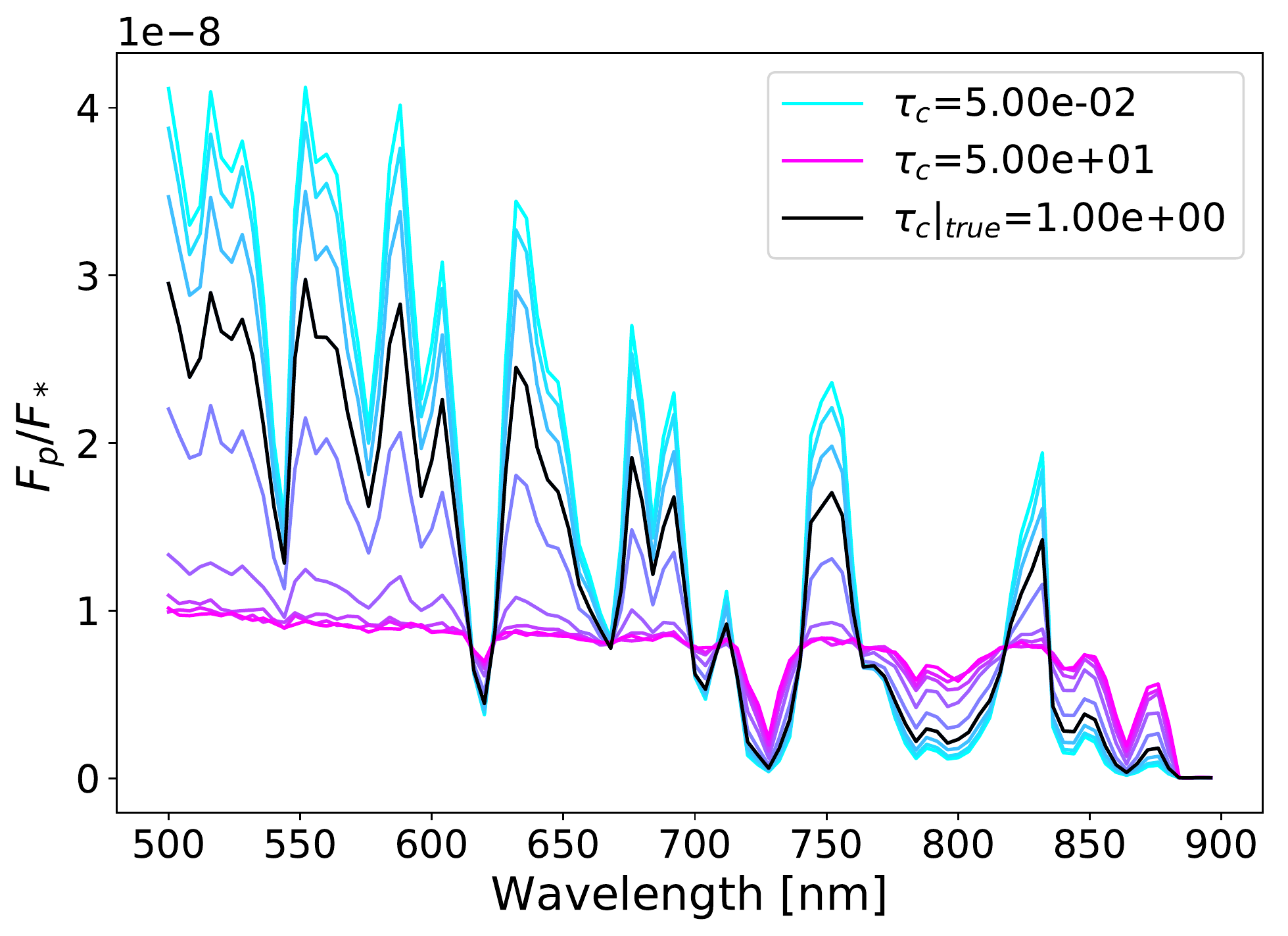}%
    \includegraphics[width=9.cm]{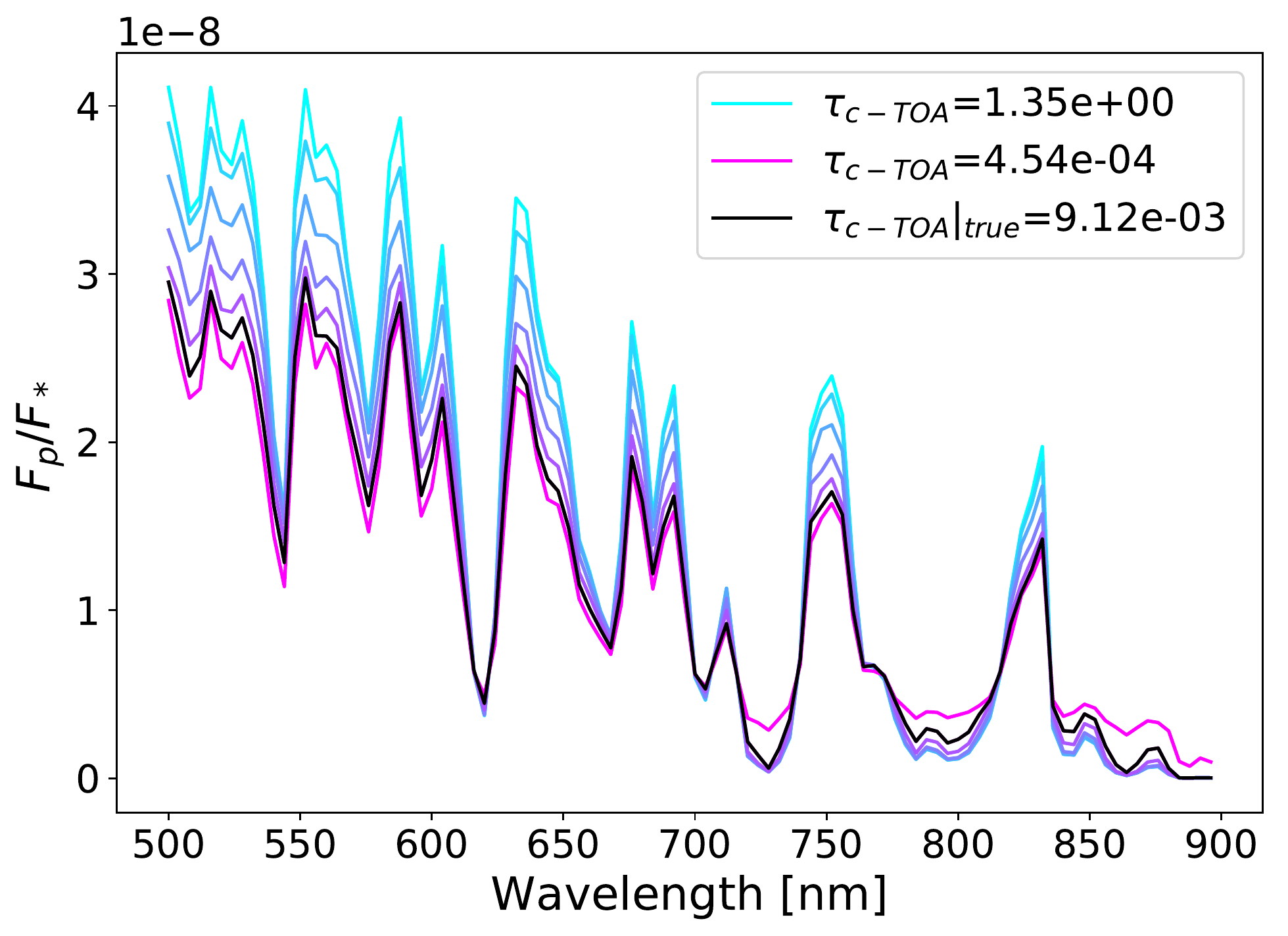}%
    \\
    \includegraphics[width=9.cm]{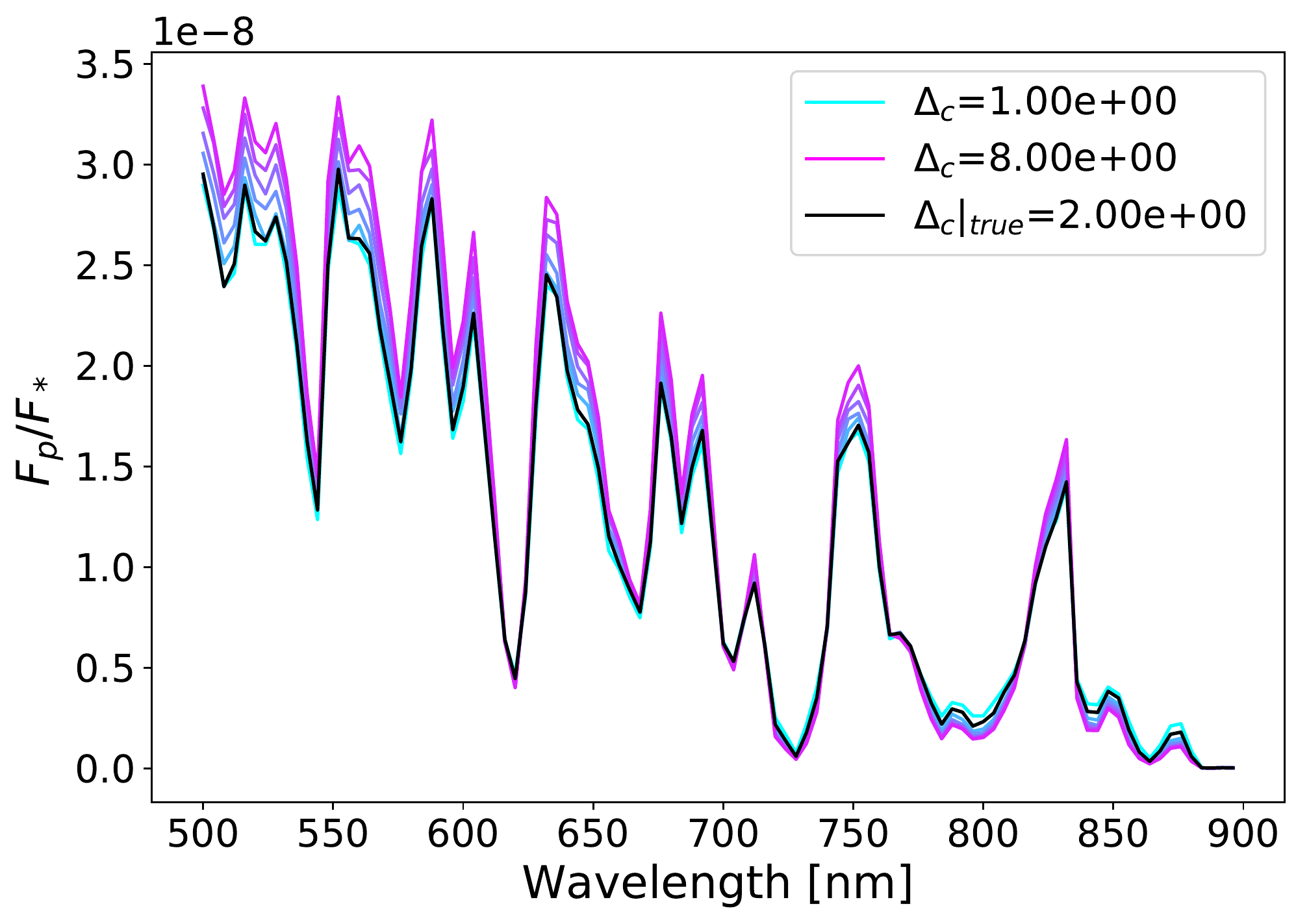}%
    \includegraphics[width=9.cm]{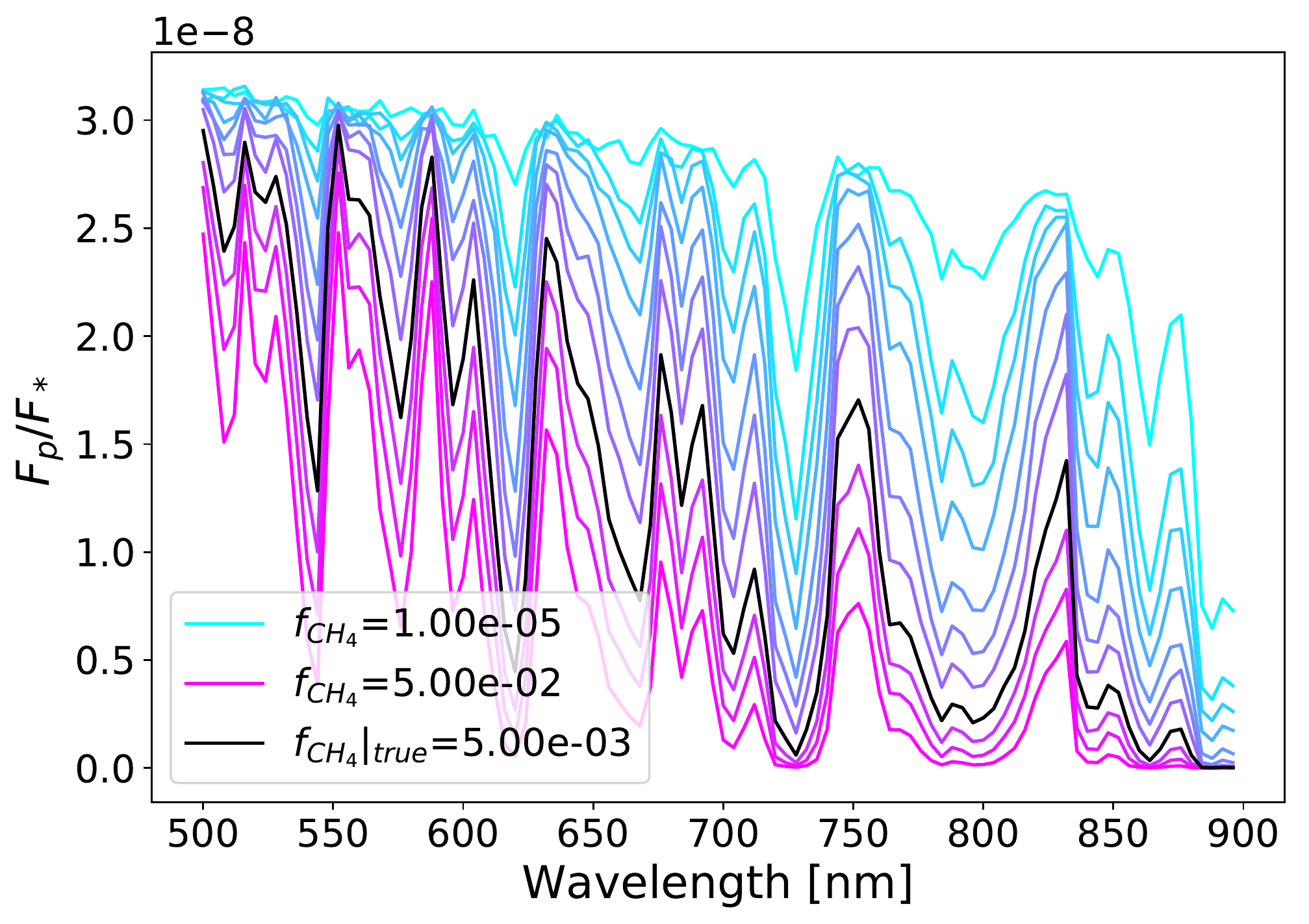}%
    \\
    \includegraphics[width=9.cm]{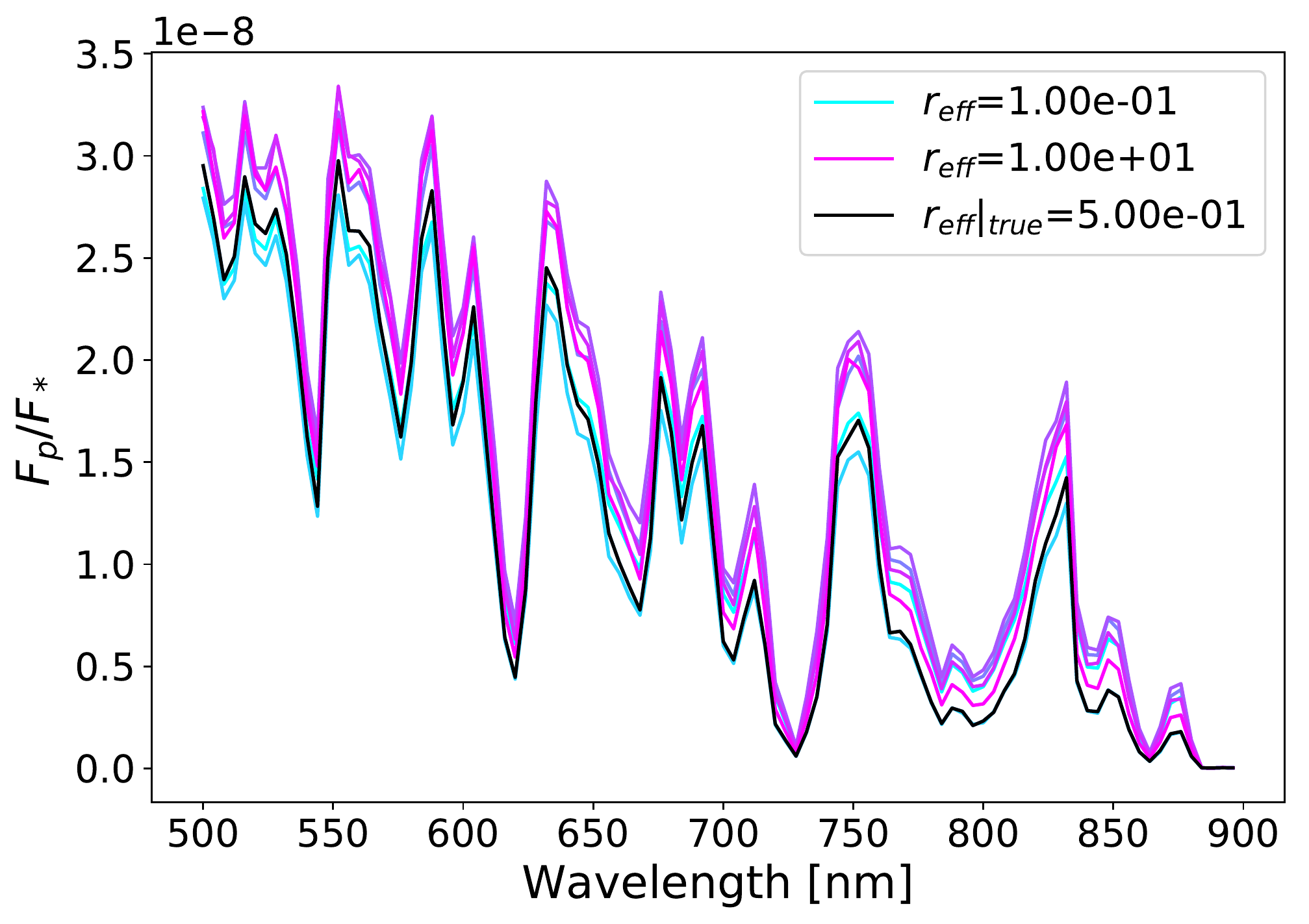}%
    \includegraphics[width=9.cm]{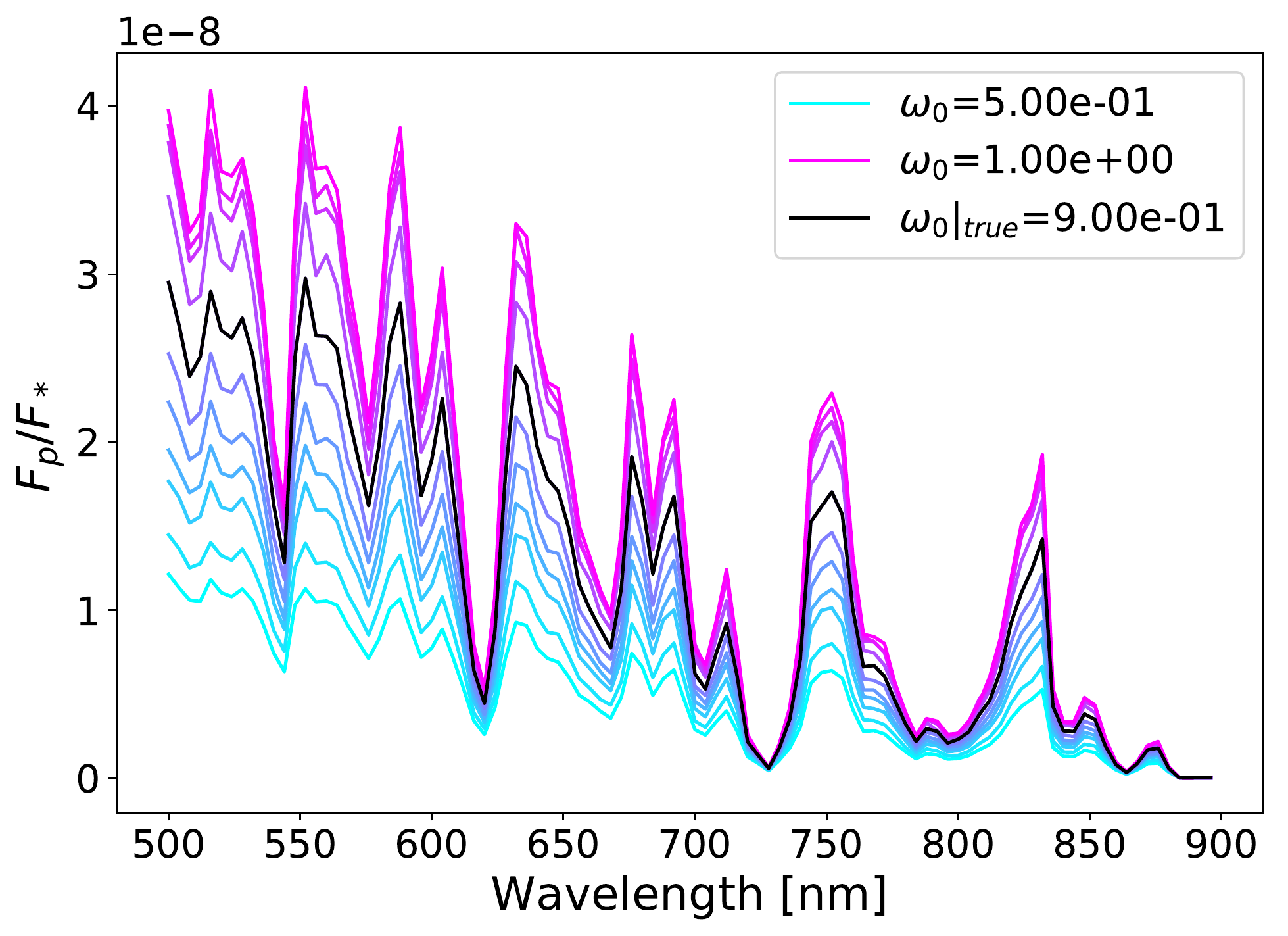}%
    \caption{Solid black lines: Synthetic spectra for the \textit{thin-cloud} configuration (see Table \ref{table:truth}).
    Colour lines: synthetic spectra for the configurations that result from perturbing each of the elements of the six-dimensional atmospheric vector $\boldsymbol{p}=$\{$\tau_{c}$, $\Delta_{c}$, $\tau_{c\rightarrow TOA}$, $r_{\rm{eff}}$, $\omega_0$, $f_{\rm{CH_4}}$ \} one at a time and according to the values listed in Table 
    \ref{table:grid}.
    The colour code is described in the legends for the lowermost and uppermost values of each parameter, with the other colours corresponding to the intermediate parameter values.
    {$\Delta_{c}$} is given in H$_g$ units and {$r_{\rm{eff}}$} in $\mu$m.
    }%
    \label{fig:completeness_variations_all_param}%
    \end{figure*}

Our description of atmospheric stratification relies on optical thickness, rather than altitude or pressure. 
Although the actual implementation into the BMC model uses altitude and therefore a scale height, 
the value of  H$_g$ disappears from the description of $\tau_k$ (see above) and in turn from the description of the atmosphere. 
In other words, the value of H$_g$ is irrelevant for small and moderate phase angles, and our implementation is essentially independent of this parameter.
This may not be true at large phase angles \citep{garciamunozetal2017,garciamunoz-cabrera2018}, but this scenario is impractical in direct imaging due to the visibility restrictions imposed by the technique.
As a check, we solved in a few hundred cases the radiative transfer equation for different H$_g$ values while keeping the rest of parameters unaltered. 
No differences were observed in the generated spectra, thereby confirming that H$_g$ plays no role in our simulations.

For each combination of the parameters in the atmospheric vector $\boldsymbol{p}$ we computed synthetic spectra in the range $500-900$ nm. 
Figure \ref{fig:completeness_variations_all_param} shows (solid black lines) the spectrum of a particular atmospheric configuration referred to as \textit{thin-cloud} (see Table \ref{table:truth}).
Also plotted are the spectra corresponding to varying the elements of $\boldsymbol{p}$ one at a time through the grid of simulations of Table \ref{table:grid}.
We used a spectral bin $\Delta \lambda=$4 nm over the entire spectral range, which was deemed appropriate because the absorption features by CH$_4$ are broad. 
This translates into resolving powers $R\sim125$ at 500 nm and $R\sim225$ at 900 nm.

Our approach does not include photochemistry or microphysics of cloud formation. Instead, 
we parameterize the optical properties of the gas and aerosols 
and explore the influence of a few key properties on the synthetic spectra.
This is complementary 
to other approaches \citep[e.g.][]{ackerman-marley2001, ohno-okuzumi2017, hellingetal2017, damiano-hu2019}  
that investigate the photochemical and microphysical mechanisms resulting in cloud formation. 
Scanning through all six parameters in our atmospheric model gives us a broad view of possibilities that can eventually be tested against more physically-based approaches.

\section{Retrieval procedure}
\label{sec:retrieval}

Observing exoplanets in direct imaging will allow us to estimate the planet radius and atmospheric composition from retrieval exercises of the measured spectra.
One of the aims of this work is to quantify the uncertainties in the estimates of the retrieved properties.
Since there are no measurements of this kind yet, we conduct the exercise through simulated measurements.
From these, we attempt to extract information through the systematic comparison with synthetic spectra.

\subsection{Adopted \textit{true} configurations} \label{subsec:retrieval_trueconfigs}

We adopt three of the atmospheric configurations in our grid of Table \ref{table:grid}
as \textit{true}, meaning that they represent the true 
atmospheric properties of the planet that is being observed.
These configurations are identical in all the parameter values except $\tau_{c}$, so we can specifically explore the transition from cloud-free to thick-cloud conditions.
We will refer to the three atmospheric configurations as the \textit{no-}, \textit{thin-} and \textit{thick-cloud} scenarios.
They are summarized in Table \ref{table:truth}.

\begin{figure*}[t]
   \centering
   \includegraphics[width=18.cm]{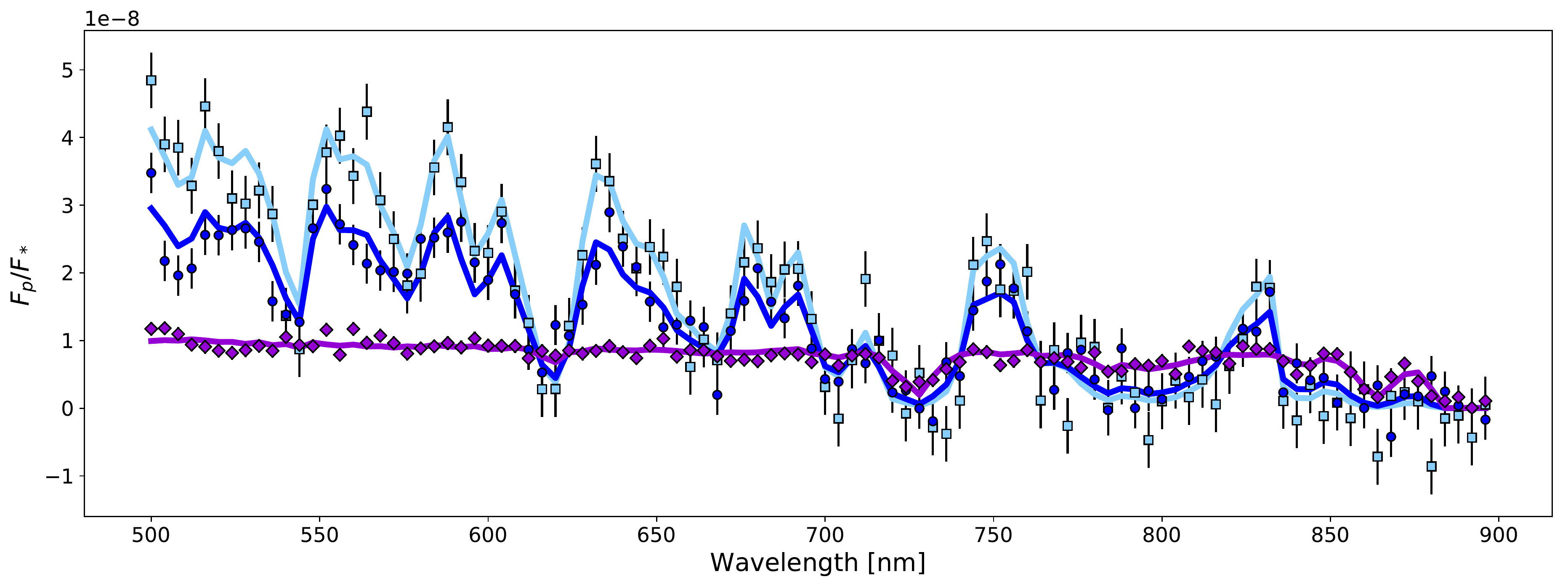}
      \caption{\label{fig:trueVSnoisy_spectra} 
      Solid lines: \textit{True} (noiseless) spectra for the three  atmospheric configurations in Table \ref{table:truth}  at phase angle $\alpha$=0$^\circ$. 
      Symbols: Corresponding \textit{measured} spectra, with error bars specific to S/N=10. 
      The size of the error bars ($\sigma_m$; see text) is defined on the basis of the brightest spectral bin (which typically occurs near 500 nm) and is the same at all wavelengths for each spectrum.
      The spectra correspond to \textit{no-cloud} (light blue), \textit{thin-cloud} (dark blue) and \textit{thick-cloud} (purple) atmospheric configurations.
     }
   \end{figure*}

\begin{table}[ht]
\caption{Adopted \textit{true} values for the model parameters. We note that in the \textit{no-cloud} scenario a negligible, but non-zero, amount of aerosols is present to allow the use of log($\tau_{c}$) in the retrieval.}
\label{table:truth}
\begin{center}
 \begin{tabular}{ c  c  c  c } 
 \hline \hline
  Parameter                     &  \textit{No-cloud}  &  \textit{Thin-cloud}  &  \textit{Thick-cloud} \\
 \hline
  $\tau_{c}$                    &   0.05              &   1.0                 &   20.0 \\
  $\Delta_{c}$/H$_g$            &   2                 &   2                   &   2 \\
  $\tau_{c\rightarrow TOA}$     &   $9.1\cdot10^{-3}$ &   $9.1\cdot10^{-3}$   &   $9.1\cdot10^{-3}$ \\
  {$r_{\rm{eff}}$} [$\mu$m]     &   0.50              &   0.50                &   0.50 \\
  $\omega_0$                    &   0.90              &   0.90                &   0.90 \\
  $f_{\rm{CH_4}}$               &   $5\cdot10^{-3}$   &   $5\cdot10^{-3}$     &   $5\cdot10^{-3}$\\
  $R_p/R_{N}$                   &   0.6               &   0.6                 &   0.6 \\
 \hline
\end{tabular}
\end{center}
\end{table}

We include the planet radius as an optional free parameter. 
All these (6+1) properties will be unknown to the observer, with the possible exception of the radius if the planet happens to transit. 
In the particular case of Barnard b, none of these properties is known for the time being.
Hence, the \textit{true} atmospheric configurations are hypothetical, although physically reasonable as described below.
The probability of transit for a randomly oriented orbit is $\sim$ $R_{\star}$/$r$, where $R_{\star}$ is the stellar radius \citep{boruckisummers1984}. 
This amounts to 0.2\% ($R_{\star}$=0.178$R_{\sun}$) for Barnard b. 
The transit probability is small but not negligible, 
and thus it is important to consider both situations in which the planet radius is either known or unknown.

The \textit{true} parameter values of Table \ref{table:truth} are motivated by 
 atmospheric models of the Solar System gas planets. 
For instance, $\tau_{c}$=1.0 is comparable to the optical thickness of 0.5 used by \citet{schmidetal2011} to model a set of spectropolarimetric measurements of Jupiter at wavelengths 520-935 nm.
For our \textit{no-cloud} and \textit{thick-cloud} configurations, we use $\tau_{c}$=0.05 and $\tau_{c}$=20 respectively.
These stand for lower and upper extremes of the values that {$\tau_{c}$} may take, in order to explore the influence of cloud prevalence on exoplanet characterisation.

Following \citet{smith-tomasko1984}, \citet{schmidetal2011} also considered that the cloud is overlaid by a gas layer of optical thickness $\tau_{gas}$=0.011, 
which is consistent with the $\tau_{c\rightarrow TOA}$=$9.1\cdot10^{-3}$ used here. 
A cloud with a geometric extension $\Delta_{c}$=2H$_g$ is rather narrow, a property also expected for the upper clouds of the Solar System gas giants \citep{sanchezlavegaetal2004}. 
Our $r_{\rm{eff}}$=0.5 $\mu$m choice for the effective radius of the aerosols corresponds to moderately small particles. 
\citet{mishchenko1989} examined ground-based spectropolarimetric observations of Jupiter at wavelengths of 423-798 nm. Their best-fit model consisted of tropospheric aerosols of $r_{\rm{eff}}$=0.39$\pm$0.08$\mu$m,on the order of our assumed $r_{\rm{eff}}$.
Sizes of {$r_{\rm{eff}}$}=0.4 $\mu$m were also reported by \citet{morozhenko-yanovitskij1973} for polarization measurements of Jupiter between 373-800 nm, and \citet{perezhoyosetal2012} modeled the tropospheric aerosols of Jupiter with $r_{\rm{eff}}$=0.75 $\mu$m.
Our single-scattering albedo $\omega_0$=0.90 is slightly low compared to e.g. \citet{schmidetal2011}, but within the ranges usually considered in the literature \citep[e.g.][]{satohetal2000, perezhoyosetal2012}.
The methane abundance that we adopt ($f_{\rm{CH_4}}$=$5\cdot10^{-3}$) is comparable to that of Jupiter ($2.1\cdot10^{-3}$) and Saturn ($4.5\cdot10^{-3}$) but somewhat lower than for Uranus or Neptune ($2.4\cdot10^{-2}$ and $3.5\cdot10^{-2}$, respectively) \citep{sanchezlavega2010}.

Our choice for the planetary radius, $R_p/R_{N}$=0.6, corresponds to a density of Barnard b equal to that of Neptune and the mass for an edge-on orbit ($M_p$=$M_{min}/sin(90^\circ)$).
According to \citet{barrosetal2017}, this mass and radius would correspond to a gaseous exoplanet.
Any other orbital inclination would suggest a more massive and probably larger planet size, entailing better conditions for the planet to retain a H$_2$-He atmosphere.
This would also yield more favourable planet-to-star contrast ratios at a given phase angle.
Our choice is therefore somewhat conservative in terms of planet-to-star contrasts.

\begin{table*}[t]
\caption{Priors used for each of the model parameters.} 
\label{table:priors}
\begin{center}
 \begin{tabular}{ c  c  c  c  c  c  c  c } 
 \hline \hline
   & log($R_p/R_{N}$)  &  log($\tau_{c}$)  &  $\Delta_{c}$ [H$_g$]  &  log($\tau_{c\rightarrow TOA}$)  &  $r_{\rm{eff}}$ [$\mu m$]   &  $\omega_0$  &  log($f_{\rm{CH_4}}$) \\
 \hline
  \textbf{Known R$_p$} & $-$ & [-1.30, 1.70] & [1, 8] & [-3.34, 0.13] & [0.10, 10.0] & [0.5, 1.0] & [-5.0, -1.30]  \\
  \textbf{Unknown R$_p$} & [-1.30, 0.70]  &  [-1.30, 1.70] & [1, 8] & [-3.34, 0.13] & [0.10, 10.0] & [0.5, 1.0] & [-5.0, -1.30]  \\
 \hline
\end{tabular}
\end{center}
\end{table*}

\subsection{Noise model} \label{subsec:retrieval_noisemodel}

We produce the \textit{measured} spectra by adding noise to the \textit{true} spectra. 
Our noise model is described by a single parameter, namely the signal-to-noise ratio.
The added noise is given by a normal distribution of zero mean and standard deviation:
\begin{equation} \label{eq:sigma}
\sigma_m= \frac{(F_p/F_{\star})_{max}}{S/N}, 
\end{equation} 
where $(F_p/F_{\star})_{max}$ is the maximum contrast over the filter band for the \textit{true} spectrum, which typically occurs in the continuum near 500 nm. 
As defined here, $\sigma_m$ is wavelength-independent and budgets in a variety of noise sources that are not explicitly considered such as dark current, speckle noise, and leakage of stray starlight \citep{maroisetal2000, wahhajetal2015, robinsonetal2016}. 
S/N can be viewed as the maximum signal-to-noise ratio over the whole measured spectrum. 
In our retrievals, we adopt S/N=10.
Figure \ref{fig:trueVSnoisy_spectra} shows both the \textit{true} (synthetic, without noise) and \textit{measured} (after adding noise) spectra for the three cloud scenarios considered.

Admittedly, our approach to the noise model does not account for all the complexity in the noise of a real observation from a particular telescope. 
Future work will update our noise model to a more realistic one which considers that the S/N of a real observation may deteriorate more in the stronger absorption bands, and that the sensitivity of a real instrument varies with wavelength.
We also note that the angular separation between planet and star will affect the S/N, as positions closer to the IWA of the instrument will result in noisier observations \citep{lacyetal2019}.
Although the final specifications of future space-based coronagraphs are not yet defined, works such as \citet{robinsonetal2016} or \citet{lacyetal2019} have started to consider instrument-specific noise models and their implications.

Further, we note that a simplified noise model allows us to focus on the difficulties in the interpretation of reflected starlight spectra which are fundamental to the radiative transfer problem, sidestepping instrument-specific difficulties.
In this respect, we describe in Section \ref{subsec:results_Rpknown} that our retrievals for a known planetary radius $R_p$ show behaviours similar to those reported in previous works and that used more complex noise models. 
This consistency check confirms that our findings are not critically affected by the adopted noise model.
In other words, our simplified noise model (see Section \ref{subsec:results_Rpunknown}) serves to set theoretical limits to how much information can be retrieved from a measured spectrum.

\subsection{Exploring the parameter space}
\label{subsec:retrieval_exploringparamspace}

Once the \textit{measured} spectrum is simulated for each of the \textit{true} configurations (Table \ref{table:truth}), we address the retrieval.
We assess whether from such an observation we are able to correctly infer the properties of the exoplanet.
This requires a thorough exploration of the parameter space, testing multiple combinations of the parameter vector \textbf{p}, and checking whether they produce spectra similar to the \textit{measured} one.
The vector \textbf{p} includes the atmospheric model parameters (see Table \ref{table:grid}) and, when the planet radius is assumed unconstrained, also $R_p/R_{N}$.

To have a continuous representation of the parameter space, we opt to sample \textbf{p} continuously.
For that, we use the Markov-Chain Monte Carlo sampler \texttt{emcee} \citep{foremanmackeyetal2013}.
This MCMC sampler proposes test points (\textbf{p$_{test}$}) in the multi-dimensional parameter space \textbf{p} and evaluates the likelihood of such a test configuration to have produced the \textit{measured} spectrum.
It does so by comparing the synthetic spectrum corresponding to that \textbf{p$_{test}$} with the \textit{measured} spectrum.
Test spectra are generated through linear interpolation within the pre-computed grid of synthetic spectra (Table \ref{table:grid}).
This approach is much faster than solving the multiple-scattering radiative transfer equation at each test configuration \textbf{p$_{test}$} sampled by \texttt{emcee}.
In \textit{Appendix \ref{sec:appendix_grid_completeness_interpol}} we prove the accuracy of producing the test spectra through interpolation from the grid, compared to computing each of them by solving the RT equation.
Fig. \ref{fig:appendix_interpolation_works} shows that we produce virtually the same spectra in both cases.

When comparing a test spectrum to the \textit{measured} one, we quantify how similar they are through the figure of merit:
\begin{equation}
 \label{eq:chisq}
 \chi^2 =
\sum_{i=1}^N\left(\frac{{F_p}/{F_{\star}}(\alpha,\lambda_i)_{test}-{F_p}/{F_{\star}}(\alpha,\lambda_i)_{measured}}
 {\sigma_m} 
 \right)^2  
\end{equation}
where $F_p/F_{\star}$ is the planet-to-star contrast ratio (Eq. \ref{eq:contrast}) and 
$\sigma_m$ is the standard deviation for the noise (Eq. \ref{eq:sigma}) \citep{bevington-robinson2003}.

Although the sampler tests the whole space of parameters, at the time of proposing a new point to test, those regions of \textbf{p} with lower values of $\chi^2$ will be favoured.
Hence, it will tend to test more frequently such regions since they are considered more likely to have produced the \textit{measured} spectrum and therefore more interesting to study.
This sampling process is carried out simultaneously by several chains (or \textit{walkers}) so that the space of parameters is sampled multiple times independently, to avoid falling in local minima of $\chi^2$.

The limits of the parameter space exploration are set by the box priors described in Table \ref{table:priors}.
We note that in retrieval cases where the radius of the exoplanet is known, R$_p$ is not explored and thus the sampler moves in a 6-dimension parameter space. 
On the other hand, if R$_p$ is considered unknown, the sampler searches in a 7-dimensional parameter space.
As shown in Table \ref{table:priors}, some parameters ($\tau_{c}$, $\tau_{c\rightarrow TOA}$, $f_{\rm{CH_4}}$, $R_p/R_{N}$) are sampled by \texttt{emcee} in logarithmic scale.
We verified that this ensures an optimal performance both during sampling and interpolation.

The MCMC sampling is performed using a total of 500 walkers. 
The ensemble of walkers runs for a maximum of 10$^5$ steps, producing a total set of 5$\cdot$10$^7$ samples.
We apply a convergence criterion to stop the run if the number of iterations surpasses 50 times the autocorrelation time. 
The autocorrelation time provides a measure of the sampler performance and whether it is obtaining independent samples from the space of parameters \citep{goodman-weare2010}. 
With this convergence criterion, we make sure that the sampling is complete enough, regardless of local minima \citep{foremanmackeyetal2013}. 
A burn-in phase is considered, discarding the samples from the first $n$ iterations, where $n$ corresponds to the autocorrelation time.

Other related works have approached the retrieval by sampling the parameter space from a uniform grid instead of randomly choosing the sampling points \citep[e.g.][]{madhusudhan-seager2009,vonparisetal2013, garciamunoz-isaak2015}.
On the other hand, in e.g. \citet{lupuetal2016, nayaketal2017, damiano-hu2019} the sampling points are selected by MCMC or multimodal nesting methods and the synthetic spectrum generated by solving the radiative-transfer equation for that particular configuration.
Compared to interpolating, solving the multiple-scattering problem at each test point \textbf{p$_{test}$} of the sampling results in a much slower performance of the retrieval process.
Hence, our approach combines an MCMC sampling with the efficiency of interpolating from a pre-computed grid of synthetic spectra.

\section{Results}
\label{sec:results}
We will explore here the retrieval results for the three atmospheric scenarios described in Table \ref{table:truth}.
As an initial exercise, we assume that the planet radius $R_p$ is known.
In this case, the MCMC sampler is run as described in Sect. \ref{sec:retrieval} with the six free parameters of the atmospheric model.
Afterwards, we repeat the retrieval analysis but considering that the planet radius is unconstrained.
In the latter case, {$R_p/R_{N}$} will be considered a free parameter and therefore the MCMC sampler will explore a 7D parameter space.
We finally discuss how the retrieval results change when the planetary radius is known to within some degree of uncertainty.

\subsection{Retrieval if $R_p$ is known.}
\label{subsec:results_Rpknown}

\begin{figure}
       \centering
    \includegraphics[width=8.cm]{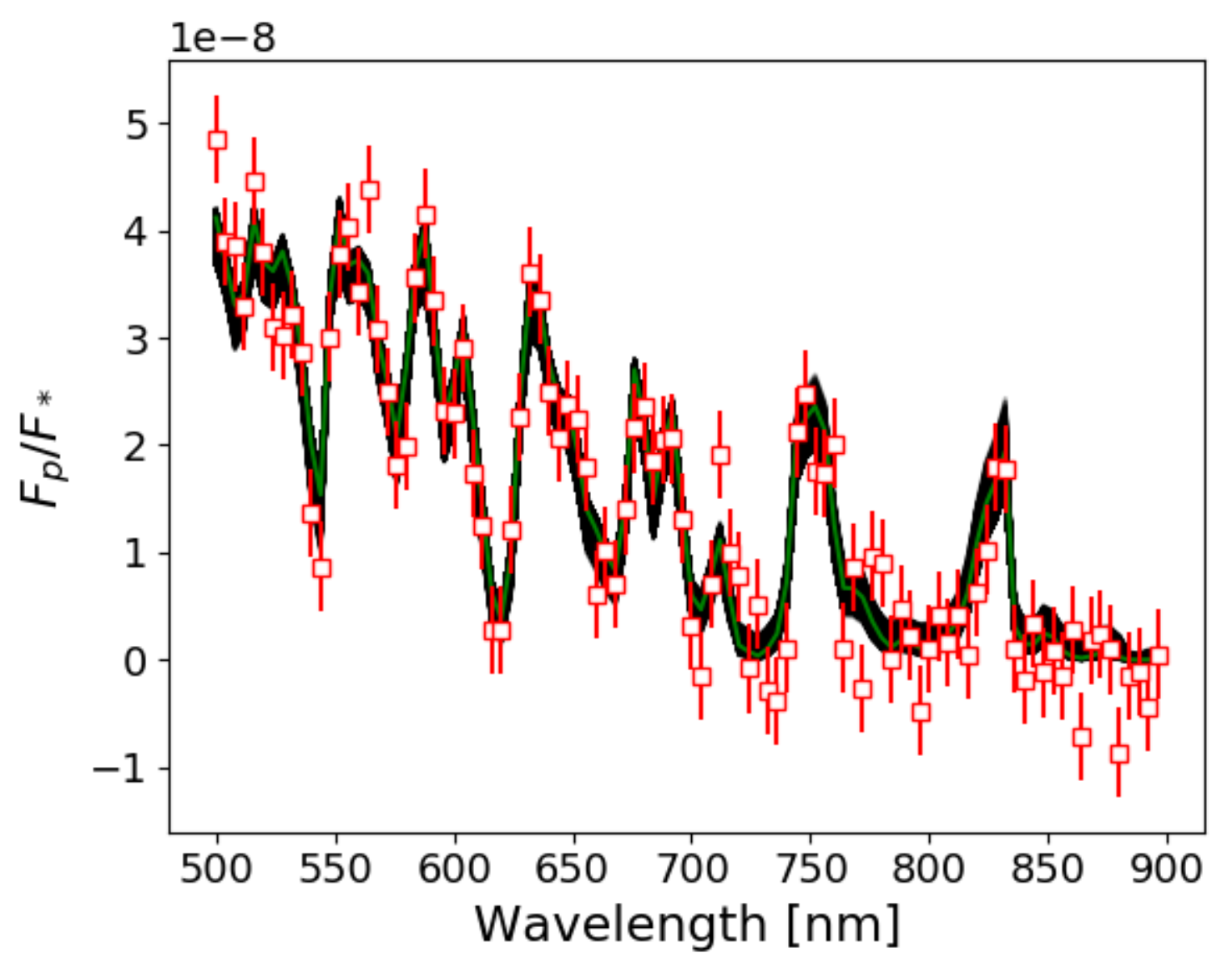}%
        \\
    \includegraphics[width=8.cm]{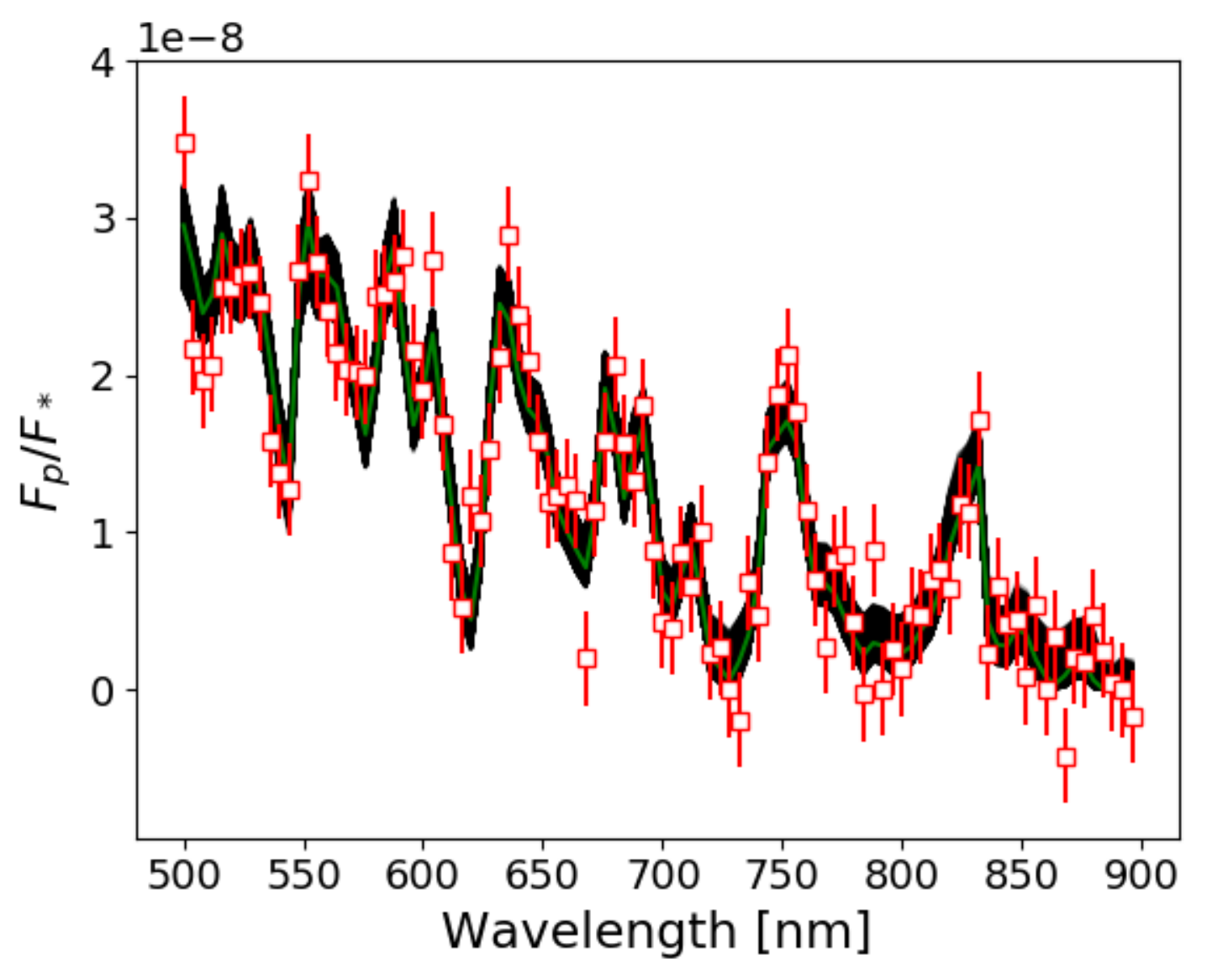}%
    \\
    \includegraphics[width=8.cm]{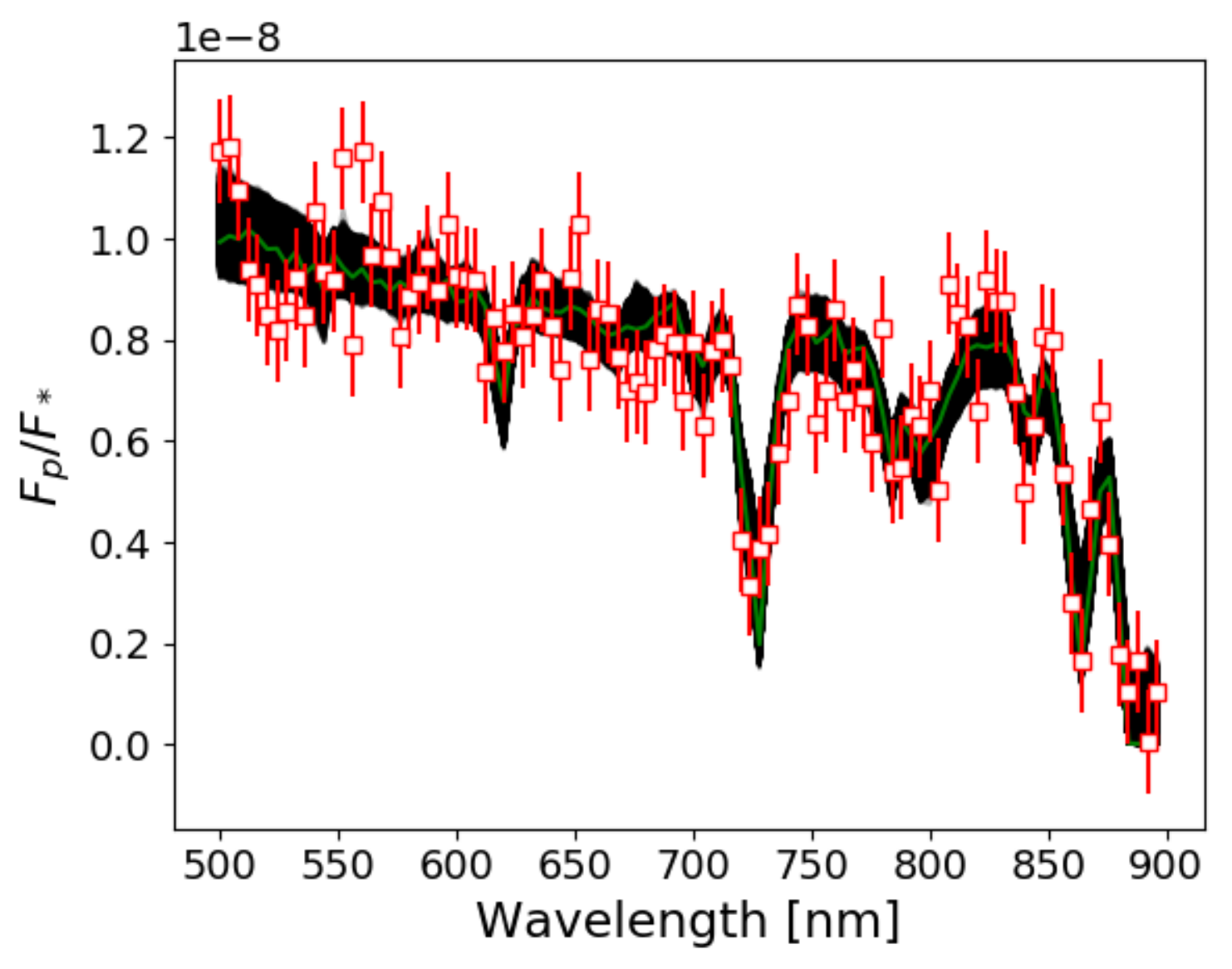}%
    \caption{
    \label{fig:results_Rpknown_goodfits}
    Solid green line: \textit{true} spectrum for the \textit{true} atmospheric configuration.
    Red symbols: simulated \textit{measured} spectrum, with error bars corresponding to S/N=10.
    Solid black lines: Synthetic spectra that meet the condition $\Delta \chi^2 < 15.1$, generated during the MCMC retrieval for a known R$_p$.
    Top: \textit{no-cloud} configuration; middle: \textit{thin-cloud}; bottom: \textit{thick-cloud}.
   }%
\end{figure}

\begin{table*}
\caption{Retrieval results for the case of known $R_p$.} 
\label{table:results_Rpknown_results}
\centering
 \begin{tabular}{ c  c  c  c  c  c  c } 
 \hline \hline
    &  log($\tau_{c}$)  &  $\Delta_{c}$ [$H_g$]  &  log($\tau_{c\rightarrow TOA}$)  &  $r_{\rm{eff}}$ [$\mu m$]  &  $\omega_0$  &  log($f_{\rm{CH_4}}$) \\
  \hline
  \textbf{\textit{No cloud}} & $-0.54^{+0.80}_{-0.52}$  &  $3.03^{+2.25}_{-1.45}$  &  $-1.90^{+1.20}_{-0.96}$  &  $5.04^{+3.38}_{-3.35}$  &  $0.79^{+0.15}_{-0.19}$  &  $-2.32^{+0.45}_{-0.52}$ \\
  True values  &  $-1.30$  &  $2$  &  $-2.04$  &  $0.50$  &  $0.90$  &  $-2.30$  \\
  \hline
  \textbf{\textit{Thin cloud}}  &  $-0.14^{+0.84}_{-0.66}$  &  $3.08^{+2.20}_{-1.46}$  &  $-1.93^{+1.09}_{-0.94}$  &  $4.99^{+3.39}_{-3.34}$  &  $0.76^{+0.16}_{-0.17}$  &  $-2.18^{+0.47}_{-0.70}$  \\
  True values  &  $0.0$  &  $2$  &  $-2.04$  &  $0.50$  &  $0.90$  &  $-2.30$  \\
  \hline
  \textbf{\textit{Thick cloud}}  &  $0.91^{+0.54}_{-0.49}$  &  $3.64^{+2.01}_{-1.81}$  &  $-2.31^{+0.65}_{-0.65}$  &  $4.91^{+3.35}_{-3.13}$  &  $0.72^{+0.09}_{-0.12}$  &  $-3.07^{+0.98}_{-1.14}$  \\
  True values  &  $1.30$  &  $2$  &  $-2.04$  & $0.50$  &  $0.90$  &  $-2.30$  \\
  \hline
\end{tabular}
\end{table*}

\begin{table*}
\caption{Retrieval results for the case of unknown $R_p$.} 
\label{table:results_Rpunknown_results}
\begin{center}
 \begin{tabular}{ c  c  c  c  c  c  c  c } 
 \hline \hline
    &  log($R_p/R_{N}$)  &  log($\tau_{c}$)  &  $\Delta_{c}$ [$H_g$]  &  log($\tau_{c\rightarrow TOA}$)  &  $r_{\rm{eff}}$ [$\mu m$]  &  $\omega_0$  &  log($f_{\rm{CH_4}}$) \\
  \hline
  \textbf{\textit{No cloud}} &  $-0.19^{+0.11}_{-0.10}$  &  $-0.29^{+0.96}_{-0.69}$  &  $3.09^{+2.16}_{-1.49}$  &  $-1.89^{+1.05}_{-0.95}$  &  $5.05^{+3.39}_{-3.42}$  &  $0.75^{+0.17}_{-0.17}$  &  $-2.18^{+0.57}_{-0.83}$ \\
  True values  &  $-0.22$  &  $-1.30$  &  $2$  &  $-2.04$  & $0.50$  &  $0.90$  &  $-2.30$  \\
  \hline
  \textbf{\textit{Thin cloud}}  &  $-0.26^{+0.13}_{-0.10}$  &  $-0.18^{+0.96}_{-0.76}$  &  $3.15^{+2.16}_{-1.52}$  &  $-1.99^{+1.06}_{-0.89}$  &  $5.11^{+3.35}_{-3.45}$  &  $0.75^{+0.17}_{-0.17}$  &  $-2.49^{+0.73}_{-0.95}$  \\
  True values  &  $-0.22$  &  $0.0$  &  $2$  &  $-2.04$  &  $0.50$  &  $0.90$  &  $-2.30$  \\
  \hline
  \textbf{\textit{Thick cloud}}  &  $-0.41^{+0.23}_{-0.13}$  &  $0.35^{+0.88}_{-1.07}$  &  $3.35^{+2.16}_{-1.66}$  &  $-2.21^{+1.00}_{-0.74}$  &  $5.05^{+3.36}_{-3.34}$  &  $0.75^{+0.17}_{-0.17}$  &  $-3.86^{+1.30}_{-0.75}$  \\
  True values  &  $-0.22$  &  $1.30$  &  $2$  &  $-2.04$  &  $0.50$  & $0.90$  &  $-2.30$  \\
  \hline
\end{tabular}
\end{center}
\end{table*}

\begin{figure*}[t]
   \centering
   \includegraphics[width=18.cm]{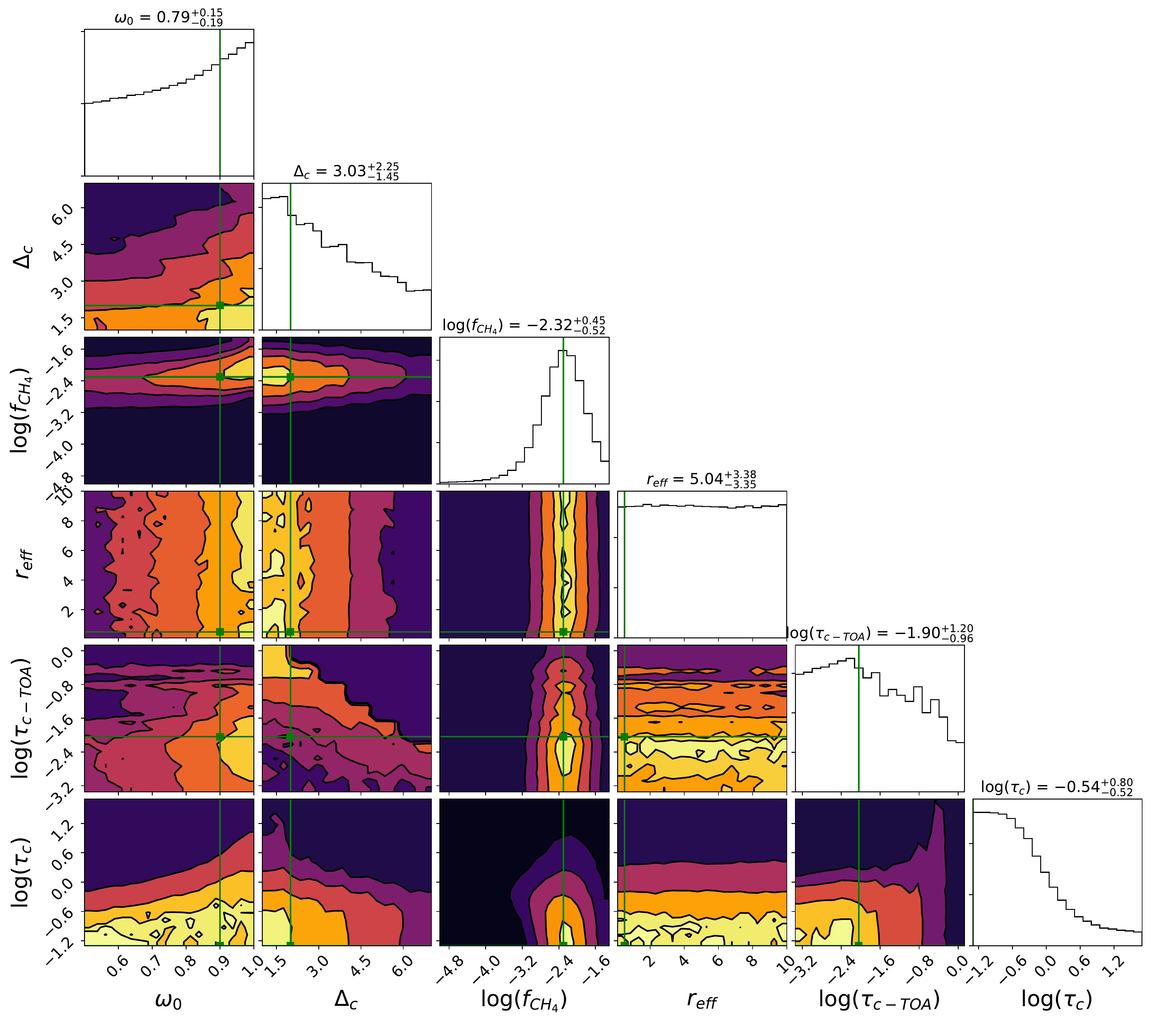}
      \caption{\label{fig:results_rpknown_nocloud_S/N10}
      Posterior probability distributions of the model parameters for a simulated observation of the \textit{no-cloud} atmospheric configuration at S/N=10.
      The planetary radius is assumed known and $R_p/R_{N}$=0.6.
      Green lines mark the \textit{true} values of the model parameters (see Table \ref{table:truth}) for this observation.
      2-D subplots show the correlations between pairs of parameters.
      Contour lines correspond to the 0.5, 1, 1.5, and 2 $\sigma$ levels.
      The median of each parameter's distribution is shown on top of their 1-D probability histogram.
      Upper and lower uncertainties correspond to the 84\% and 16\% quantiles.
      The figures corresponding to the other scenarios discussed in Section \ref{subsec:results_Rpknown} and \ref{subsec:results_Rpunknown} are shown in the Appendix \ref{sec:appendix_retrieval_cornerplots}.
      }
   \end{figure*}

%\label{fig:results_rpknown_thincloud_S/N10}
%\label{fig:results_rpknown_thickcloud_S/N10}

For a signal-to-noise ratio S/N=10, we simulated the \textit{measured} spectra for the \textit{no-cloud}, \textit{thin-cloud} and \textit{thick-cloud} configurations (Table \ref{table:truth}) and run the MCMC sampler.
After discarding the initial burn-in samples, we are left with $\sim$6 million samples in each exploration.
Each sample corresponds to a particular atmospheric configuration with its corresponding
synthetic spectrum and $\chi^2$.

We confirmed that many different atmospheric configurations produce spectra that are nearly identical to the \textit{measured} one.
Figure \ref{fig:results_Rpknown_goodfits} shows all the spectra that meet the condition $\Delta \chi^2 = \chi^2-{\chi^2}_{min} < 15.1$ for each of the cloud scenarios.
For Gaussian distributions, the set of spectra meeting this criterion contains the best-fitting configuration with a 99.99\% probability \citep{pressetal2003}.
Fig. \ref{fig:results_Rpknown_goodfits} also shows the \textit{true} and \textit{measured} spectra for observations at S/N=10 with a known value of $R_p$.
Fig. \ref{fig:results_rpknown_nocloud_S/N10} and Figs. \ref{fig:results_rpknown_thincloud_S/N10}-\ref{fig:results_rpknown_thickcloud_S/N10} show the retrieval results for the three cloud scenarios and, together with Fig. \ref{fig:results_Rpknown_goodfits}, give a sense of the parameter correlations and their effects on the spectra.
The two-dimensional posterior probability distributions in Fig. \ref{fig:results_rpknown_nocloud_S/N10} and Figs. \ref{fig:results_rpknown_thincloud_S/N10}-\ref{fig:results_rpknown_thickcloud_S/N10} identify correlations between each pair of model parameters.
Contour lines indicate the 0.5, 1, 1.5, and 2 $\sigma$ levels that bracket the 12, 39, 68 and 86{\%} confidence intervals, respectively.
The marginalized one-dimensional probability distributions depict the computed likelihood of finding the solution for each parameter in a certain range of values.
For comparison, the green points and lines in the figures mark the \textit{true} parameter values. 
Table \ref{table:results_Rpknown_results} summarizes the retrieval results. 
The quoted values correspond to the median of the marginalized probability distributions.
Upper and lower limits define the intervals within the 84\% and 16\% quantiles, respectively.
This corresponds to the 68\% confidence interval.
The meaning of the confidence intervals is lost in cases where only upper or lower limits can be set on the
model parameters, as described below.

In all cases, for observations at S/N=10, the retrievals can discriminate reasonably well between cloudy and cloud-free atmospheres. 
However, the retrievals do a poorer job at estimating the optical thickness of the cloud, regardless of the \textit{true} {$\tau_{c}$}.
In the \textit{no-cloud} case (Fig. \ref{fig:results_rpknown_nocloud_S/N10}), the probability distribution of log($\tau_{c}$) peaks towards negative values, with an estimated  log($\tau_{c}$)=$-0.54^{+0.80}_{-0.52}$. 
Even if the retrieval overestimates the \textit{true} log({$\tau_{c}$})=$-1.30$, it shows strong evidence of an atmosphere with no cloud or with an extremely thin one. 
Indeed, the probability distribution indicates that the 
values of $\tau_{c}$ quoted from the retrieval must be interpreted as upper limits, and that the quoted uncertainties lose their usual meaning.
In the \textit{thin-cloud} scenario (Fig. \ref{fig:results_rpknown_thincloud_S/N10}) the estimated  log($\tau_{c}$)=$-0.14^{+0.84}_{-0.66}$ is lower than the true one, log($\tau_{c}$)=$0.0$.
However, the marginalized probability distribution of log($\tau_{c}$) suggests the presence of a thin cloud, and tends to rule out both the \textit{no-cloud} and \textit{thick-cloud} configurations.
Finally, the \textit{thick-cloud} configuration (Fig. \ref{fig:results_rpknown_thickcloud_S/N10}) produces a probability distribution for log($\tau_{c}$) that rules out any cloud-free atmospheric configuration.
This sets a sharp lower limit for {$\tau_{c}$}, meaning that in this case the detection of a cloud is evident, although its optical thickness remains poorly constrained.

We find that the retrieved {$f_{\rm{CH_4}}$} differs more from the \textit{true} abundances as the optical thickness of the cloud increases.
We interpret this as a consequence of scattering and absorption within the cloud that interferes with CH$_4$ absorption, thereby modifying the band-to-continuum contrast in the spectra. 
Indeed, the 2D probability distributions show a complex coupling of {$f_{\rm{CH_4}}$} with cloud properties, primarily log($\tau_{c}$) and {$\tau_{c\rightarrow TOA}$}.
In our retrievals, the estimates for both {$\tau_{c}$} and {$\tau_{c\rightarrow TOA}$} tend to deviate from the true values, affecting the retrieved value of {$f_{\rm{CH_4}}$}.
Consistently, our best-fitting estimate for log($f_{\rm{CH_4}}$) is obtained for the \textit{no-cloud} scenario (see Table \ref{table:results_Rpknown_results}).

The location of the cloud top {$\tau_{c\rightarrow TOA}$} is poorly constrained in the \textit{no-cloud} and \textit{thin-cloud} cases.
This is not surprising as the impact of a cloud on the spectral appearance in these cases is weak or moderate. 
In the \textit{thick-cloud} scenario (see Fig. \ref{fig:results_rpknown_thickcloud_S/N10}), the marginalized probability of {$\tau_{c\rightarrow TOA}$} tends to rule out configurations in which the cloud is located either very deep or very high.
The effects of having such an optically-thick cloud at those positions would be apparent in the spectrum.
For instance, a deep cloud is concealed by the gas in the upper atmospheric layers.
Hence, the resulting spectrum would be effectively equivalent to that of a thin-cloud or even a cloud-free configuration.
As discussed above, the retrieval at S/N=10 tends to discard such scenarios for the \textit{thick-cloud} configuration.
In turn, a high-altitude cloud entails that CH$_4$ absorption diminishes to levels inconsistent with the \textit{measured} spectrum, a situation that is also easily identified.
Regarding the geometrical extension of the cloud {$\Delta_{c}$}, it is virtually unconstrained in all cases for this S/N.

The single-scattering albedo of the aerosols {$\omega_0$} is systematically underestimated in the retrievals, especially for the \textit{thick-cloud} case.
This has implications on prospective efforts to identify the nature of the cloud particles through their refractive index. 
Furthermore, the inferred {$\omega_0$} shows strong correlations with both {$\tau_{c}$} and {$r_{\rm{eff}}$}. 
The {$\omega_0$}--{$r_{\rm{eff}}$} correlation is explained by the fact that {$\omega_0$} and the scattering phase function $p$($\theta$) multiply each other in the radiative transfer equation. 
Indeed, in the limit of single scattering, the amount of scattered light is proportional to {$\omega_0$}$\cdot p$($\theta$). 
The fact that $p$($\theta$) is dependent on {$r_{\rm{eff}}$} explains the two branches observed in the 2D probability distribution for {$\omega_0$}--{$r_{\rm{eff}}$} (see Fig. \ref{fig:results_rpknown_thickcloud_S/N10}).
Figure \ref{fig:Miescattering} shows that for $\theta$=180$^\circ$ (backward scattering, 
dominating at $\alpha$=0$^\circ$), $p(\theta)$ presents a minimum around {$r_{\rm{eff}}$}=0.50 $\mu$m.
Thus, in this viewing geometry, a similar amount of photons would be reflected for any other {$r_{\rm{eff}}$} (which means that $p$($\theta$=180$^{\circ}$) increases) if {$\omega_0$} is reduced. 
This creates numerous possibilities for reproducing the \textit{measured} spectrum with {$\omega_0$} values smaller than the \textit{true} one.
This degeneracy depends on the shape of the scattering phase function of the aerosols, whose value changes with the viewing geometry.
In future work we will explore the prospects for breaking this degeneracy with observations at multiple phases.

The {$\omega_0$}--{$\tau_{c}$} correlation reveals the impact that cloud absorption has on the spectrum.
In the \textit{thick-cloud} scenario, such a thick cloud (estimated log($\tau_{c}$)=$0.91^{+0.54}_{-0.49}$), with aerosols of estimated $\omega_0$=$0.72^{+0.09}_{-0.12}$, absorbs a certain amount of incoming photons.
This absorption shapes the reflected-light spectrum, specially in the continuum. 
However, a similar spectral shape may result from an optically thinner cloud composed of darker aerosols (see Fig. \ref{fig:results_rpknown_thickcloud_S/N10}).
In other words, both parameters can compensate each other and yield a similar amount of continuum absorption.
The same idea applies to the \textit{thin-cloud} case (Fig. {\ref{fig:results_rpknown_thincloud_S/N10}}), although the correlation there is not so strong due to the smaller optical thickness.

Our retrieval results for a known value of R$_p$ are consistent with previous literature.
Although the model setups differ between the different studies we observe that, for those parameters which are comparable, our results behave similarly.
For instance, analogous correlations between gaseous-species absorption and cloud pressure level were discussed by e.g. \citet{heidinger-stephens2000, lupuetal2016, damiano-hu2019}.
Correlations between cloud properties (scattering and absorption properties; optical thickness and pressure level) were also observed in e.g. \citet{heidinger-stephens2000, lupuetal2016}. 
As the noise models in all those cases differ, these similarities drive us to conclude, on the one hand, that our noise model approach is not blurring the physical fundamentals of the exercise.
On the other hand, this ensures that our results for a known R$_p$ are well-founded in order to move on and compare them to the case of an unknown R$_p$.

\subsection{Retrieval if $R_p$ is unknown.}
\label{subsec:results_Rpunknown}

\begin{figure}
       \centering
    \includegraphics[width=8.cm]{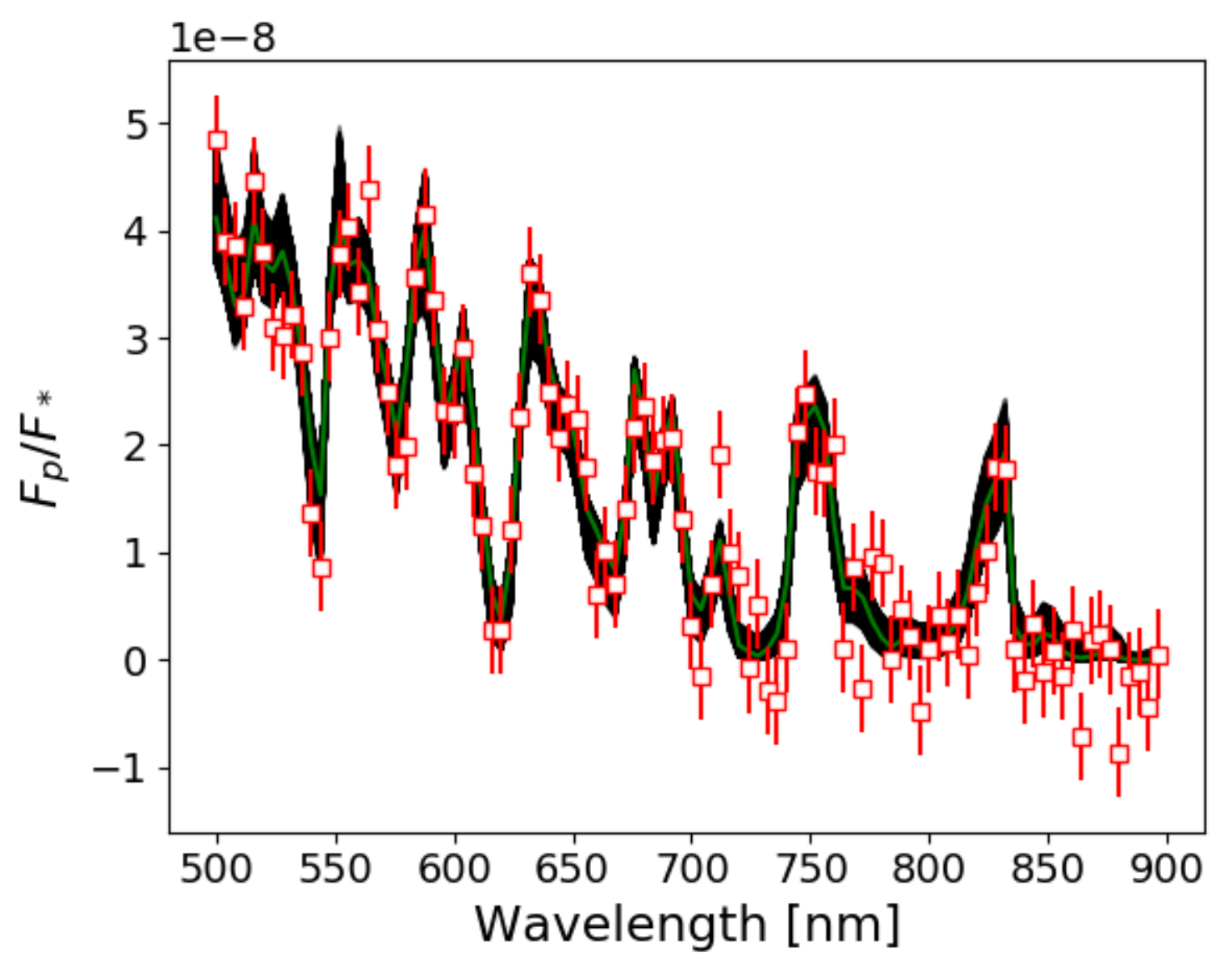}%
    \\
    \includegraphics[width=8.cm]{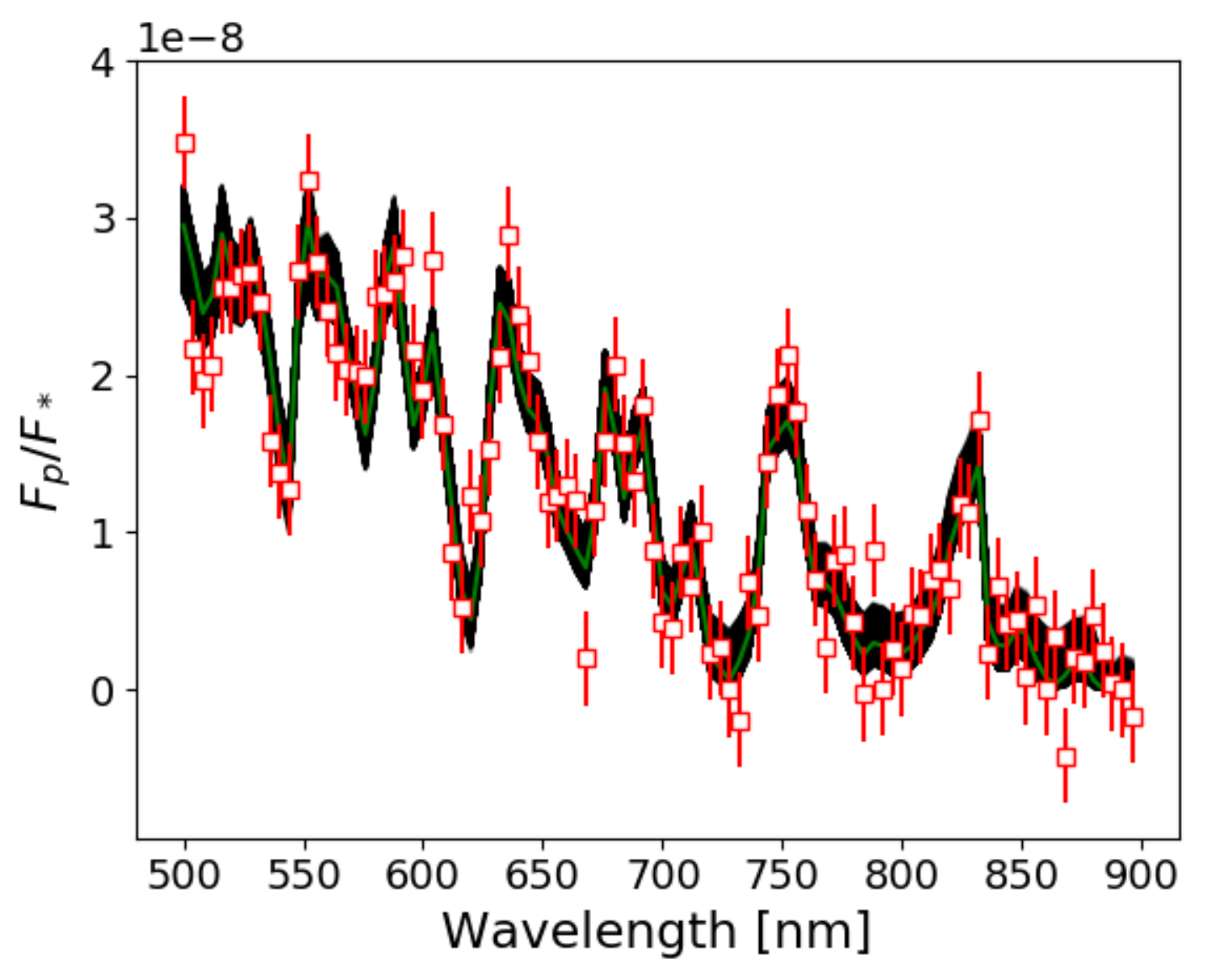}%
    \\
    \includegraphics[width=8.cm]{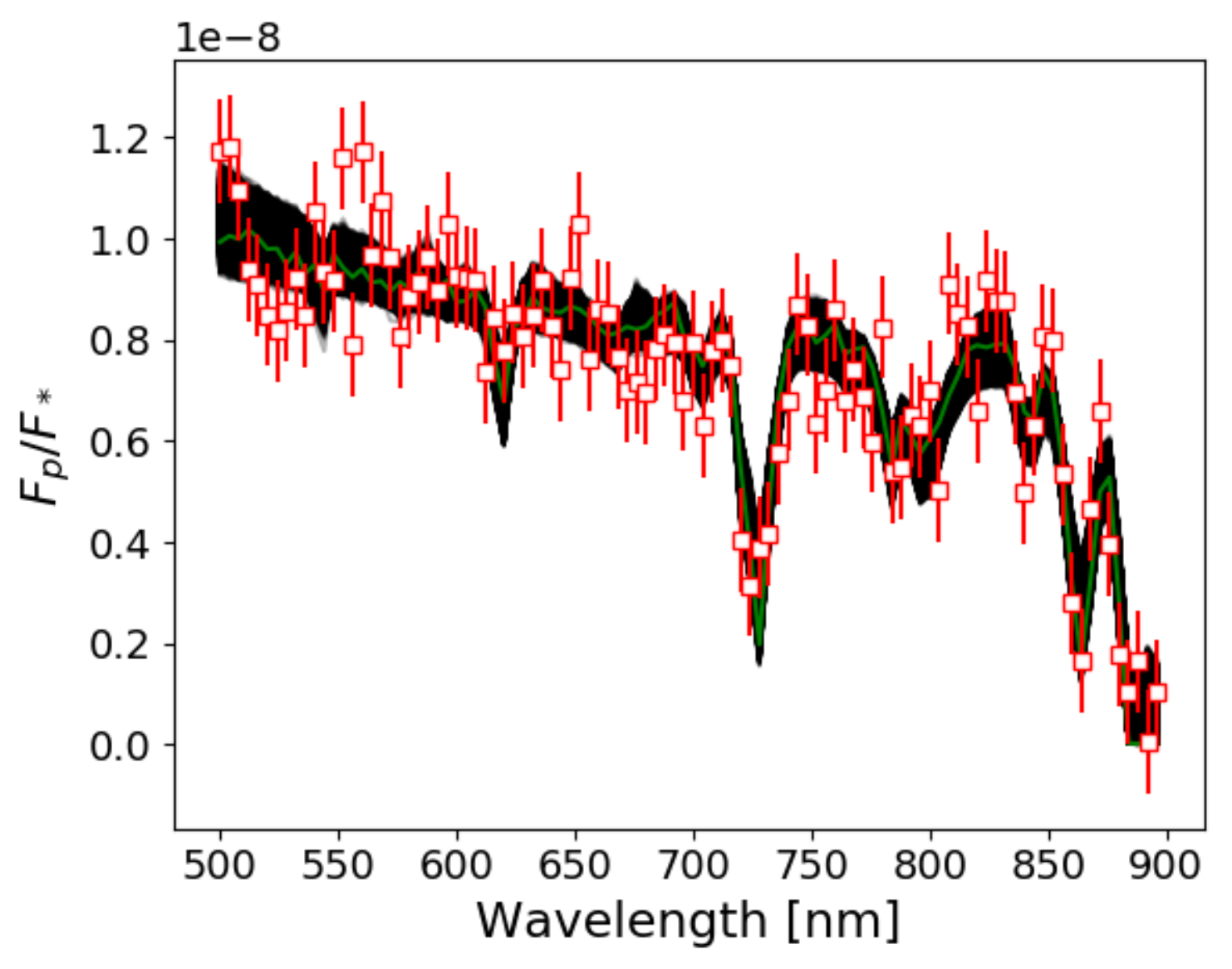}%
    \caption{\label{fig:results_Rpunknown_goodfits} As Fig. \ref{fig:results_Rpknown_goodfits}, but for retrievals where the planetary radius is unknown. Top: \textit{no-cloud} configuration; middle: \textit{thin-cloud}; bottom: \textit{thick-cloud}.
    }%
    \end{figure}
    
%\label{fig:results_rpunknown_nocloud_S/N10}
%\label{fig:results_rpunknown_thincloud_S/N10}
%\label{fig:results_rpunknown_thickcloud_S/N10}

We repeated the previous study but now including the planetary radius {$R_p/R_{N}$} as an additional free parameter to be explored with the MCMC sampler.
The priors for this parameter (see Table \ref{table:priors}) mean that the value of R$_p$ is completely unconstrained.
Again, in this exercise, we adopt a signal-to-noise ratio S/N=10.
We display in Fig. \ref{fig:results_Rpunknown_goodfits} the spectra meeting the $\Delta \chi^2 < 15.1$ criterion for the three cloud cases. 
The posterior probability distributions are shown in Figures \ref{fig:results_rpunknown_nocloud_S/N10}--\ref{fig:results_rpunknown_thickcloud_S/N10} for the \textit{no-cloud}, \textit{thin-cloud} and \textit{thick-cloud} scenarios, respectively.
The corresponding retrieval results are summarized in Table \ref{table:results_Rpunknown_results}.

Interestingly,  it is feasible to constrain reasonably well the planet radius for all cloud scenarios.
The estimates of {$R_p/R_{N}$} for the \textit{no-cloud} (Fig. \ref{fig:results_rpunknown_nocloud_S/N10}) and  \textit{thin-cloud} configurations (Fig. \ref{fig:results_rpunknown_thincloud_S/N10}) are in agreement with the true value $R_p/R_{N}$=0.6, even at a S/N=10 (see Table \ref{table:results_Rpunknown_results}).
The upper and lower boundaries of these estimates allow us to constrain {$R_p/R_{N}$} to within less than a factor of 2 from the true value.
Namely, for the \textit{no-cloud} atmosphere the confidence interval of $R_p/R_{N}$ ranges between 0.51 and 0.83.
This compares well with the true $R_p/R_{N}$=0.6.
In the \textit{thin-cloud} case, {$R_p/R_{N}$} is constrained between 0.44 and 0.74.
However, if the atmosphere contains a thick cloud (Fig. \ref{fig:results_rpunknown_thickcloud_S/N10}), the retrieved {$R_p/R_{N}$} is significantly less accurate. 
Here, the median of the probability distribution (log($R_p/R_{N}$)=$-0.41$) deviates more from the true value (log($R_p/R_{N}$)=$-0.22$ than in the other cloud scenarios. 
Still, in this \textit{thick-cloud} case the planet radius is constrained to within a factor of $\sim$2 (ranging from $R_p/R_{N}$=0.29 to 0.66).

Our finding that it is possible to constrain $R_p$ to within a factor of 2 is consistent with the results of \citet{nayaketal2017}, who studied the correlations between phase angle $\alpha$ and R$_p$, with both parameters unknown \textit{a priori}.
Their study, however, did not focus on analysing which new correlations between atmospheric parameters are generated by including R$_p$ as a free parameter.
By comparing fixed-R$_p$ and free-R$_p$ retrievals, we show that there are indeed important degeneracies involved between R$_p$ and atmospheric parameters such as $\tau_{c}$.
Here we prove that constraining R$_p$ to within a factor of 2 is a result that holds valid regardless of the presence or absence of clouds or their optical thickness.
As the cloud coverage of an exoplanet will not be known beforehand, this sets an encouraging perspective to estimate the radius of non-transiting exoplanets.

Detecting the presence or absence of clouds is not evident in any case if the planet radius is unknown. 
For a \textit{no-cloud} atmosphere, the estimated log($\tau_{c}$) value is $-0.29^{+0.96}_{-0.69}$ (Table \ref{table:results_Rpunknown_results}), larger than the result obtained for a known R$_p$ ($-0.54^{+0.80}_{-0.52}$, see Table \ref{table:results_Rpknown_results}).
We note in the upper error of these results that the confidence interval in case of an unknown R$_p$ reaches higher values of {$\tau_{c}$}.
Graphically, this is shown in the 1D probability distribution of log($\tau_{c}$) from Fig. \ref{fig:results_rpunknown_nocloud_S/N10}, in comparison with that in Fig. \ref{fig:results_rpknown_nocloud_S/N10} for a known R$_p$.
The log($\tau_{c}$) probability distribution for the \textit{thin-cloud} scenario is also broader if R$_p$ is unknown.
Indeed, the marginalized probability distribution (see Fig. \ref{fig:results_rpunknown_thincloud_S/N10}) shows in this case the same likelihood for a cloud with optical thickness log($\tau_{c}$)=0.0 (the true value) and for a log($\tau_{c}$)=$-1.2$ (that effectively corresponds to a cloud-free atmosphere).
Both in Fig. \ref{fig:results_rpunknown_thincloud_S/N10} and Fig. \ref{fig:results_rpunknown_thickcloud_S/N10} we observe practically equivalent upper limits for the value of $\tau_{c}$.
Thus, the \textit{no-cloud} and \textit{thin-cloud} scenarios would turn indistinguishable if $R_p$ is unknown.

The change in the log($\tau_{c}$) probability distribution with respect to the case of known R$_p$ is more dramatic in the \textit{thick-cloud} retrieval (Fig. \ref{fig:results_rpunknown_thickcloud_S/N10}).
Here, the 1D probability distribution becomes almost flat and its median presents wide upper and lower errors ($0.35^{+0.88}_{-1.07}$).
Hence, the log($\tau_{c}$) posterior distribution reveals comparable probabilities of having observed a planet with thick clouds or with a cloud-free atmosphere.
This is in great contrast with the retrieval for a known R$_p$, in which case the \textit{thick-cloud} retrieval shows an unambiguous cloud detection (Fig. \ref{fig:results_rpknown_thickcloud_S/N10}). 

The methane abundance is also worse constrained when R$_p$ is unknown, with larger ranges of possible values consistent with the \textit{measured} spectrum (Table \ref{table:results_Rpunknown_results}).
As was the case for known R$_p$, the \textit{thick-cloud} scenario shows the poorest {$f_{\rm{CH_4}}$} estimates (Fig. \ref{fig:results_rpunknown_thickcloud_S/N10}). 
In this case, {$f_{\rm{CH_4}}$} is strongly underestimated, demonstrating again the physical degeneracy between {$f_{\rm{CH_4}}$} and {$\tau_{c}$}, which is also underestimated in this \textit{thick-cloud} retrieval.

This degeneracy also affects the \textit{no-cloud} and \textit{thin-cloud} retrievals (Figures \ref{fig:results_rpunknown_nocloud_S/N10} and \ref{fig:results_rpunknown_thincloud_S/N10}).
We note that in both cases the 1D posterior probability distributions of {$\tau_{c}$} are very similar.
This is a consequence of the R$_p$ uncertainties, which have a great effect on the spectrum.
Compared to such a strong influence of $R_p/R_{N}$, the impact on the spectrum of having a cloud with $\tau_{c}$=0.05 or $\tau_{c}$=1.0 becomes negligible. 
This makes it challenging to distinguish between a \textit{no-cloud} and a \textit{thin-cloud} atmosphere.
Nevertheless, the deviation of these 1D posterior distributions with respect to the true values is much larger in the \textit{no-cloud} scenario.
Overestimating {$\tau_{c}$}, therefore, produces an increase in the resulting estimate of {$f_{\rm{CH_4}}$}.
That is the reason why, if R$_p$ is unconstrained, the \textit{no-cloud} atmospheric configurations do not necessarily yield the best estimates of $f_{\rm{CH_4}}$.

\begin{figure}
   \centering
   \includegraphics[width=8.5cm]{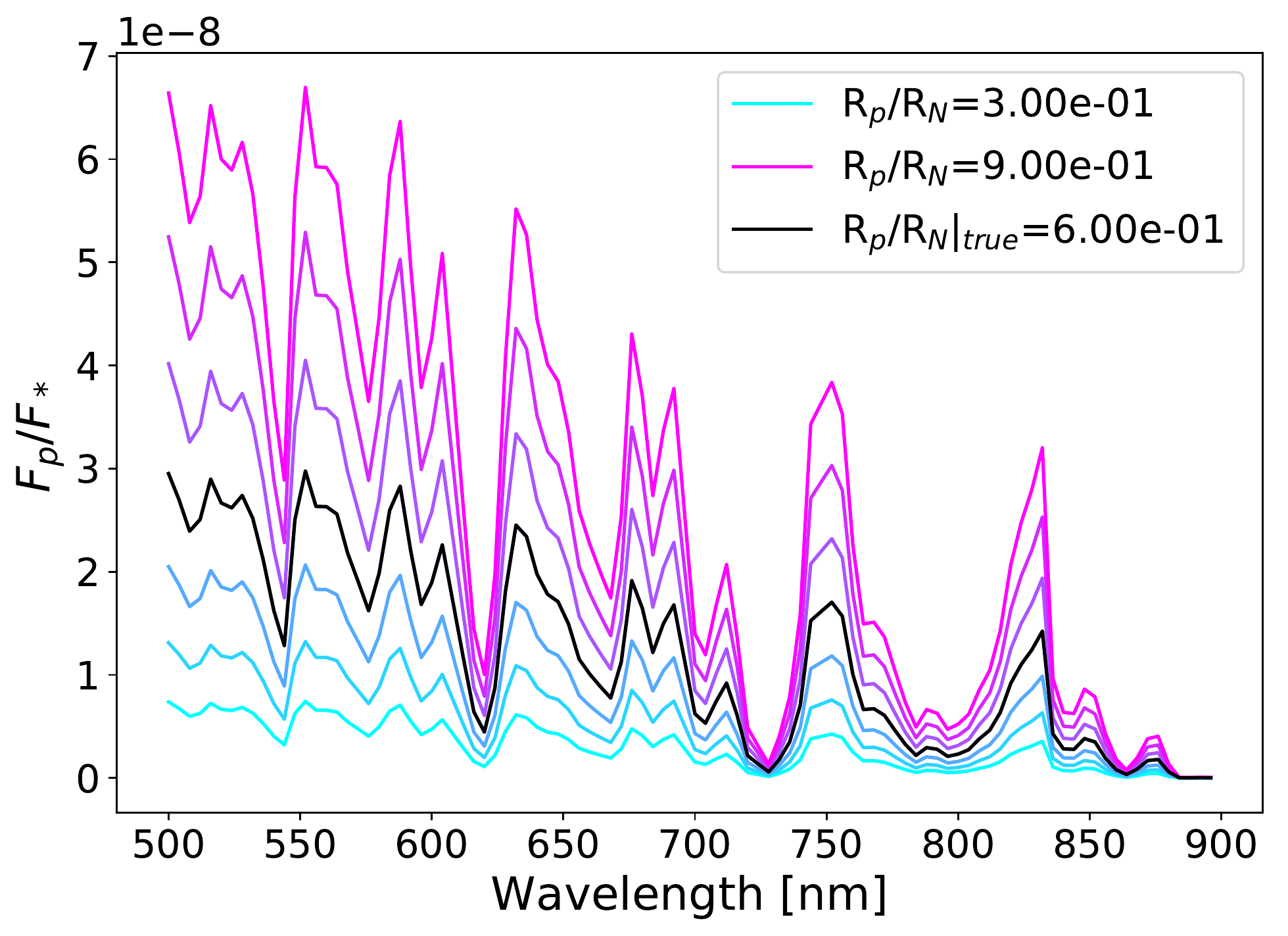}%
      \caption{\label{fig:completeness_grid_Rp}
      Black line: Synthetic spectrum for the \textit{thin-cloud} configuration (Table \ref{table:truth}).
    Colour lines: synthetic spectra for the configurations that result from perturbing $R_p/R_{N}$.
    The colour code is described in the legend for lowermost and uppermost values.
    Intermediate colours span over the values $R_p/R_{N}$=\{0.40, 0.50, 0.60, 0.70, 0.80\}.
    This corresponds to a radius uncertainty $\Delta R_p$=33\%.
    }%
   \end{figure}

The rest of properties of the cloud ($\tau_{c\rightarrow TOA}$, $\Delta_{c}$) and its aerosols ($r_{\rm{eff}}$, $\omega_0$) are generally unconstrained in all three scenarios.
Figure \ref{fig:completeness_variations_all_param} shows, for the \textit{thin-cloud} atmosphere, the effects that modifying each of the atmospheric model parameters produces on the reflected-light spectrum.
Fig. \ref{fig:completeness_grid_Rp} repeats this exercise for {$R_p/R_{N}$}.
From the comparison of both figures, we observe that $F_p$/$F_{\star}$ undergoes larger variations when {$R_p/R_{N}$} is modified than when the other parameters are modified.
This suggests that introducing {$R_p/R_{N}$} as a free parameter makes this parameter a dominating source of uncertainty in the retrieval.

\subsection{Intermediate uncertainties in R$_p$} \label{subsec:results_intermediateRpuncertainties}

\begin{table*}[t]
\caption{Retrieval results for different levels of uncertainty in the value of $R_p/R_{N}$. We note that in the cases with no uncertainty ($\Delta R_p$=$\pm$0\%), {$R_p/R_{N}$} was indeed fixed and was not explored by the MCMC retrieval sampler.} 
\label{table:results_Rpunknown_different_uncertainties}
\begin{center}
 \begin{tabular}{ c  c  c  c  c  c  c  c  c } 
 \hline \hline
   & $\Delta R_p$ &  log($R_p/R_{N}$)  &  log($\tau_{c}$)  &  $\Delta_{c}$ [$H_g$]  &  log($\tau_{c\rightarrow TOA}$)  &  $r_{\rm{eff}}$ [$\mu m$]  &  $\omega_0$  &  log($f_{\rm{CH_4}}$) \\
  \hline
  \multirow{7}{*}{\rotatebox{90}{\textit{\textbf{No-cloud}}}} & Unconstrained &  $-0.19^{+0.11}_{-0.10}$  &  $-0.29^{+0.96}_{-0.69}$  &  $3.09^{+2.16}_{-1.49}$  &  $-1.89^{+1.05}_{-0.95}$  &  $5.05^{+3.39}_{-3.42}$  &  $0.75^{+0.17}_{-0.17}$  &  $-2.18^{+0.57}_{-0.83}$ \\
  & $\pm$50\% &  $-0.20^{+0.08}_{-0.09}$  &  $-0.40^{+0.87}_{-0.61}$  &  $3.06^{+2.23}_{-1.46}$  &  $-1.90^{+1.11}_{-0.96}$  &  $5.09^{+3.34}_{-3.40}$  &  $0.77^{+0.16}_{-0.18}$  &  $-2.27^{+0.58}_{-0.80}$ \\
  & $\pm$25\% &  $-0.21^{+0.06}_{-0.07}$  &  $-0.49^{+0.84}_{-0.56}$  &  $3.00^{+2.25}_{-1.42}$  &  $-1.88^{+1.15}_{-0.98}$  &  $5.08^{+3.36}_{-3.39}$  &  $0.79^{+0.15}_{-0.19}$  &  $-2.29^{+0.53}_{-0.70}$ \\
  & $\pm$10\% &  $-0.22^{+0.03}_{-0.03}$  &  $-0.54^{+0.81}_{-0.52}$  &  $3.01^{+2.26}_{-1.42}$  &  $-1.90^{+1.19}_{-0.96}$  &  $5.03^{+3.37}_{-3.35}$  &  $0.79^{+0.15}_{-0.19}$  &  $-2.32^{+0.47}_{-0.56}$ \\
  & $\pm$1.6\% &  $-0.22^{+0.00}_{-0.00}$  &  $-0.51^{+0.82}_{-0.54}$  &  $3.02^{+2.27}_{-1.43}$  &  $-1.91^{+1.19}_{-0.96}$  &  $5.04^{+3.36}_{-3.36}$  &  $0.79^{+0.15}_{-0.19}$  &  $-2.36^{+0.47}_{-0.54}$ \\
  & $\pm$0\% &  $-$  & $-0.54^{+0.80}_{-0.52}$  &  $3.03^{+2.25}_{-1.45}$  &  $-1.90^{+1.20}_{-0.96}$  &  $5.04^{+3.38}_{-3.35}$  &  $0.79^{+0.15}_{-0.19}$  &  $-2.32^{+0.45}_{-0.52}$ \\
  & True values  &  $-0.22$  &  $-1.30$  &  $2$  &  $-2.04$  &  $0.50$  &  $0.90$  &  $-2.30$  \\
  \hline
  \multirow{7}{*}{\rotatebox{90}{\textit{\textbf{Thin-cloud}}}} & Unconstrained  &  $-0.26^{+0.13}_{-0.10}$  &  $-0.18^{+0.96}_{-0.76}$  &  $3.15^{+2.16}_{-1.52}$  &  $-1.99^{+1.06}_{-0.89}$  &  $5.11^{+3.35}_{-3.45}$  &  $0.75^{+0.17}_{-0.17}$  &  $-2.49^{+0.73}_{-0.95}$  \\
  & $\pm$50\% &  $-0.26^{+0.11}_{-0.10}$  &  $-0.22^{+0.92}_{-0.73}$  &  $3.11^{+2.18}_{-1.49}$  &  $-1.94^{+1.07}_{-0.93}$  &  $4.99^{+3.39}_{-3.33}$  &  $0.76^{+0.16}_{-0.18}$  &  $-2.55^{+0.73}_{-0.96}$ \\
  & $\pm$25\% &  $-0.25^{+0.07}_{-0.06}$  &  $-0.23^{+0.86}_{-0.68}$  &  $3.11^{+2.20}_{-1.49}$  &  $-1.95^{+1.11}_{-0.93}$  &  $5.00^{+3.40}_{-3.34}$  &  $0.77^{+0.16}_{-0.18}$  &  $-2.40^{+0.60}_{-0.74}$ \\
  & $\pm$10\% &  $-0.23^{+0.03}_{-0.03}$  &  $-0.17^{+0.84}_{-0.67}$  &  $3.09^{+2.18}_{-1.46}$  &  $-1.94^{+1.08}_{-0.94}$  &  $5.02^{+3.38}_{-3.36}$  &  $0.76^{+0.16}_{-0.17}$  &  $-2.23^{+0.50}_{-0.69}$ \\
  & $\pm$1.6\% &  $-0.22^{+0.01}_{-0.00}$  &  $-0.20^{+0.86}_{-0.66}$  &  $3.06^{+2.18}_{-1.44}$  &  $-1.88^{+1.07}_{-0.97}$  &  $5.08^{+3.35}_{-3.38}$  &  $0.76^{+0.16}_{-0.17}$  &  $-2.06^{+0.42}_{-0.61}$ \\
  & $\pm$0\% &  $-$  &  $-0.14^{+0.84}_{-0.66}$  &  $3.08^{+2.20}_{-1.46}$  &  $-1.93^{+1.09}_{-0.94}$  &  $4.99^{+3.39}_{-3.34}$  &  $0.76^{+0.16}_{-0.17}$  &  $-2.18^{+0.47}_{-0.70}$  \\
  & True values  &  $-0.22$  &  $0.0$  &  $2$  &  $-2.04$  &  $0.50$  &  $0.90$  &  $-2.30$  \\
  \hline\hline
  \multirow{7}{*}{\rotatebox{90}{\textit{\textbf{Thick-cloud}}}} & Unconstrained  &  $-0.41^{+0.23}_{-0.13}$  &  $0.35^{+0.88}_{-1.07}$  &  $3.35^{+2.16}_{-1.66}$  &  $-2.21^{+1.00}_{-0.74}$  &  $5.05^{+3.36}_{-3.34}$  &  $0.75^{+0.17}_{-0.17}$  &  $-3.86^{+1.30}_{-0.75}$  \\
  & $\pm$50\% &  $-0.36^{+0.18}_{-0.13}$  &  $0.53^{+0.77}_{-0.83}$  &  $3.35^{+2.15}_{-1.65}$  &  $-2.21^{+0.93}_{-0.73}$  &  $4.95^{+3.40}_{-3.25}$  &  $0.74^{+0.16}_{-0.16}$  &  $-3.53^{+1.22}_{-0.87}$ \\
  & $\pm$25\% &  $-0.24^{+0.08}_{-0.07}$  &  $0.86^{+0.56}_{-0.51}$  &  $3.58^{+2.03}_{-1.77}$  &  $-2.30^{+0.70}_{-0.66}$  &  $4.89^{+3.39}_{-3.13}$  &  $0.72^{+0.12}_{-0.13}$  &  $-3.17^{+1.04}_{-1.10}$ \\
  & $\pm$10\% &  $-0.23^{+0.03}_{-0.03}$  &  $0.91^{+0.53}_{-0.50}$  &  $3.59^{+2.02}_{-1.77}$  &  $-2.31^{+0.67}_{-0.65}$  &  $4.87^{+3.39}_{-3.10}$  &  $0.72^{+0.09}_{-0.12}$  &  $-3.09^{+0.99}_{-1.13}$ \\
  & $\pm$1.6\% &  $-0.22^{+0.00}_{-0.00}$  &  $0.91^{+0.53}_{-0.50}$  &  $3.63^{+2.02}_{-1.81}$  &  $-2.31^{+0.65}_{-0.65}$  &  $4.82^{+3.41}_{-3.07}$  &  $0.73^{+0.09}_{-0.12}$  &  $-3.04^{+0.96}_{-1.14}$ \\
  & $\pm$0\% &  $-$  &  $0.91^{+0.54}_{-0.49}$  &  $3.64^{+2.01}_{-1.81}$  &  $-2.31^{+0.65}_{-0.65}$  &  $4.91^{+3.35}_{-3.13}$  &  $0.72^{+0.09}_{-0.12}$  &  $-3.07^{+0.98}_{-1.14}$  \\
  & True values  &  $-0.22$  &  $1.30$  &  $2$  &  $-2.04$  &  $0.50$  &  $0.90$  &  $-2.30$  \\
  \hline
\end{tabular}
\end{center}
\end{table*}

\begin{figure*}
       \centering
    \includegraphics[width=5.cm]{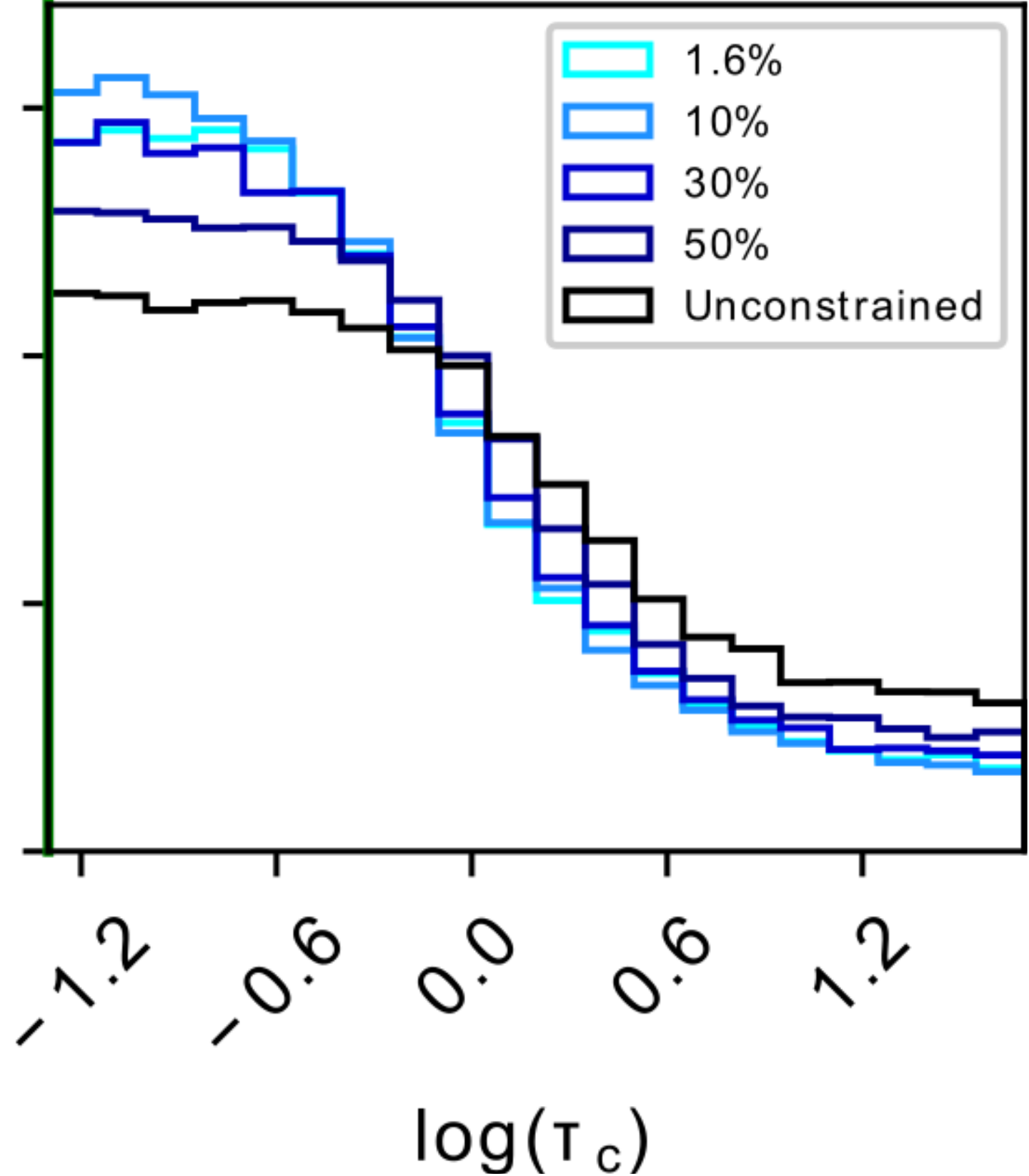}%
    \includegraphics[width=5.cm]{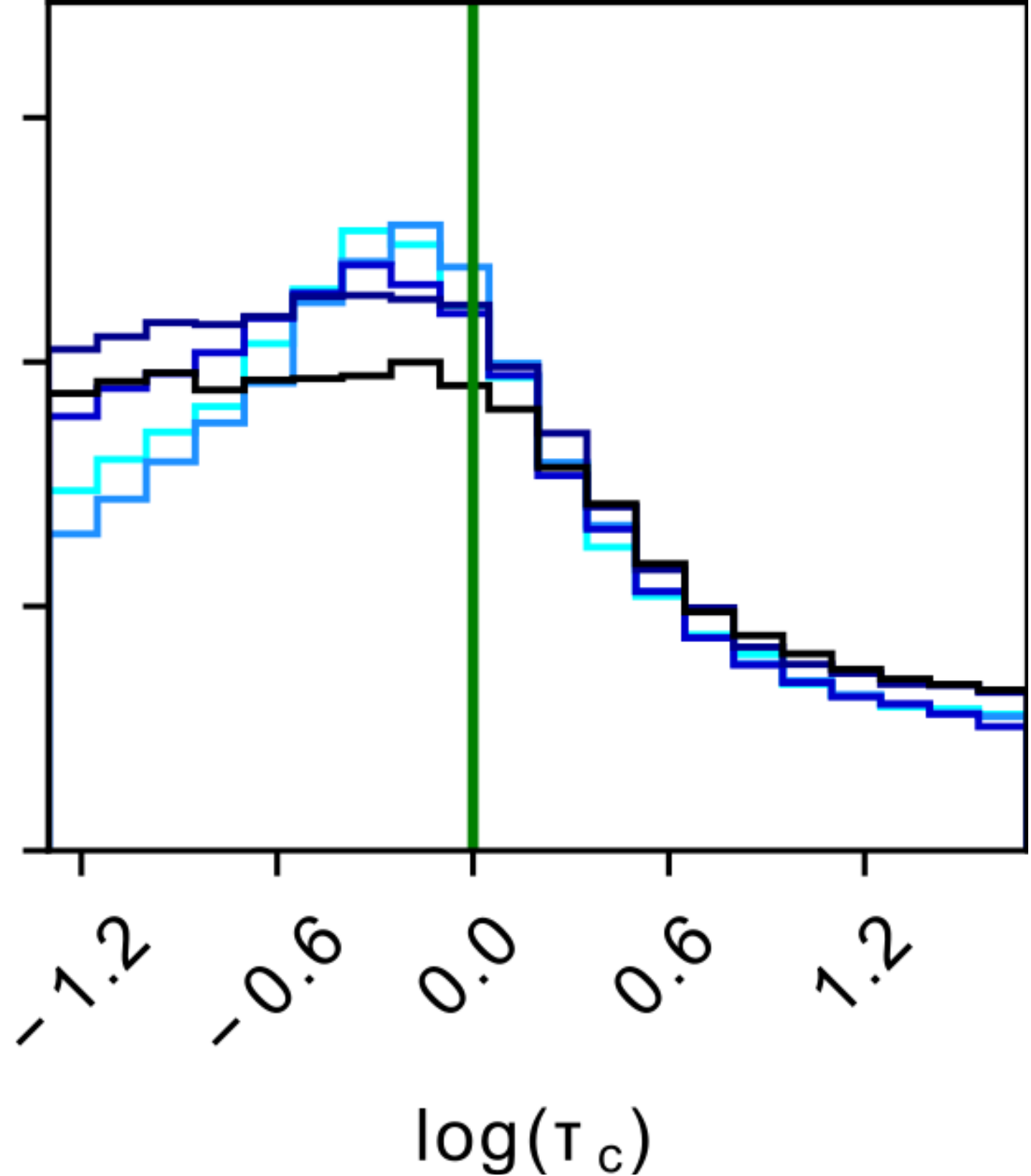}%
    \includegraphics[width=5.cm]{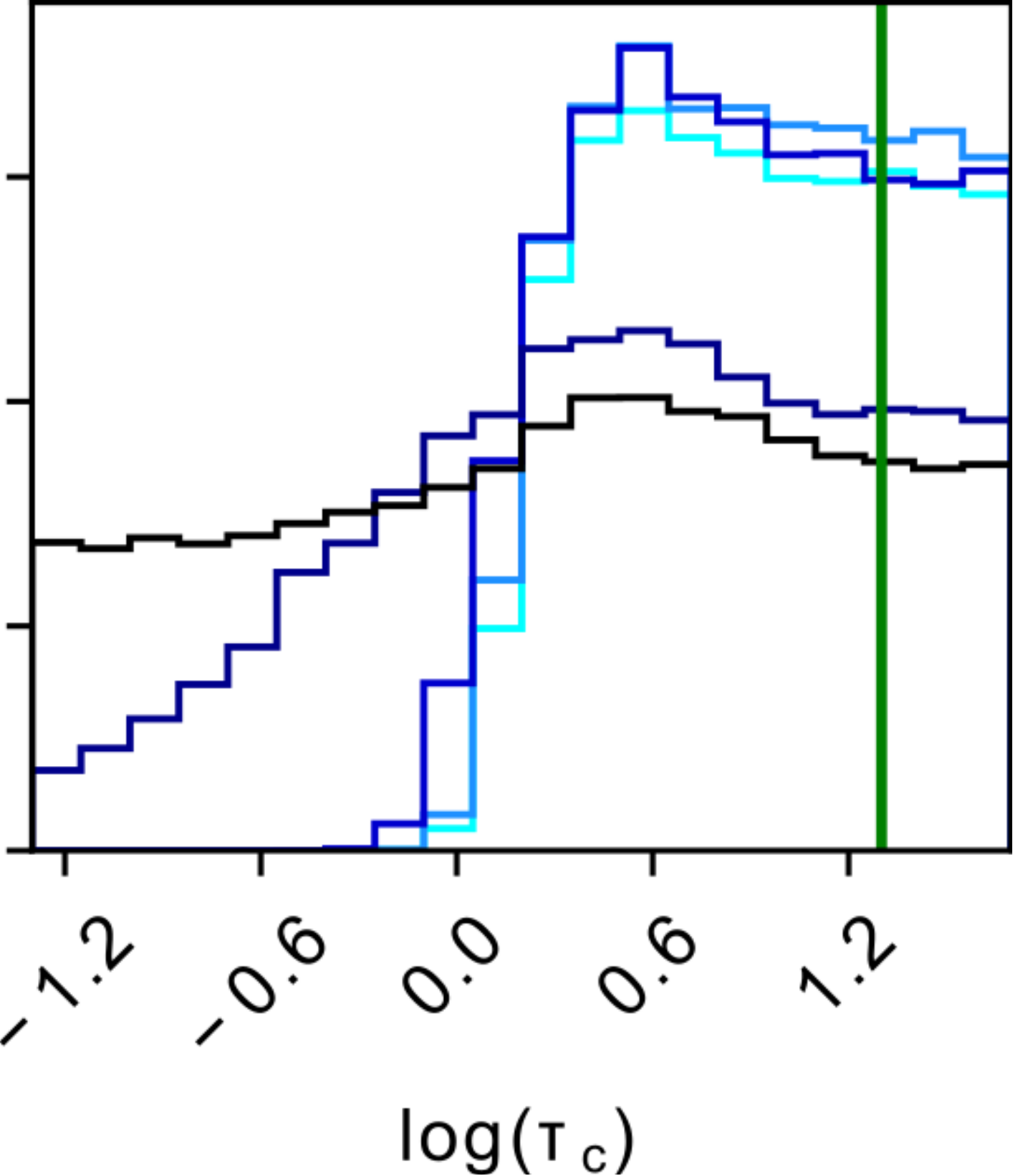}%
    \\
    \vspace{1cm}
    \includegraphics[width=5.cm]{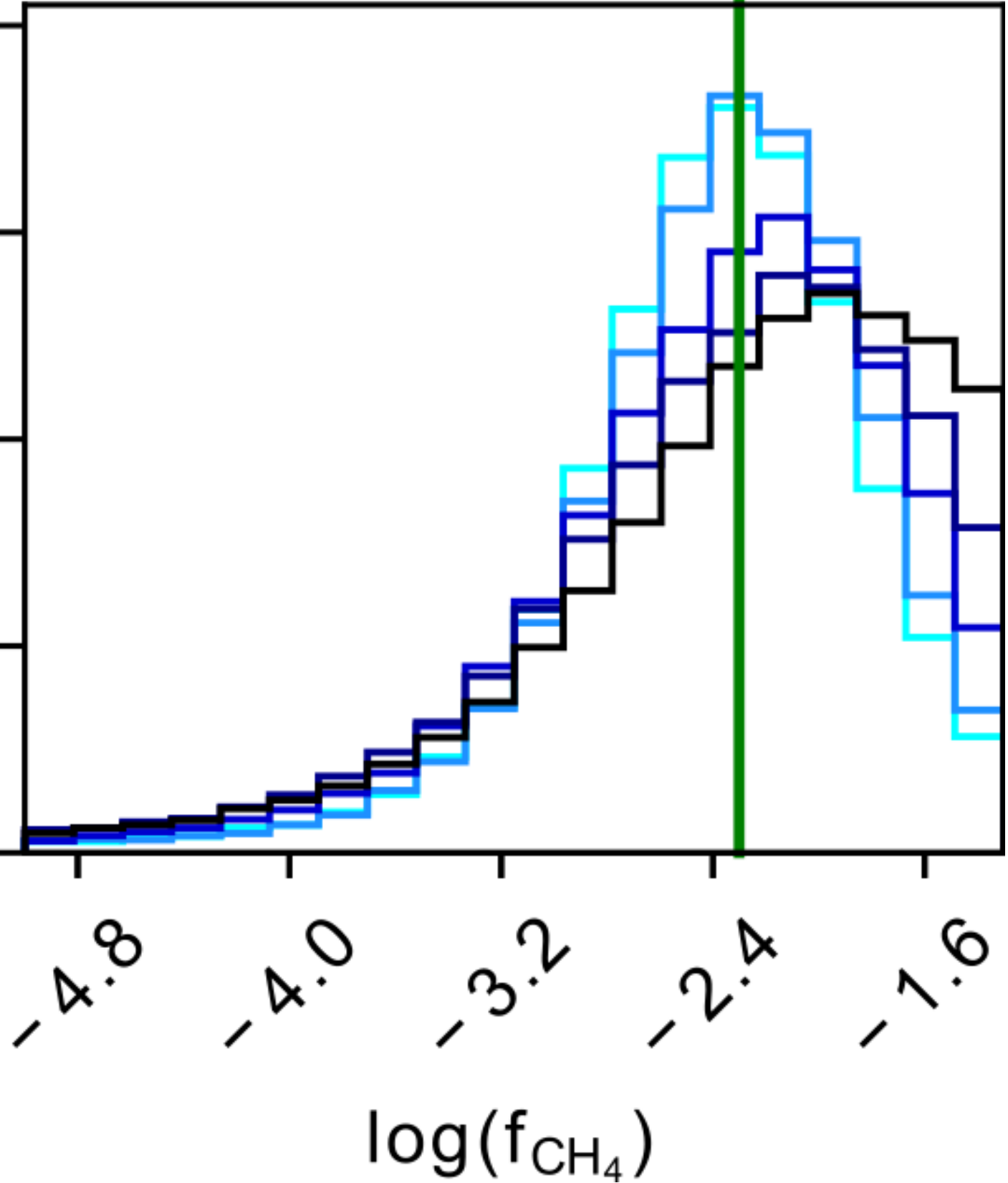}%
    \includegraphics[width=5.cm]{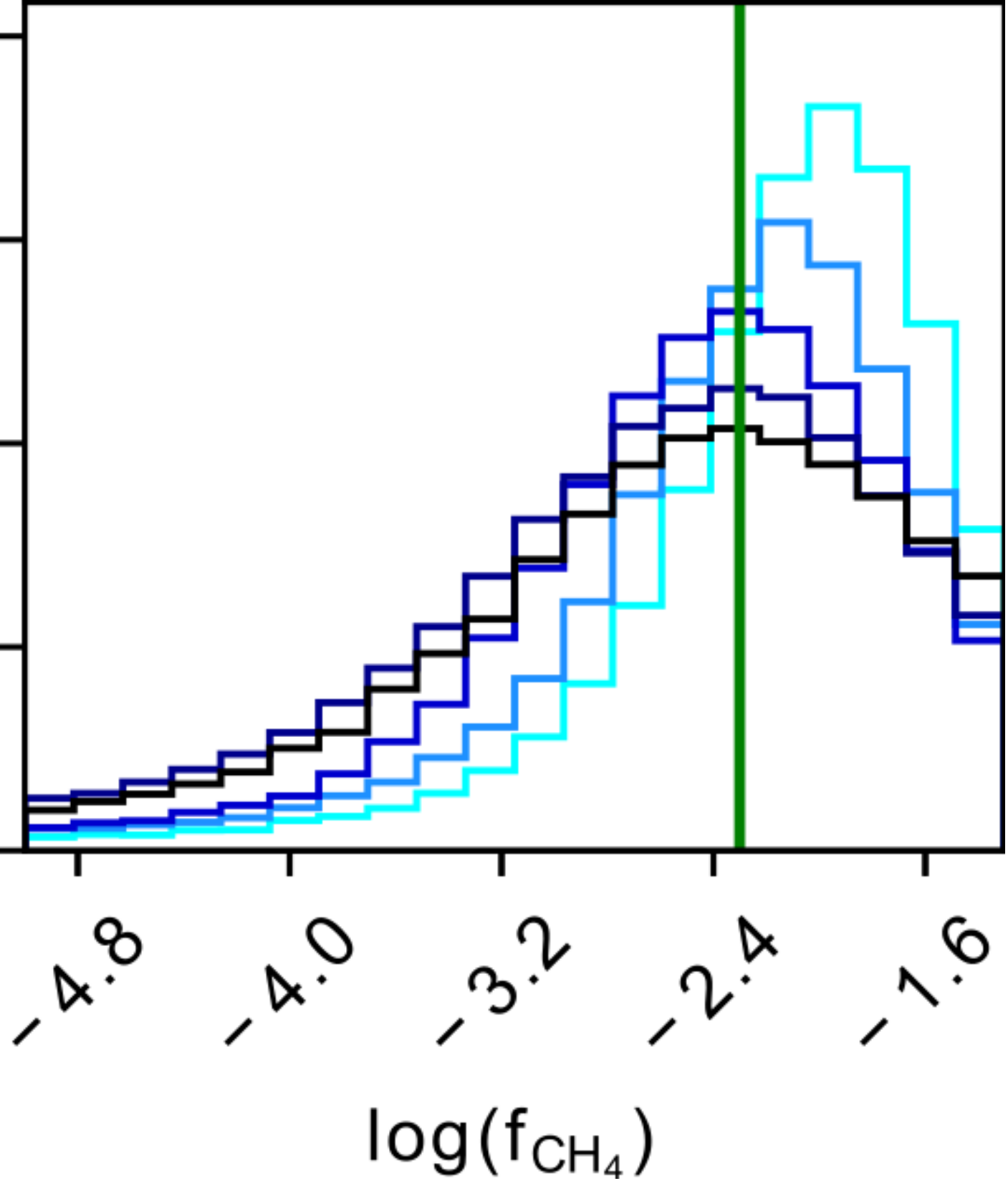}%
    \includegraphics[width=5.cm]{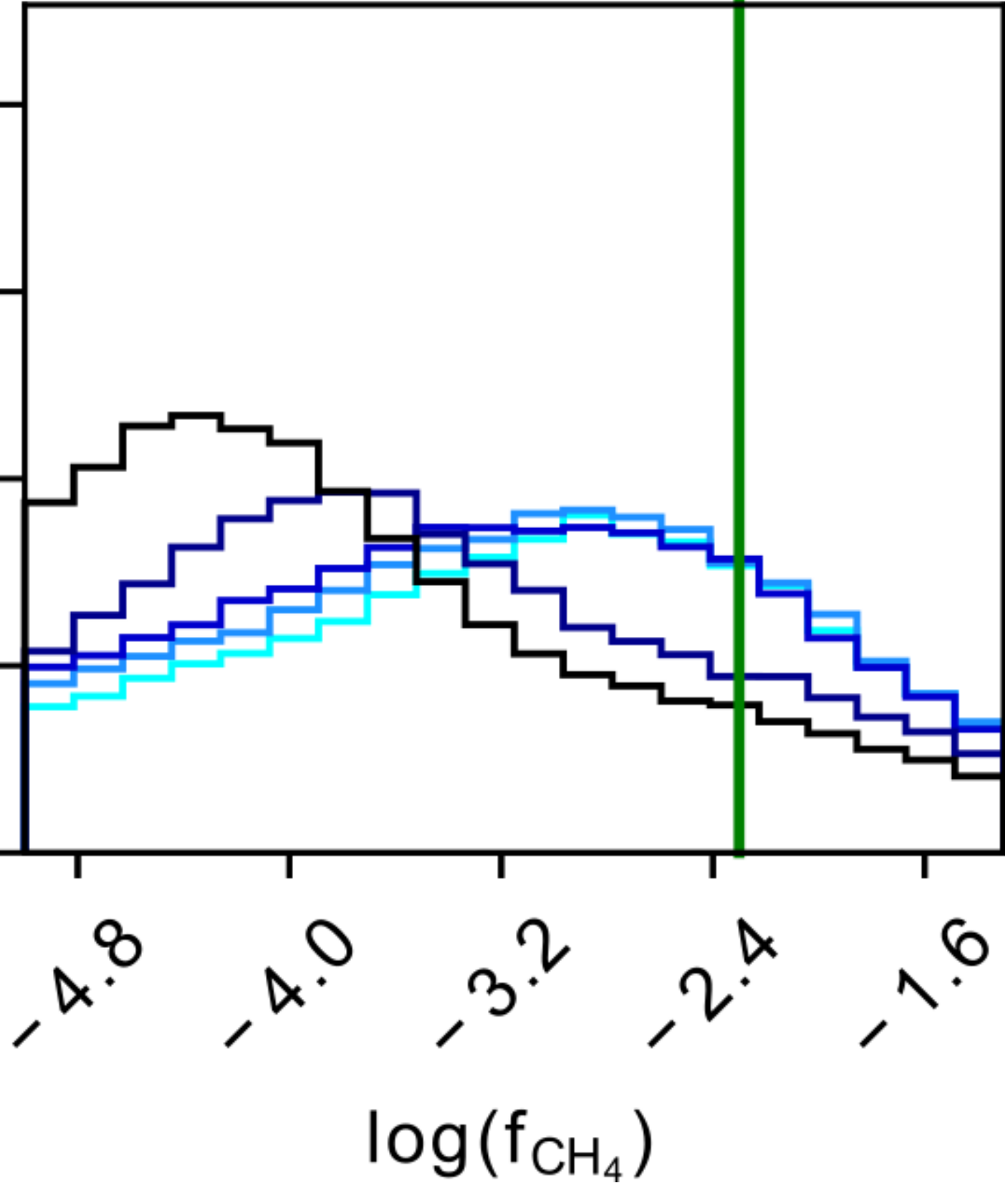}%
    \caption{ Marginalized probability distributions for {$\tau_{c}$} (top) and {$f_{\rm{CH_4}}$} (bottom). Different colours correspond to different levels of uncertainty in the \textit{a priori} estimate of R$_p$ ($\Delta R_p$), as stated in the legend. All panels share the same legend. From left to right: \textit{no-cloud}, \textit{thin-cloud} and \textit{thick-cloud} scenarios. 
    As discussed in the text, a plateau in the marginalized probability distribution entails that only an upper or lower limit can be set.
    }%
    \label{fig:results_several_DeltaRp}%
    \end{figure*}

So far we have considered two limiting cases for our retrievals: one with a known, fixed value of the planet radius; the other one, with {$R_p/R_{N}$} as a free parameter.
This exercise allowed us to understand the influence of $R_p$ on the retrieval outcome.
However, it is possible that we encounter exoplanets whose radius is not completely unconstrained, but rather known to some extent.

For instance, this would be the case for exoplanets discovered in radial-velocity surveys. 
In combination with other techniques such as astrometry, the planet mass could be estimated.
With that, mass-radius relationships would yield a range of values for $R_p$.
We note the inherent inaccuracy of such mass-radius relationships, as the planet composition and density will not be known \textit{a priori}.
The suitability of these relations needs to be discussed for each specific planet.
However, even with these caveats, mass-radius relationships will offer at the very least a range of possibilities for the planet density and, hence, a hypothetical range of possible R$_p$ values.
Furthermore, uncertainties in the estimates of $R_p$ from transit observations might reach values close to 10\% depending on the planet size and the stellar parameters \citep{raueretal2014}.
In these cases, $R_p$ would not be completely unconstrained.
This motivates us to study how different radius uncertainties would impact the characterisation of the exoplanet via direct-imaging.

In this setting, we explored how the outcomes of the retrieval are affected by different levels of uncertainty in the \textit{a priori} estimates of {$R_p/R_{N}$} ($\Delta R_p$).
Table \ref{table:results_Rpunknown_different_uncertainties} shows the retrieval results for the three cloud scenarios and $\Delta R_p$=0, 1.6, 10, 25 and 50\%, as well as for an unconstrained value of R$_p$.
Here, the box priors for {$R_p/R_{N}$} in the MCMC retrieval sampler correspond to each $\Delta R_p$, while the rest of parameters had the same box priors as in Table \ref{table:priors}.
As above, we set S/N=10.

Figure \ref{fig:results_several_DeltaRp} shows that reducing $\Delta R_p$ tends to provide better estimates of {$\tau_{c}$} and {$f_{\rm{CH_4}}$}.
The conclusions on the other model parameters do not change appreciably with respect to the retrievals for both {$R_p/R_{N}$} known or unknown.
We argue that reducing $\Delta R_p$ partially breaks some of the correlations between parameters, which are boosted if {$R_p/R_{N}$} is set as a free parameter.
Namely, reducing the {R$_p$}-{$\tau_{c}$} degeneracy affects other parameter correlations, thus improving the estimates of {$f_{\rm{CH_4}}$}.
In accordance with this, the improvement in the retrievals as the uncertainty in {$R_p/R_{N}$} decreases is more evident in the \textit{thick-cloud} scenario.
Given the large optical thickness of the cloud in this scenario, the {R$_p$}-{$\tau_{c}$} correlation had a bigger impact on the retrievals than for \textit{no-cloud} or \textit{thin-cloud} cases.

\subsection{The impact of noise realizations} 
In Sections \ref{subsec:results_Rpknown} and \ref{subsec:results_Rpunknown}, we have presented retrieval results based on the \textit{measured} spectra of Fig. \ref{fig:trueVSnoisy_spectra}, and therefore on single noise realizations.
To confirm that these conclusions do not strongly depend on the randomly-generated noise, 
we repeated each of the retrievals nine more times for different noise realizations at S/N=10. 
Figure \ref{fig:retrievals_several_noises} summarizes the outcome of this exercise, confirming that the findings reported above are generally robust.

\section{Conclusions}
\label{sec:conclusions}

WFIRST will be the first direct imaging space mission to obtain spectra of reflected starlight from cold exoplanets. Scheduled to be launched in 2025, it will open a new field of research on exoplanetary atmospheres.
Due to constraints in contrast and angular separation, WFIRST will mostly observe long-period giant planets.
However, for special cases it might also reach the interesting range of super-Earth to mini-Neptune exoplanets, not existent in the Solar System. 
Many of those targets are expected to have H$_2$-He atmospheres and clouds with a diverse range of optical properties.
Understanding how each atmospheric parameter affects the reflected-light spectra of those planets will pave the way for future direct imaging missions such as LUVOIR or HabEx.
These will be focused on characterising the atmospheres of Earth-like planets.

In this work we have explored what properties of an atmosphere would be possible to infer from a direct-imaging spectroscopic measurement.
For that, we have created an atmospheric model with six free parameters which was outlined in Sect. \ref{sec:model}.
This description is consistent with the structure of gas- and ice-giants in the Solar System.
Then, using a previously validated radiative transfer code, we computed the synthetic spectra for $\sim$300,000 atmospheric configurations at phase angle $\alpha$=0$^\circ$.
We analysed how different atmospheric configurations may generate similar spectra with a set of retrieval exercises.
In each retrieval, we simulated a \textit{measured} spectrum by adding noise at a certain S/N to the spectrum computed for an assumed \textit{true} atmospheric configuration.
Afterwards, we used a Markov-Chain Monte Carlo sampler to explore the space of parameters in search for atmospheric configurations yielding a spectrum consistent with the \textit{measured} one.
We also tested the influence of including the planet radius as a free parameter in the retrieval, and thus sampling a 7D space instead of a 6D one.
This is a relevant question since most of the long-period exoplanets that will be observable with direct-imaging techniques will not transit and thus their radius will remain uncertain.

We specified this analysis to the planet candidate Barnard b \citep{ribasetal2018}. 
This is a realistic case as the proximity of its host star and its orbital parameters make it a very promising candidate for direct-imaging observations.
Indeed, we showed that both the angular separation and contrast of this exoplanet are in the operating range of WFIRST \citep{spergeletal2015, traugeretal2016}. 
Therefore, this mission would be able to detect Barnard b in direct imaging and start characterising its atmosphere.
Eventual spectra taken by WFIRST will not have the full wavelength coverage between 500nm and 900nm or the resolving power used here. 
\citet{lacyetal2019} reported that the filters that will finally fly with the telescope are still to be decided, with only one filter guaranteed for spectroscopy so far.
We will quantify in future work the exact influence of partial spectral coverage on the atmospheric characterization.
However, a wide spectral coverage will surely improve the potential of WFIRST and the next-generation direct-imaging missions to analyse exoplanetary atmospheres.

The retrieval exercise was run for three different \textit{true} atmospheric configurations: one with virtually no clouds ($\tau_{c}$=0.05); another with a thin cloud layer ($\tau_{c}$=1) and finally one with an optically thick cloud ($\tau_{c}$=20).
If the value of R$_p$ is known, we found that observations at S/N=10 yielded evidence of the presence or absence of clouds in each of the cloud scenarios.
The optical thickness of the cloud, however, was not accurately constrained in any of the three cases.
The retrievals also detected the presence of methane in the atmosphere.
Besides, it was possible to constrain the CH$_4$ abundance with reasonable success in the three cloud scenarios.
The rest of the model parameters, on the other hand, remain largely unconstrained.
We identified a correlation amongst the optical properties of the cloud aerosols that prevents a reliable estimate of the aerosols’ single scattering albedo, which is potentially informative of their composition.

The main result of our study is that an unknown planetary radius will prevent detecting the presence or absence of clouds.
R$_p$ uncertainties trigger physical degeneracies between model parameters, particularly between R$_p$, the optical depth of the cloud ($\tau_{c}$), its pressure level (here given by $\tau_{c\rightarrow TOA}$) and the CH$_4$ abundance ($f_{\rm{CH_4}}$).
These correlations blur the atmospheric information that is possible to extract when $R_p$ is known.
Hence, we conclude that predictions for atmospheric characterisation built upon retrieval exercises in which $R_p$ is assumed to be known are overly optimistic.
They should be taken as ideal, best-case scenarios.

We also find that, if the radius of the exoplanet is completely unknown \textit{a priori}, an observation with S/N=10 could constrain it to within a factor of $\sim$2.
This estimate is consistent with the findings by \citet{nayaketal2017}, who focus on the R$_p$-$\alpha$ correlation.
We here showed that this result holds true regardless of the cloud coverage of the exoplanet.
This is a non-trivial finding, as we also show that R$_p$ is highly correlated with the optical thickness of the clouds in the atmosphere.
Indeed, if the cloud is very thick, we found that the most-likely estimates of R$_p$ deviate more from the true value, although still within a factor $\sim$2.
How the {$\tau_{c}$}-{R$_p$} degeneracy unfolds for several S/N levels will be a matter of future work.

Furthermore, we show that reducing the uncertainties in the \textit{a priori} estimate of R$_p$ results in more accurate retrievals of {$\tau_{c}$} and {$f_{\rm{CH_4}}$}.
With this, we outline how direct-imaging studies could benefit from other observational techniques as well as theoretical works on mass-radius relationships.

Still, constraining the radius of non-transiting exoplanets to a factor of 2 will potentially narrow the set of compatible internal compositions.
Achieving this independently of \textit{a priori} unknowns, such as the atmospheric cloud coverage, would help understanding the diversity and size distribution of the exoplanet population.
Currently, the population of exoplanets suitable for these direct-imaging measurements remains modest.
However, this population is expected to grow steadily as different techniques (radial velocity, astrometry, transit photometry) continue surveying the sky. 
It is essential to exploit the synergies between different techniques to maximize the possibilities of characterizing the atmospheres of long-period exoplanets.

\begin{acknowledgements}
      The authors acknowledge the support of the DFG priority program SPP 1992 “Exploring the Diversity of Extrasolar Planets (GA 2557/1-1)”. 
      NCS acknowledges the support by FCT - Fundação para a Ciência e a Tecnologia through national funds and by FEDER through COMPETE2020 - Programa Operacional Competitividade e Internacionalização by these grants: UID/FIS/04434/2019; UIDB/04434/2020; UIDP/04434/2020; PTDC/FIS-AST/32113/2017 \& POCI-01-0145-FEDER-032113; PTDC/FIS-AST/28953/2017 \& POCI-01-0145-FEDER-028953.
\end{acknowledgements}


\begin{thebibliography}{}

\bibitem[Ackerman \& Marley(2001)]{ackerman-marley2001}
Ackerman, A. S., \&
Marley, M. S., 2001,
\apj, 556, 872

\bibitem[Barnard(1916)]{barnard1916}
Barnard, E., 1916,
\aj, 29, 695

\bibitem[Barros et al.(2017)]{barrosetal2017}
Barros, S. C. C.,
Gosselin, H.,
Lillo-Box, J.,
Bayliss, D. et al., 2017,
\aap, 608, A25

\bibitem[Batalha et al.(2018)]{batalhaetal2018}
Batalha, N. E.,
Smith, A. J. R. W.,
Lewis, N. K.,
Marley, M. S. et al., 2018,
\apj, 156, 158

\bibitem[Bevington \& Robinson(2003)]{bevington-robinson2003}
Bevington, P. R. \&
Robinson, D. K., 2003,
McGraw Hill, New York, USA

\bibitem[Bolcar et al.(2016)]{bolcaretal2016}
Bolcar, M. R.,
Feinberg, L.,
France, K.,
Rauscher, B. J. et al., 2016,
Proc. SPIE, 9904

\bibitem[Borucki \& Summers, 1984]{boruckisummers1984}
Borucki, W.J. \& Summers, A. L., 1984, 
Icarus, 58, 121

\bibitem[Buenzli \& Schmid(2009)]{buenzli-schmid2009}
Buenzli, E. \&
Schmid, H. M., 2009,
\aap, 504, 259

\bibitem[Cahoy et al.(2010)]{cahoyetal2010}
Cahoy, K. L.,
Marley, M. S. \&
Fortney, J. J., 2010,
\apj, 724, 189

\bibitem[Damiano \& Hu(2019)]{damiano-hu2019}
Damiano, M. \&
Hu, R., 2019,
\aj, 159, 175

\bibitem[D'Angelo \& Bodenheimer(2013)]{dangelo-bodenheimer2013}
D'Angelo, G. \&
Bodenheimer, P., 2013,
\apj, 778, 77

\bibitem[Dawson \& De Robertis(2004)]{dawson-derobertis2004}
Dawson, P. C. \&
De Robertis, M. M., 2004,
\aj, 127, 2909

\bibitem[Dlugach \& Yanovitskij(1974)]{dlugach-yanovitskij1974}
Dlugach, J. M. \&
Yanovitskij, E. G., 1974,
Icarus, 22, 66

\bibitem[Fabrycky et al.(2014)]{fabryckyetal2014}
Fabrycky, D. C.,
Lissauer, J. J.,
Ragozzine, D.,
Rowe, J. F. et al., 2014,
\apj, 790, 146

\bibitem[Feng et al.(2018)]{fengetal2018}
Feng, Y. K.,
Robinson, T. D.,
Fortney, J. J.,
Lupu, R. E. et al., 2018
\aj, 155, 200

\bibitem[Foreman-Mackey et al.(2013)]{foremanmackeyetal2013}
Foreman-Mackey, D.,
Hogg, D. W.,
Lang, D. \&
Goodman, J., 2013,
\pasp, 125, 306

\bibitem[Garc\'ia Mu\~noz et al.(2012)]{garciamunozetal2012}
Garc\'ia Mu\~noz, A.,
Zapatero Osorio, M. R.,
Barrena, R.,
Monta\~n\'es-Rodr\'iguez, P. et al., 2012,
\apj, 755, 103

\bibitem[Garc\'ia Mu\~noz et al.(2014)]{garciamunozetal2014}
Garc\'ia Mu\~noz, A.,
P\'erez-Hoyos, S. \&
S\'anchez-Lavega, A., 2014,
\aap, 566, L1

\bibitem[Garc\'ia Mu\~noz \& Isaak(2015)]{garciamunoz-isaak2015}
Garc\'ia Mu\~noz, A. \&
Isaak, K. G., 2015,
PNAS, 112, 44

\bibitem[Garc\'ia Mu\~noz \& Mills(2015)]{garciamunoz-mills2015}
Garc\'ia Mu\~noz, A. \&
Mills, F. P., 2015,
\aap, 573, A72

\bibitem[Garc\'ia Mu\~noz et al.(2017)]{garciamunozetal2017}
Garc\'ia Mun\~oz, A.,
Lavvas, P. \&
West, R. A., 2017,
Nat. Astron., 1, 0114

\bibitem[Garc\'ia Mu\~noz \& Cabrera (2018)]{garciamunoz-cabrera2018}
Garc\'ia Mun\~oz, A. \&
Cabrera, J., 2018,
MNRAS, 473, 1801

\bibitem[Gatewood \& Eichhorn(1973)]{gatewood-eichhorn1973}
Gatewood, G. \&
Eichhorn, H., 1973,
\aj, 78, 769

\bibitem[Giampapa et al.(1996)]{giampapaetal1996}
Giampapa, M. S.,
Rosner, R.,
Kashyap, V.,
Fleming, T. A.,
Schmitt, J. H. M. M.  \&
Bookbinder, J. A., 1996,
\apj, 463, 707

\bibitem[Goodman \& Weare(2010)]{goodman-weare2010}
Goodman, J. \&
Weare, J., 2010,
COMM APP MATH COM SC, 5, 65

\bibitem[Greco \& Burrows(2015)]{greco-burrows2015}
Greco, J. P. \&
Burrows, A., 2015,
\apj, 808, 172

\bibitem[Guimond \& Cowan(2018)]{guimond-cowan2018}
Guimond, C. M. \&
Cowan, N. B., 2018,
\aj, 155, 230

\bibitem[Heidinger \& Stephens(2000)]{heidinger-stephens2000}
Heindinger, A. K. \&
Stephens, G. L., 2000,
JAS, 57, 1615

\bibitem[Helling et al.(2017)]{hellingetal2017}
Helling, Ch.,
Tootill, D.,
Woitke, P. \&
Lee, G., 2017,
\aap, 603, A123

\bibitem[Horak(1950)]{horak1950}
Horak, H. G., 1950,
\apj, 112, 445

\bibitem[Hu(2019)]{hu2019}
Hu, R., 2019,
\apj, 887, 166

\bibitem[Ida \& Lin(2008)]{ida-lin2008}
Ida, S. \&
Lin, D. N. C.,  2008,
\apj, 685, 584

\bibitem[Ikoma et al.(2001)]{ikomaetal2001}
Ikoma, M.,
Emori, H. \&
Nakazawa, K., 2001,
\apj, 553, 999

\bibitem[Ilic et al.(2018)]{ilicetal2018}
Ilic, N.,
Garc\'ia Mu\~noz, A.,
Beno\^it, S.,
West, R. A. et al., 2018,
EPSC Abstracts, EPSC2018-745

\bibitem[Inaba \& Ikoma(2003)]{inaba-ikoma2003}
Inaba, S. \&
Ikoma, M.,  2003,
\aap, 410, 711

\bibitem[Karkoschka(1994)]{karkoschka1994}
Karkoschka, E., 1994,
Icarus, 111, 174

\bibitem[Lacy et al.(2019)]{lacyetal2019}
Lacy, B.,
Shlivko, D. \&
Burrows, A., 2019,
\aj, 157, 132

\bibitem[Lupu et al.(2016)]{lupuetal2016}
Lupu, R. E.,
Marley, M. S.,
Lewis, N.,
Line, M. et al., 2016,
\aj, 152, 217

\bibitem[Madhusudhan \& Seager(2009)]{madhusudhan-seager2009}
Madhusudhan, N. \&
Seager, S., 2009,
\apj, 707, 24

\bibitem[Marois et al.(2000)]{maroisetal2000}
Marois, C.,
Doyon, R.,
Racine, R. \&
Nadeau, D., 2000,
\pasp, 112, 767

\bibitem[Mennesson et al.(2016)]{mennessonetal2016}
Mennesson, B.,
Gaudi, S.,
Seager, S.,
Cahoy, K. et al., 2016,
Proc. SPIE, 9904

\bibitem[Mishchenko(1989)]{mishchenko1989}
Mishchenko, M.I., 1989,
Icarus, 84, 296

\bibitem[Misra et al.(2014)]{misraetal2014}
Misra, A.,
Meadows, V. \&
Crisp, D., 2014,
\apj, 792, 61

\bibitem[Morozhenko \& Yanovitskij(1973)]{morozhenko-yanovitskij1973}
Morozhenko, A. V.,
Yanovitskij, E. G., 1973,
Icarus, 18, 583

\bibitem[Nayak et al.(2017)]{nayaketal2017}
Nayak, M.,
Lupu, R.,
Marley.,
Fortney, J. J. et al., 2017,
\pasp, 129, 973

\bibitem[Ohno \& Okuzumi(2017)]{ohno-okuzumi2017}
Ohno, K. \&
Okuzumi, S., 2017,
\apj, 835, 261

\bibitem[P\'erez-Hoyos et al.(2012)]{perezhoyosetal2012}
P\'erez-Hoyos, S.,
Sanz-Requena, J. F.,
Barrado-Izaguirre, N.,
Rojas, J. F. et al., 2012,
Icarus, 217, 256

\bibitem[Pierrehumbert \& Gaidos(2011)]{pierrehumbert-gaidos2011}
Pierrehumbert, R. \&
Gaidos, E.,  2011,
\apjl, 734, L13

\bibitem[Press et al.(2003)]{pressetal2003}
Press, W. H.,
Teukolsky, S. A.,
Vetterling, W. T. \&
Flannery, B. P., 2003,
Cambridge University Press, Cambridge, UK

\bibitem[Rauer et al.(2014)]{raueretal2014}
Rauer, H.,
Catala, C.,
Aerts, C.,
Appourchaux, T. et al., 2014,
Exp. Astron., 38, 249

\bibitem[Ribas et al.(2018)]{ribasetal2018}
Ribas, I.,
Tuomi, M.,
Reiners, A.,
Butler, R. P.,
Morales, J.C. et al., 2018,
Nature, 563, 365

\bibitem[Robinson et al.(2016)]{robinsonetal2016}
Robinson, R. D.,
Stapelfeldt, K. R. \&
Marley, M. S., 2016,
\pasp, 128, 025003

\bibitem[S\'anchez-Lavega et al.(2004)]{sanchezlavegaetal2004}
S\'anchez-Lavega, A.,
P\'erez-Hoyos, S. \&
Hueso, R., 2004,
Am. J. Phys., 72, 767

\bibitem[S\'anchez-Lavega(2010)]{sanchezlavega2010}
S\'anchez-Lavega, A., 2010,
CRC Press, Boca Raton, USA

\bibitem[Satoh et al.(2000)]{satohetal2000}
Satoh, T.,
Itoh, S.,
Kawabata, K.,
Tenma, T. et al., 2000,
\pasj, 52, 363

\bibitem[Schmid et al.(2011)]{schmidetal2011}
Schmid, H. M.,
Buenzli, F. J. E. \&
Gisler, D., 2011,
Icarus, 212, 701

\bibitem[Schmitt \& Liefke(2004)]{schmitt-liefke2004}
Schmitt, J. H. M. M. \&
Liefke, C., 2004,
\aap, 417, 651

\bibitem[Smith \& Tomasko(1984)]{smith-tomasko1984}
Smith, P. H. \&
Tomasko, M. G., 1984,
Icarus, 58, 35

\bibitem[Spergel et al.(2013)]{spergeletal2013}
Spergel, D.,
Gehrels, N.,
Breckinridge, J.,
Donahue, M. et al., 2013,
\url{https://arxiv.org/abs/1305.5422}

\bibitem[Spergel et al.(2015)]{spergeletal2015}
Spergel, D.,
Gehrels, N.,
Baltay, C.,
Bennett, D. et al., 2015,
\url{https://arxiv.org/abs/1503.03757}

\bibitem[Stark et al.(2015)]{starketal2015}
Stark, C. C.,
Roberge, A.,
Mandell, A.,
Clampin, M. et al., 2015,
\apj, 808, 149

\bibitem[Stephens \& Heidinger(2000)]{stephens-heidinger2000}
Stephens, G. L. \&
Heindinger, A. K., 2000,
JAS, 57, 1599

\bibitem[Tal-Or et al.(2019)]{tal-oretal2019}
Tal-Or, L.,
Zucker, S.,
Ribas, I.,
Anglada-Escudé, G. \&
Reiners, A., 2019,
\aap, 623, A10

\bibitem[The LUVOIR Team(2018)]{luvoirteam2018}
The LUVOIR Team, 2018,
\url{https://arxiv.org/abs/1809.09668}

\bibitem[Trauger et al.(2016)]{traugeretal2016}
Trauger, J.,
Moody, D.,
Krist, J. \&
Gordon, B., 2016,
JATIS, 2(1), 011013

\bibitem[van de Kamp(1963)]{vandekamp1963}
van de Kamp, P., 1963,
\aj, 68, 515

\bibitem[Virtanen et al.(2020)]{virtanenetal2020}
Virtanen, P.,
Gommers, R.,
Oliphant, T. E.,
Haberland, M. et al., 2020,
Nat. Methods, 17, 261

\bibitem[von Paris et al.(2013)]{vonparisetal2013}
von Paris, P.,
Hedelt, P.,
Selsis, F.,
Schreier, F. \&
Trautmann, T., 2013
\aap, 551, A120

\bibitem[Wahhaj et al.(2015)]{wahhajetal2015}
Wahhaj, Z.,
Cieza, L. A.,
Mawet, D.,
Yang, B.,
Canovas, H. et al., 2015,
\aap, 581, A24

\bibitem[Zeng et al.(2019)]{zengetal2019}
Zeng, L.,
Jacobsen, S. B.,
Sasselov, D. D.,
Petaev, M. I. et al., 2019,
PNAS, 116, 20


\end{thebibliography}

\clearpage
\begin{appendix} %First appendix

\section{Completeness and interpolation of the grid of spectra}
\label{sec:appendix_grid_completeness_interpol}
\setcounter{figure}{0} \renewcommand{\thefigure}{A.\arabic{figure}} 
\setcounter{table}{0} \renewcommand{\thetable}{A.\arabic{table}}

Our approach to the inversion problem relies on sampling continuously the six-dimensional atmospheric vector  $\boldsymbol{p}=$\{$\tau_{c}$, $\Delta_{c}$, $\tau_{c\rightarrow TOA}$, $r_{\rm{eff}}$, $\omega_0$, $f_{\rm{CH_4}}$ \} from the discreet grid summarized in Table \ref{table:grid}.
To that end, we must ensure that the grid is dense enough, and devise a way to predict the spectra for atmospheric configurations not represented in the grid.

The differences amongst consecutive spectra are small for the specific example shown in Fig. \ref{fig:completeness_variations_all_param}.
They are also small in general, which simply confirms that our grid is dense.
To sample continuously in $\boldsymbol{p}$, we opted to interpolate linearly within the multiple dimensions of the grid of spectra with the \texttt{scipy} \texttt{interpolate} routine \citep{virtanenetal2020}.
To test the accuracy of this approach, we compared the albedos interpolated at 1000 random realizations of $\boldsymbol{p}$ against the exact solutions obtained by solving the scattering problem with the PBMC model.
These 1000 random realizations of $\boldsymbol{p}$ were generated by sampling each variable from a uniform distribution between the limits set in Table \ref{table:priors}.
We define as the figure of merit for the comparison:
\begin{equation} \label{eq:appendix_variance2}
var = \sum_{i=1}^{i=N} {\frac{\left( A_g(\lambda_i)_{comput}-A_g(\lambda_i)_{interp}\right)^2}{A_g(\lambda_i)_{comput}}}. 
\end{equation}

\begin{figure}
    \centering
    \includegraphics[width=9.cm]{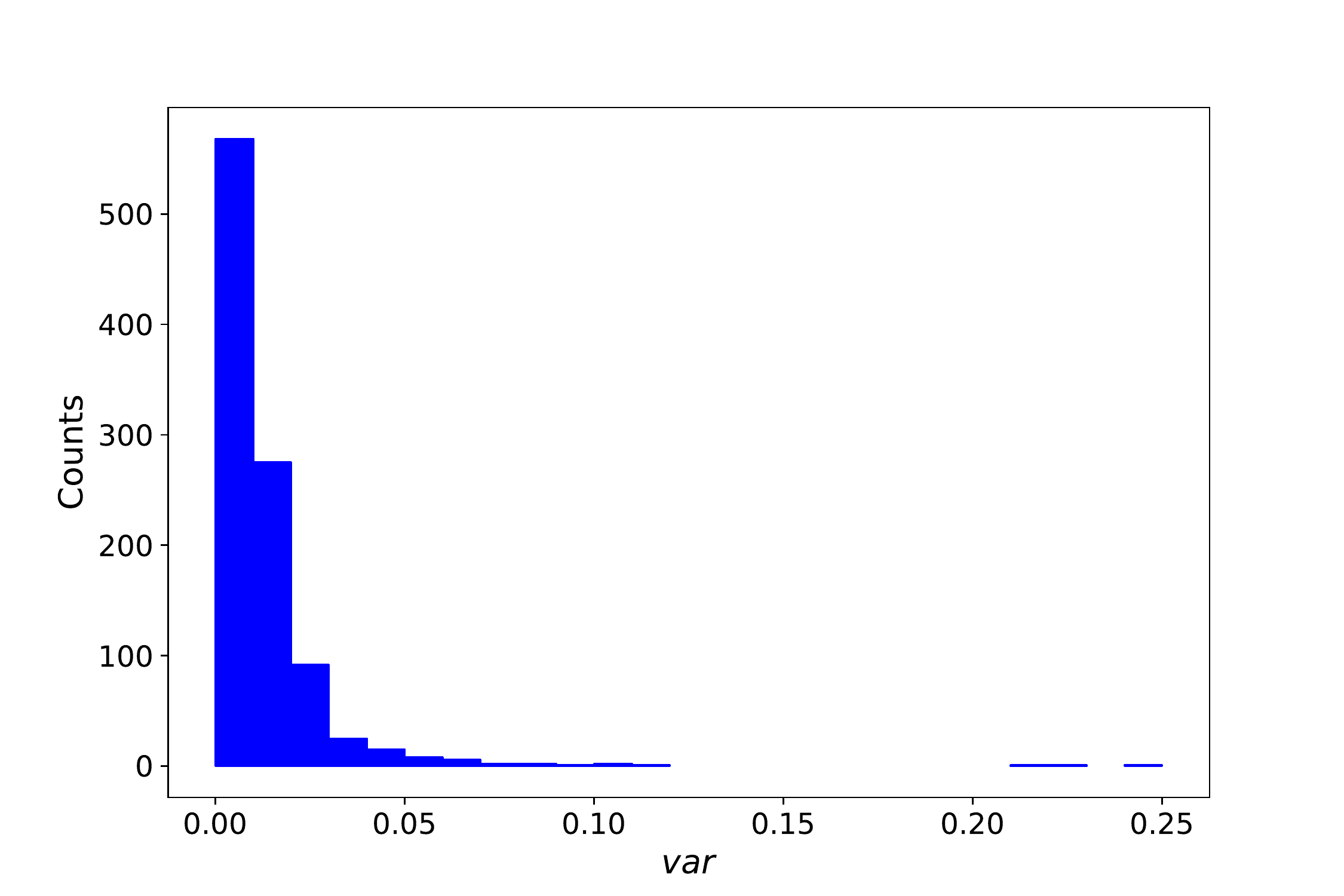}
    \\
    \includegraphics[width=9.cm]{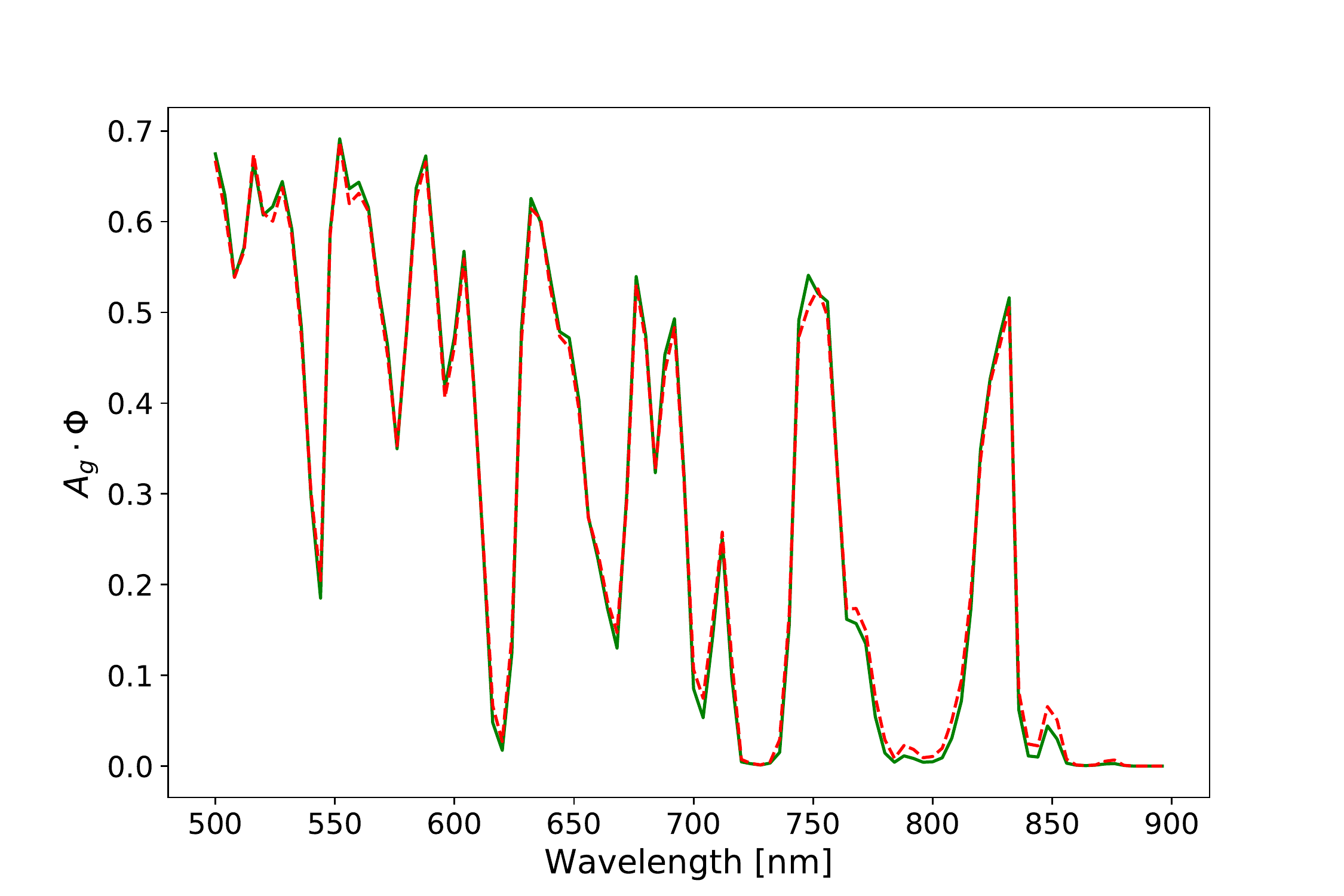}
      \caption{Top: Distribution of $var$ values for the 1000 random realizations selected to compare
      the planetary albedo obtained by solving the multiple scattering problem and that obtained by linear interpolation from the grid. 
      Bottom: Computed (solid green line) and interpolated (dashed red line) albedos for the configuration with the highest $var$ value.}
    \label{fig:appendix_interpolation_works}%
    \end{figure}

Figure \ref{fig:appendix_interpolation_works} shows the histograms for the $var$ values from the comparison.
The albedo corresponding to the highest $var$ (and poorest performance of the linear interpolator) is shown in Fig. \ref{fig:appendix_interpolation_works}.
The exercise confirms that our adopted interpolation is an accurate way of producing spectra at arbitrary $\boldsymbol{p}$ realizations provided that the grid of spectra is dense enough, as is the case here.

\clearpage
\section{Posterior probability distributions}
\label{sec:appendix_retrieval_cornerplots}
\setcounter{figure}{0} \renewcommand{\thefigure}{B.\arabic{figure}} 
\setcounter{table}{0} \renewcommand{\thetable}{B.\arabic{table}}

\subsection{Known $R_p$; \textit{thin-cloud} scenario}
\begin{figure}[h]
   \centering
   \includegraphics[width=18.cm]{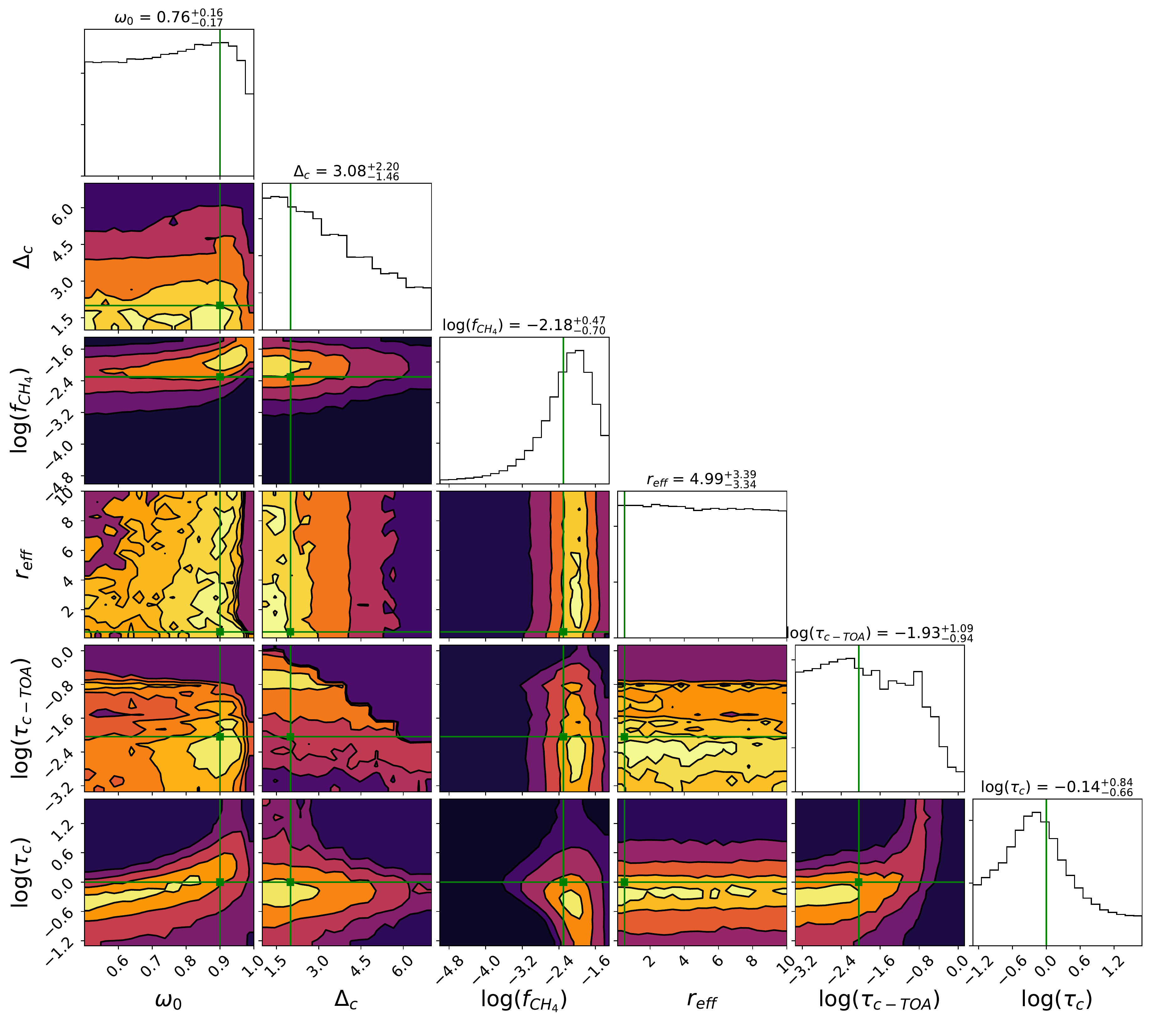}
      \caption{\label{fig:results_rpknown_thincloud_S/N10} Same as Figure \ref{fig:results_rpknown_nocloud_S/N10}, but for the \textit{thin-cloud} scenario (see Table \ref{table:truth}).
     }
   \end{figure}
\clearpage   
   
\subsection{Known $R_p$; \textit{thick-cloud} scenario}
\begin{figure}[h]
   \centering
   \includegraphics[width=18.cm]{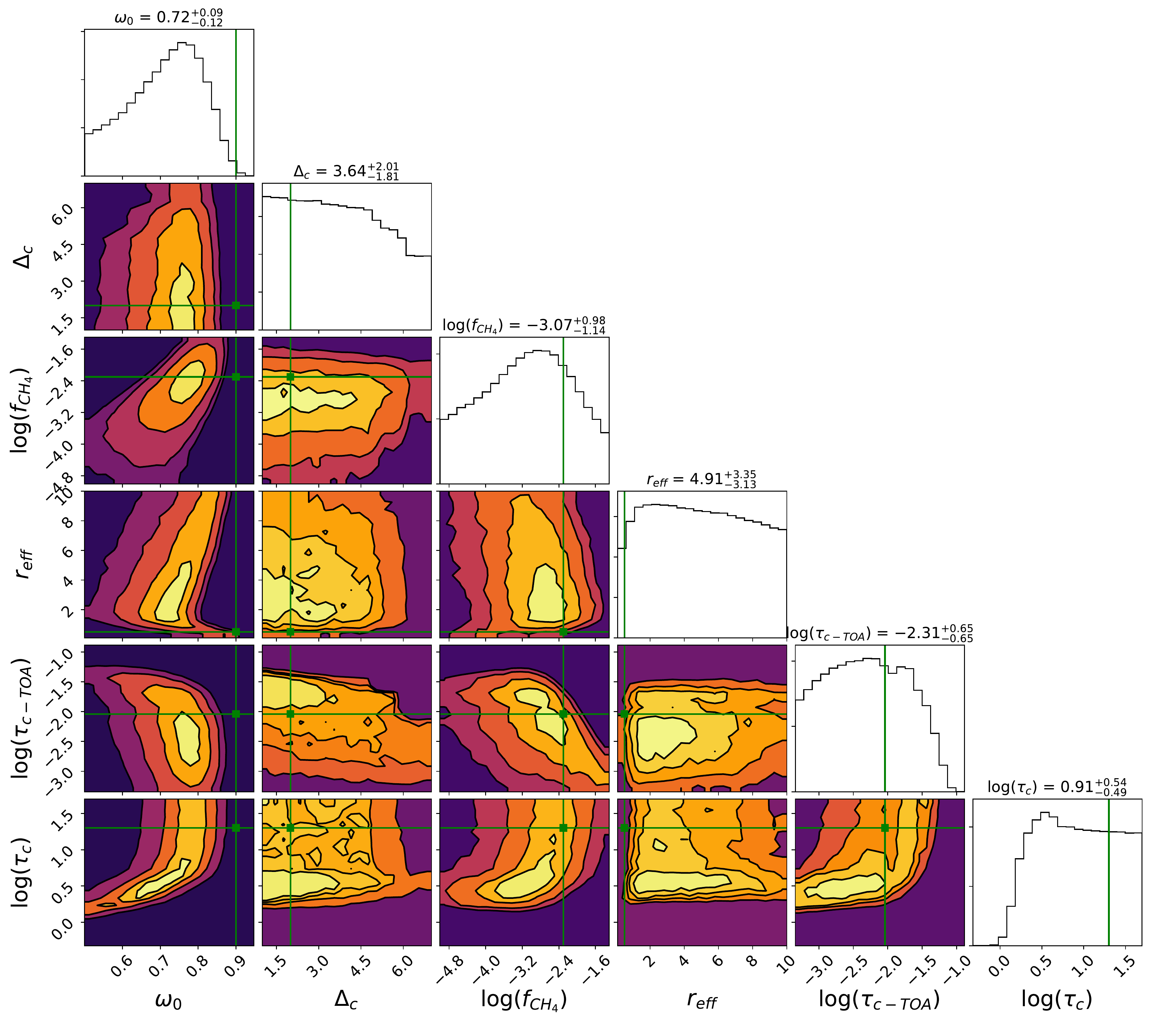}
      \caption{\label{fig:results_rpknown_thickcloud_S/N10} Same as Figure \ref{fig:results_rpknown_nocloud_S/N10}, but for the \textit{thick-cloud} scenario (see Table \ref{table:truth}).
     }
   \end{figure}
\clearpage

\subsection{Unknown $R_p$; \textit{no-cloud} scenario}
\begin{figure}[h]
   \centering
   \includegraphics[width=18.cm]{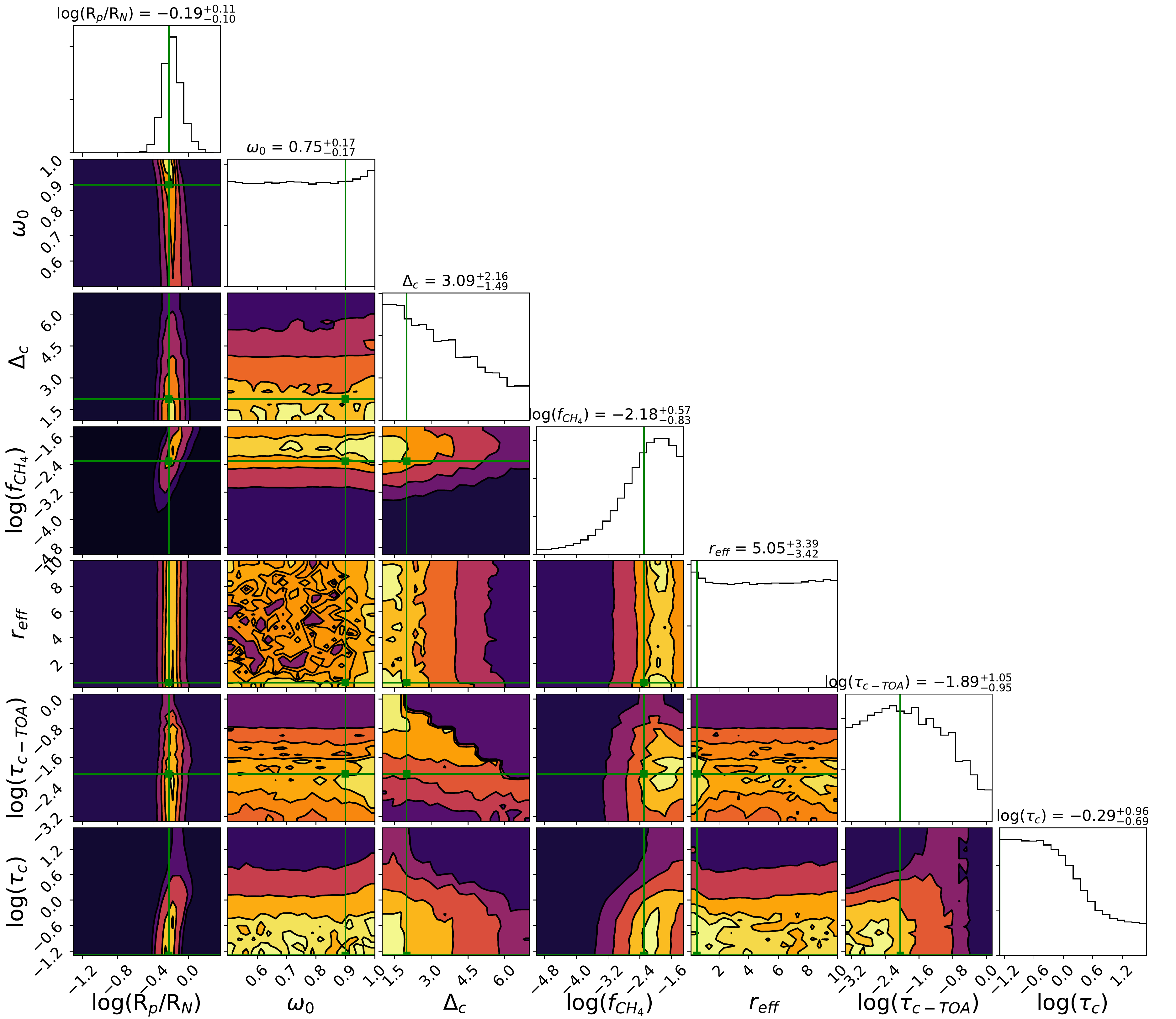}
      \caption{\label{fig:results_rpunknown_nocloud_S/N10}
      Posterior probability distributions of the model parameters for a simulated observation of the \textit{no-cloud} atmospheric configuration at S/N=10.
      The planetary radius is considered unconstrained.
      Green lines mark the \textit{true} values of the model parameters (see Table \ref{table:truth}) for this observation.
      2-D subplots show the correlations between pairs of parameters.
      Contour lines correspond to the 0.5, 1, 1.5, and 2 $\sigma$ levels.
      The median of each parameter's distribution is shown on top of their 1-D probability histogram.
      Upper and lower errors correspond to the 84\% and 16\% quantiles.
     }
   \end{figure}
\clearpage

\subsection{Unknown $R_p$; \textit{thin-cloud} scenario}
\begin{figure}[h]
   \centering
   \includegraphics[width=18.cm]{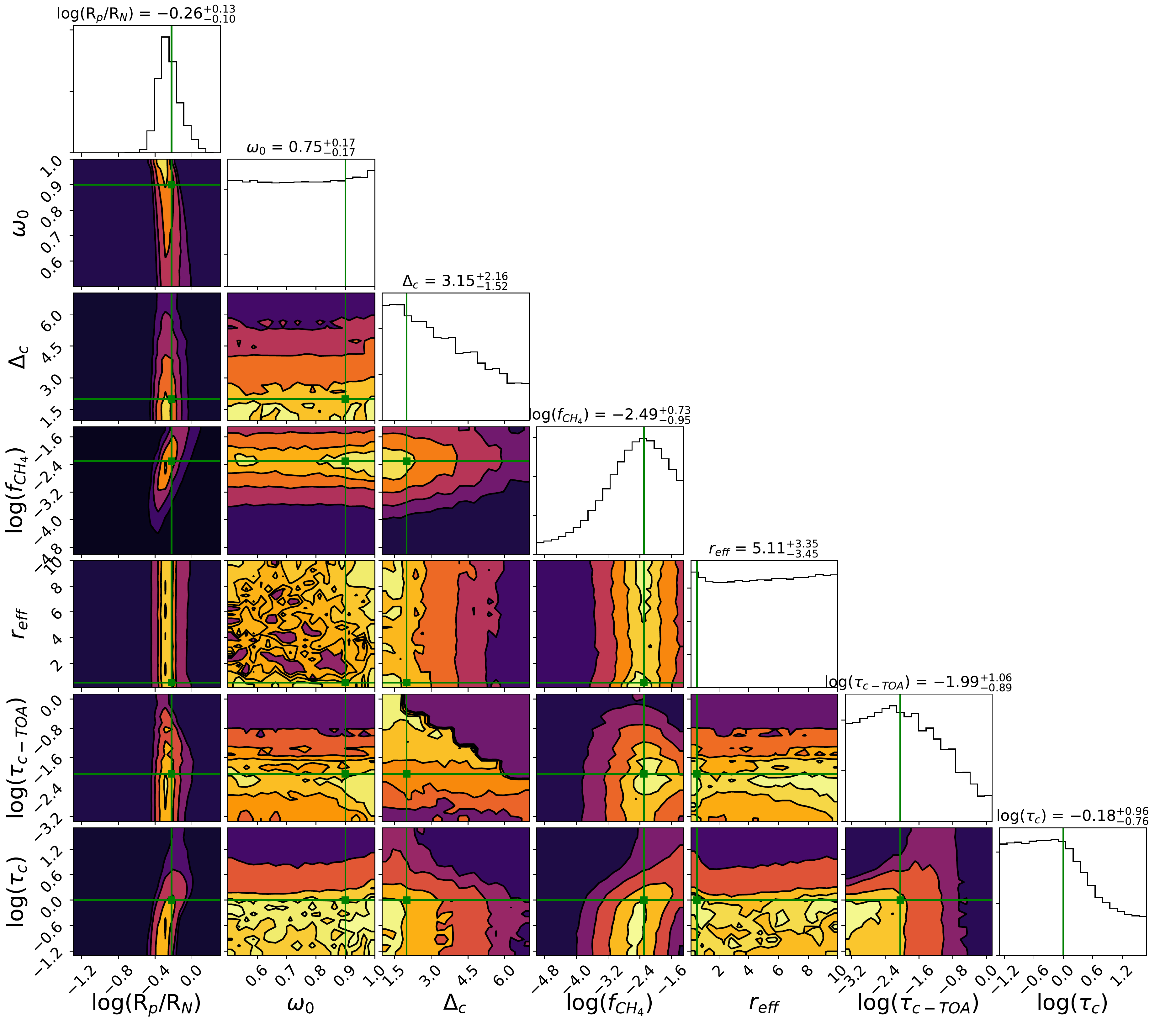}
      \caption{\label{fig:results_rpunknown_thincloud_S/N10} Same as Figure \ref{fig:results_rpunknown_nocloud_S/N10}, but for a \textit{thin-cloud} scenario (see Table \ref{table:truth}).
     }
   \end{figure}
\clearpage

\subsection{Unknown $R_p$; \textit{thick-cloud} scenario}
\begin{figure}[h]
   \centering
   \includegraphics[width=18.cm]{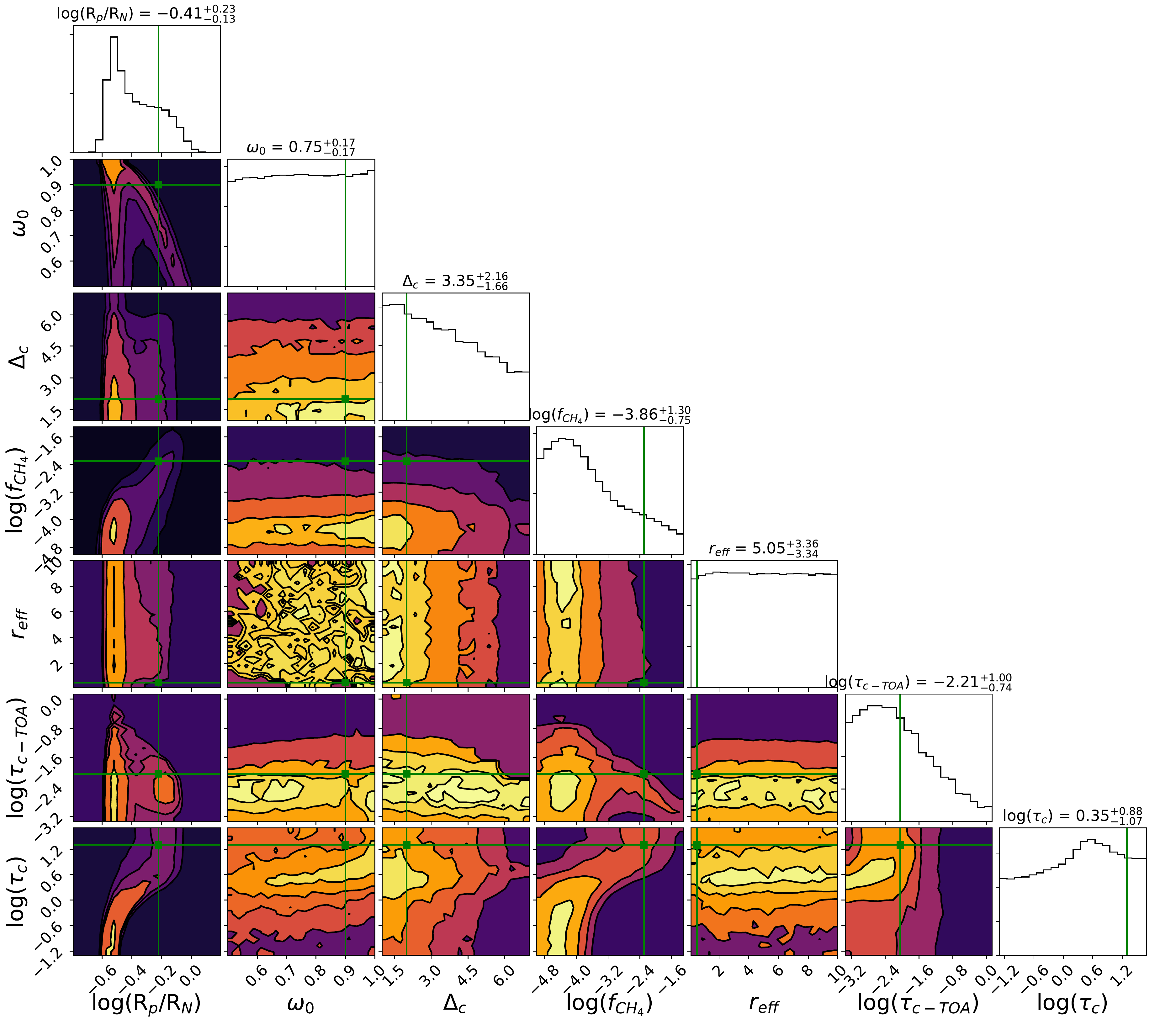}
      \caption{\label{fig:results_rpunknown_thickcloud_S/N10} Same as Figure \ref{fig:results_rpunknown_nocloud_S/N10}, but for a \textit{thick-cloud} scenario (see Table \ref{table:truth}).
     }
   \end{figure}
\clearpage

\clearpage
\section{Retrieval results for different noise realizations.}
\label{sec:appendix_retrievals_noise_realizations}
\setcounter{figure}{0} \renewcommand{\thefigure}{C.\arabic{figure}} 
\setcounter{table}{0} \renewcommand{\thetable}{C.\arabic{table}}

  \begin{figure}[h]
       %\centering
    \includegraphics[width=2.5cm]{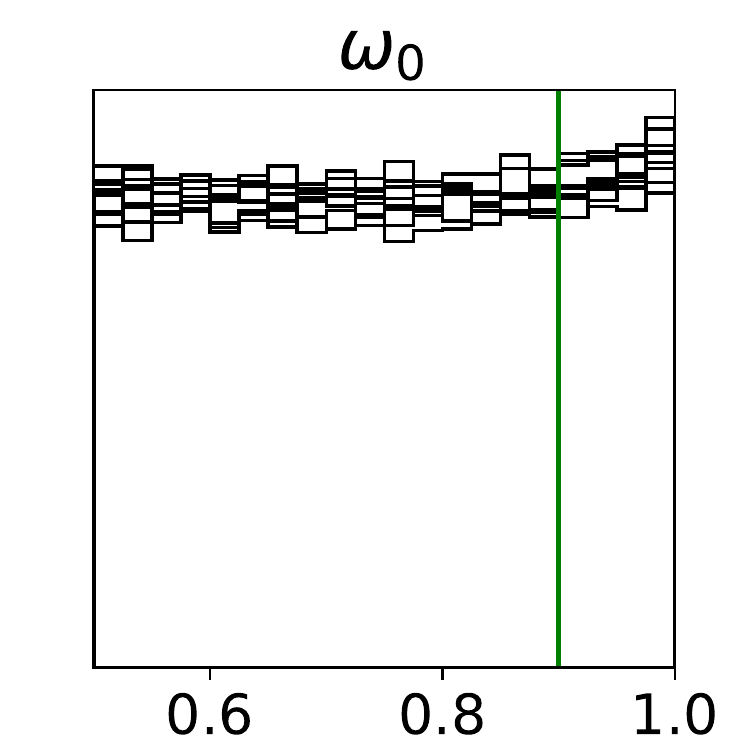}%
    \includegraphics[width=2.5cm]{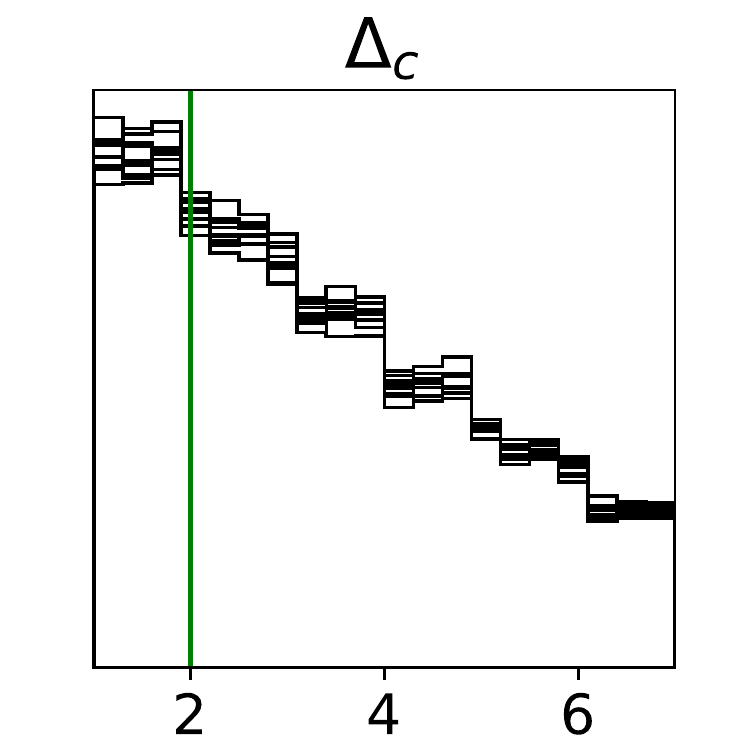}%
    \includegraphics[width=2.5cm]{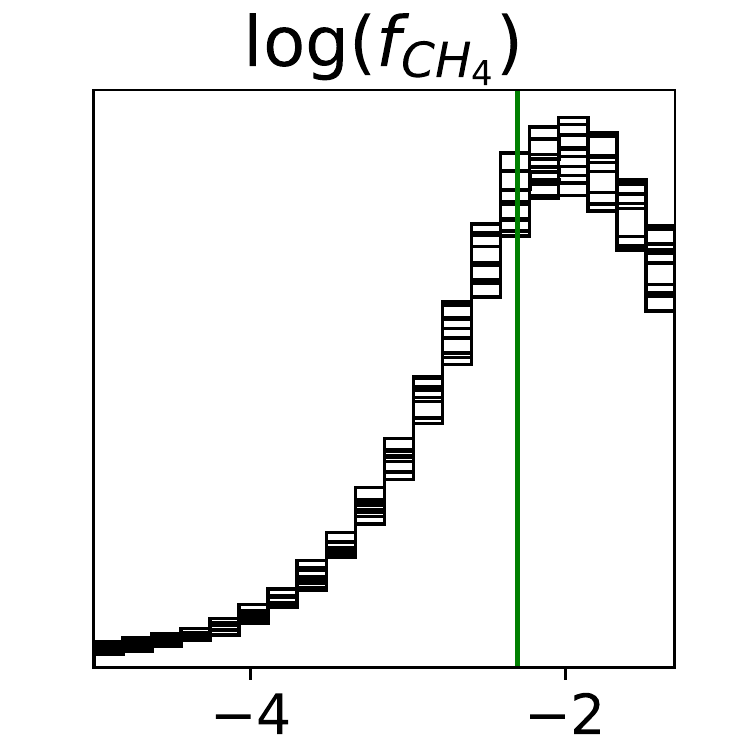}%
    \includegraphics[width=2.5cm]{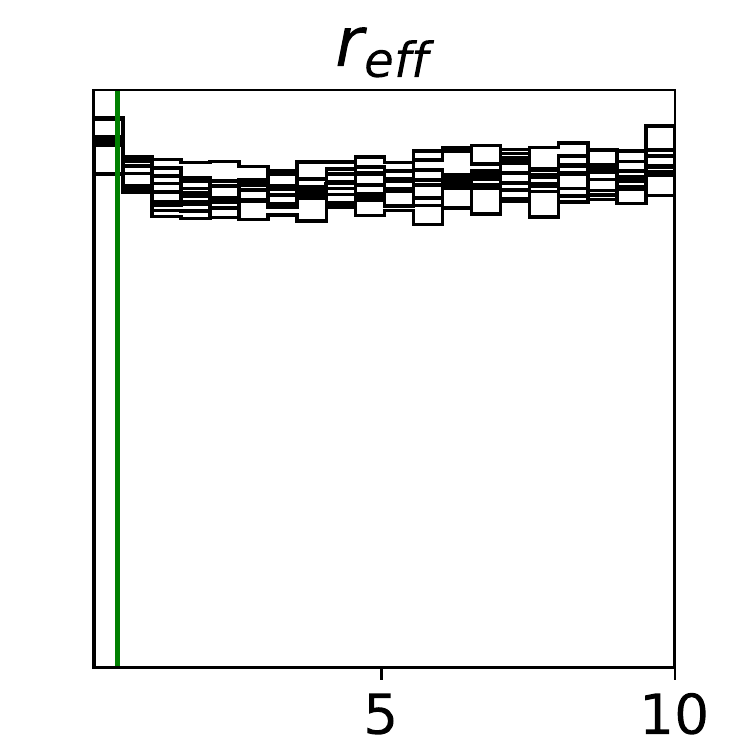}%
    \includegraphics[width=2.5cm]{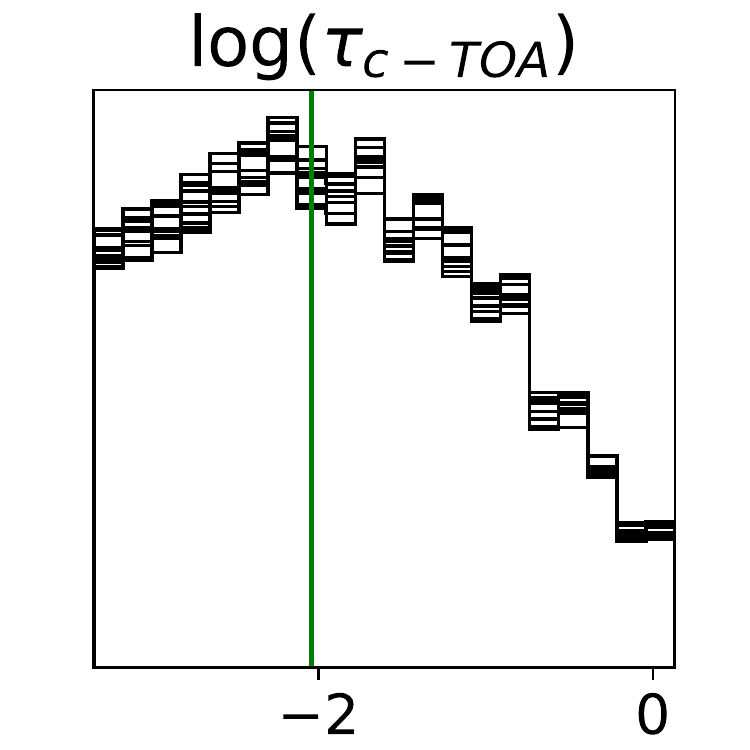}%
    \includegraphics[width=2.5cm]{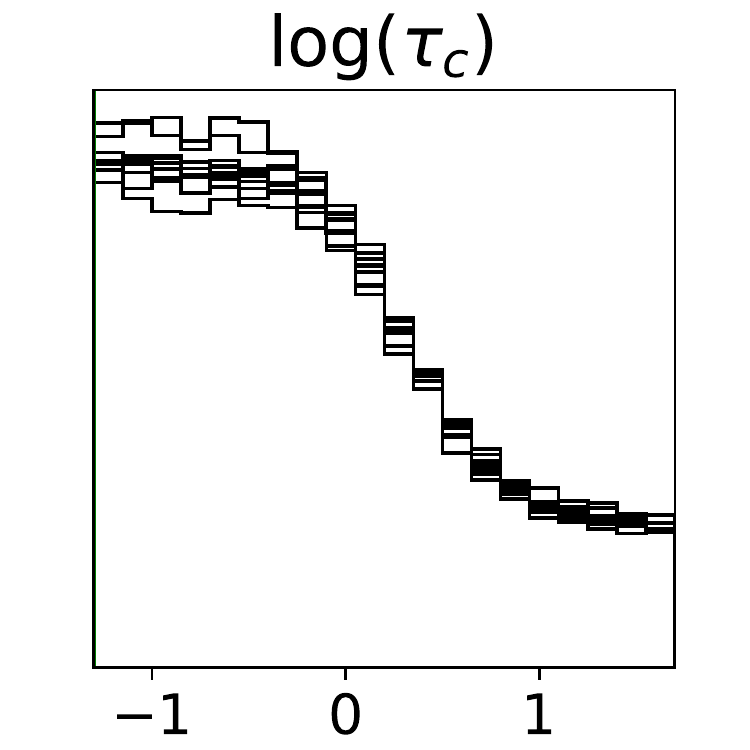}%
    \includegraphics[width=2.5cm]{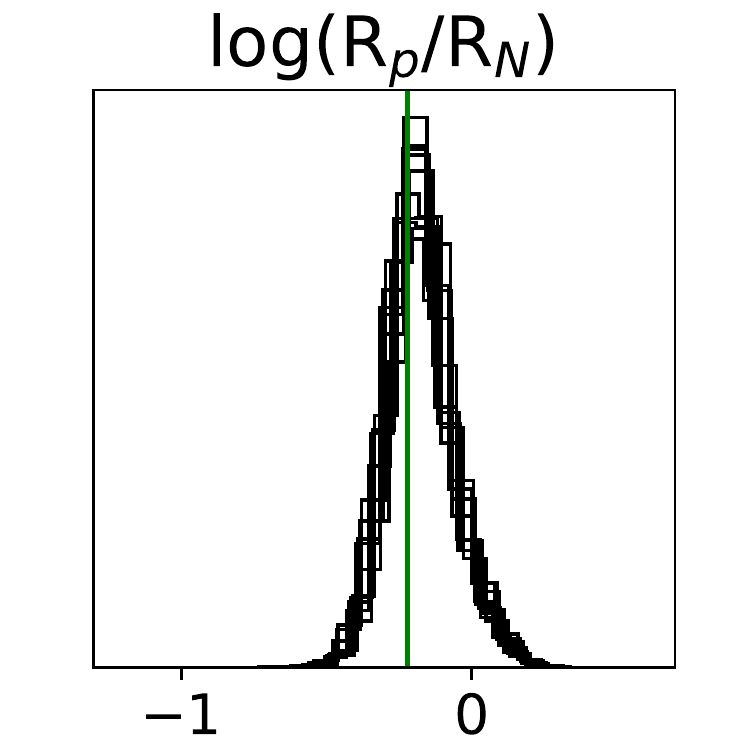}%
    \\
    \includegraphics[width=2.5cm]{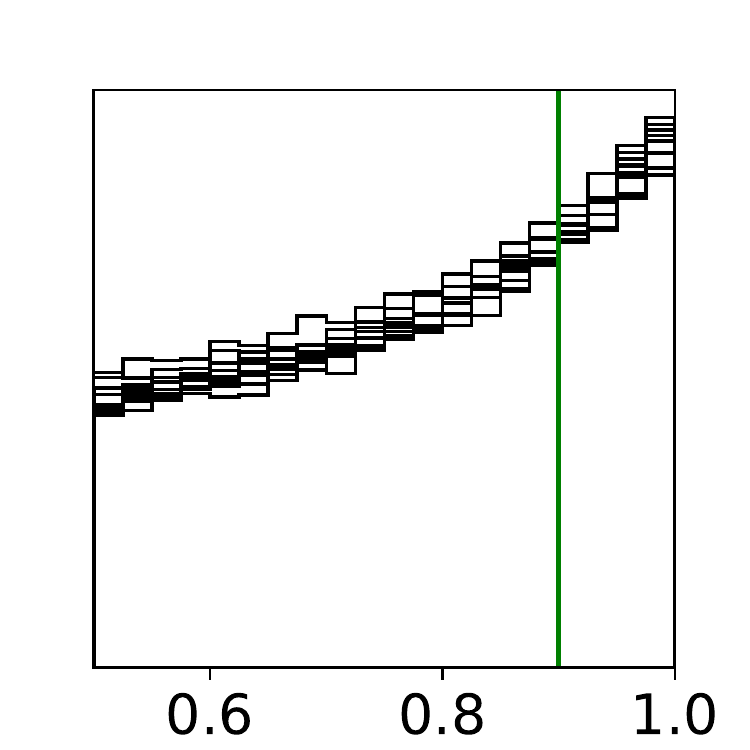}%
    \includegraphics[width=2.5cm]{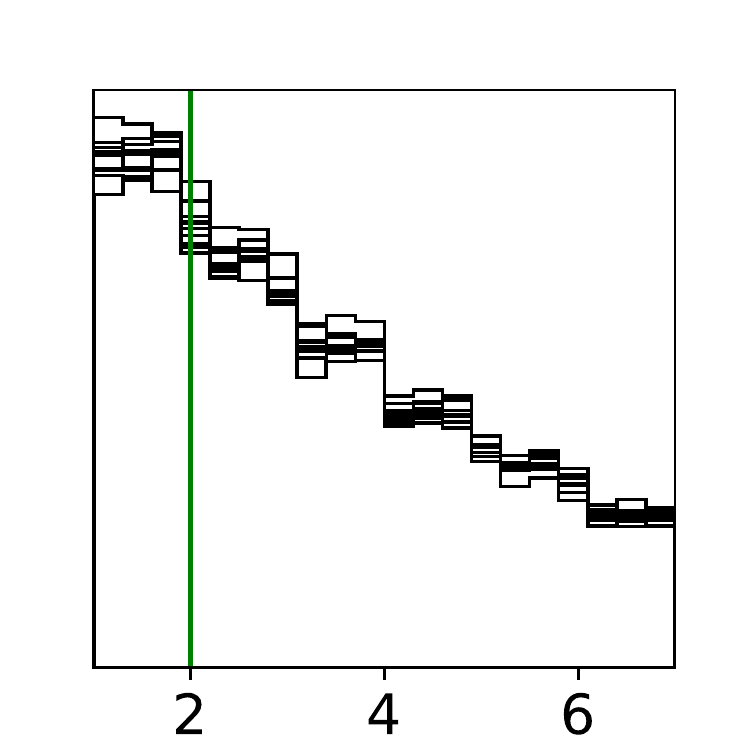}%
    \includegraphics[width=2.5cm]{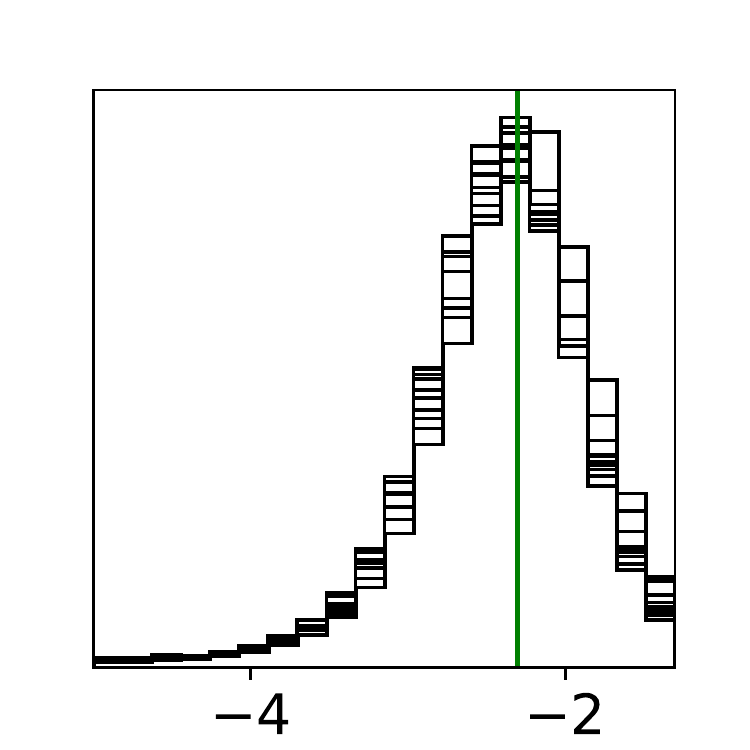}%
    \includegraphics[width=2.5cm]{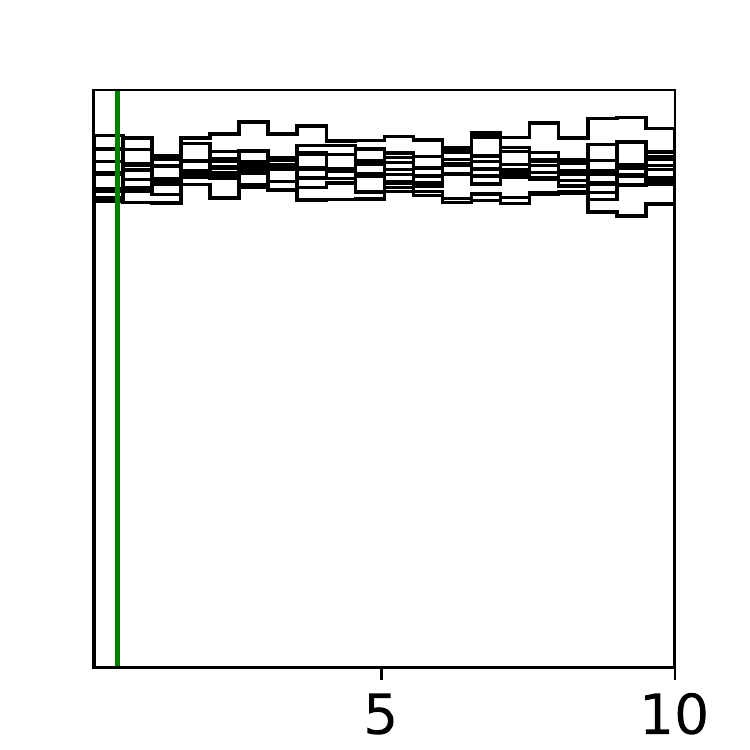}%
    \includegraphics[width=2.5cm]{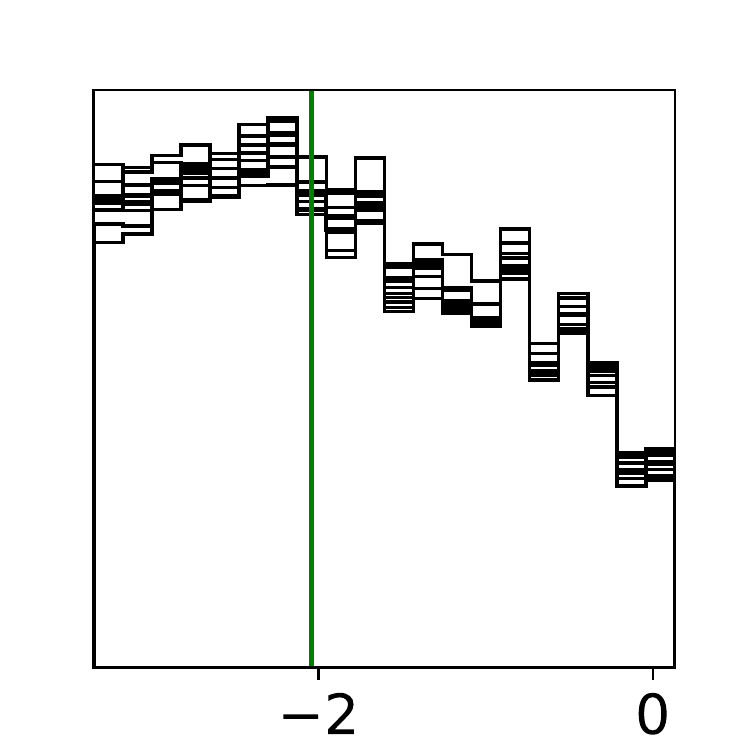}%
    \includegraphics[width=2.5cm]{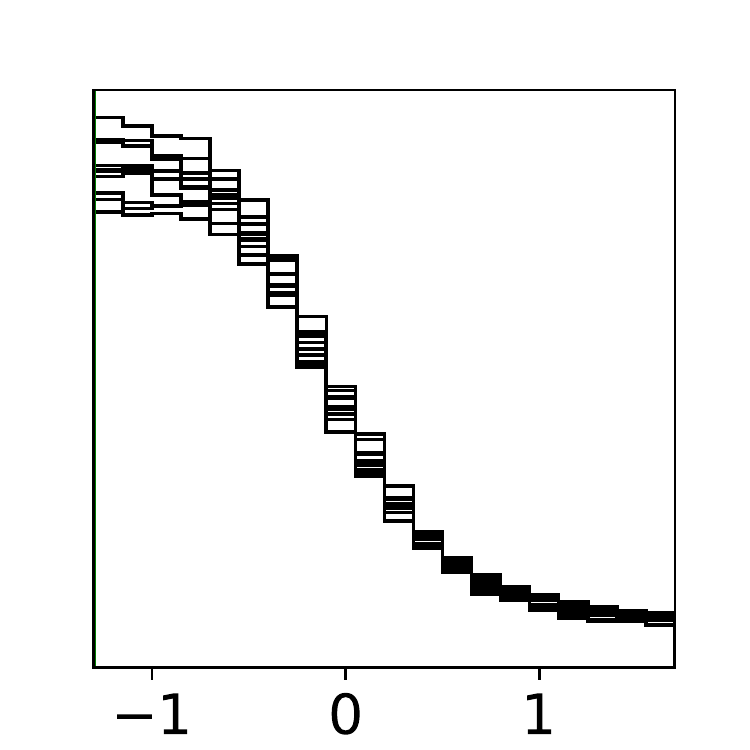}%
    \vspace{0.5cm}
    \\
    \includegraphics[width=2.5cm]{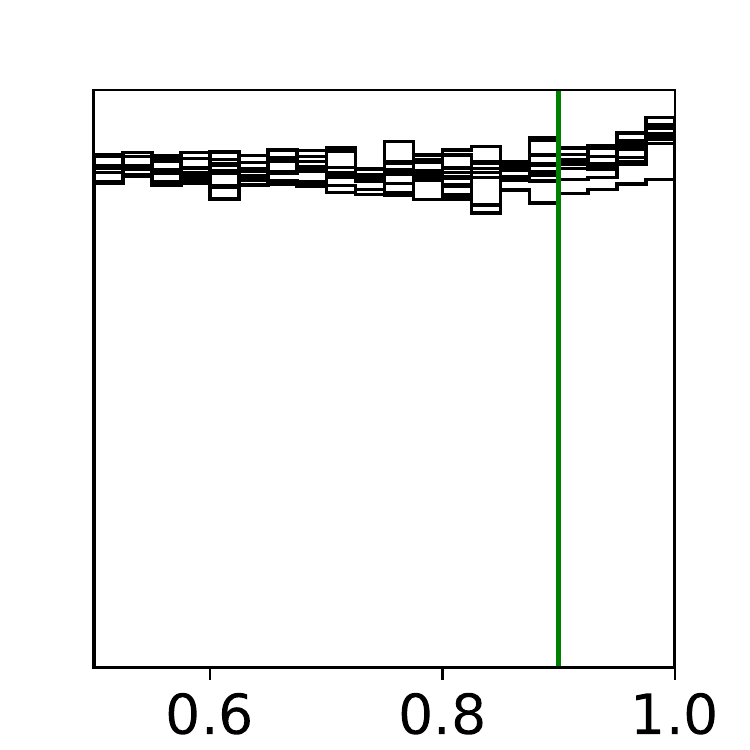}%
    \includegraphics[width=2.5cm]{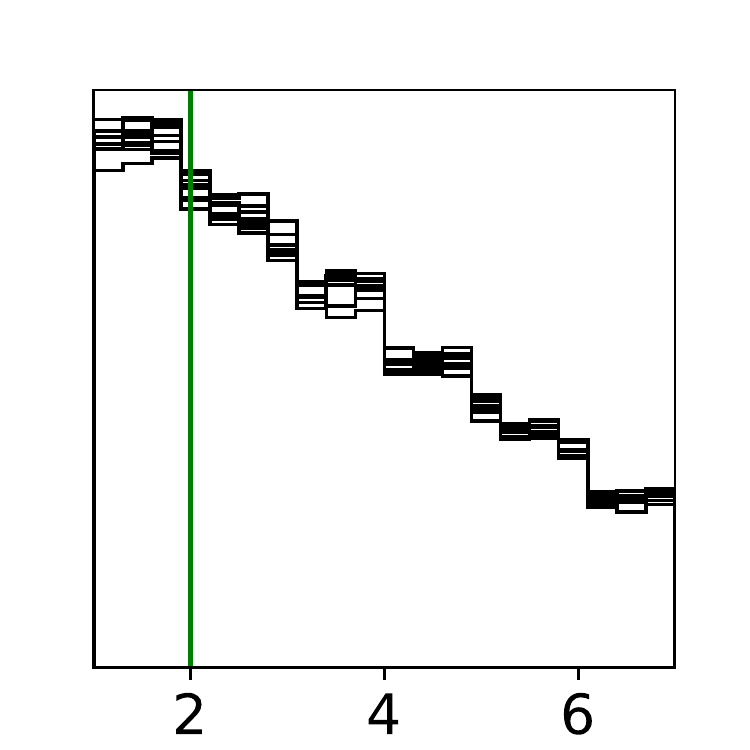}%
    \includegraphics[width=2.5cm]{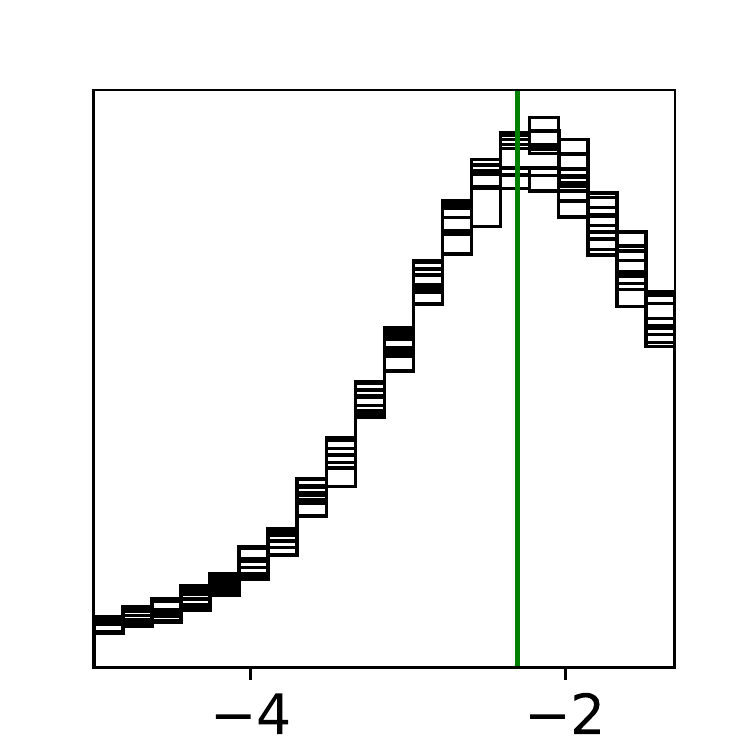}%
    \includegraphics[width=2.5cm]{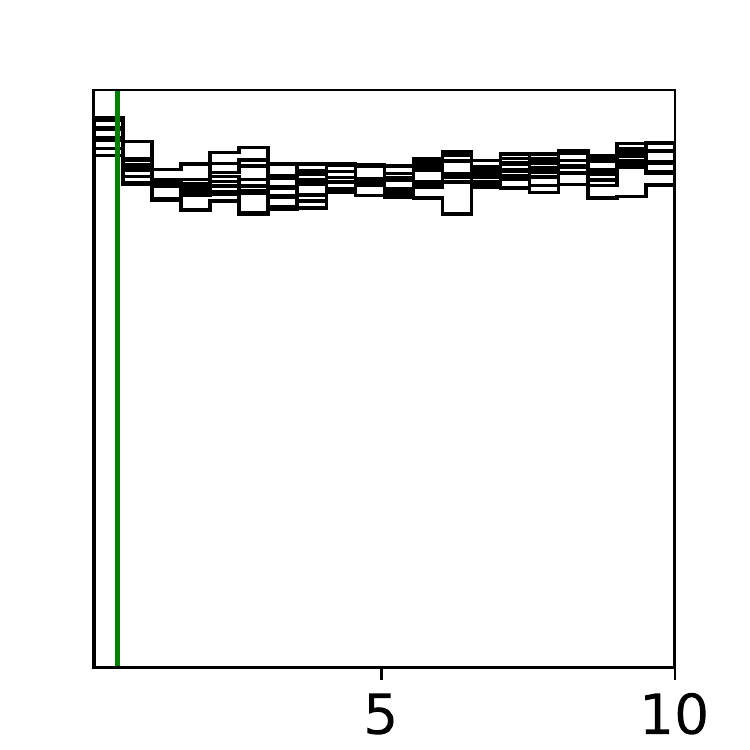}%
    \includegraphics[width=2.5cm]{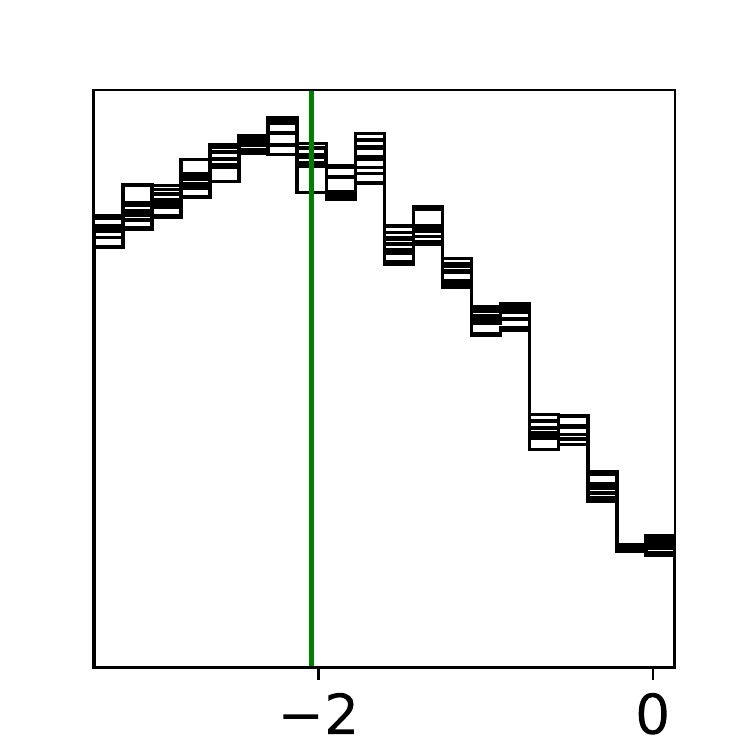}%
    \includegraphics[width=2.5cm]{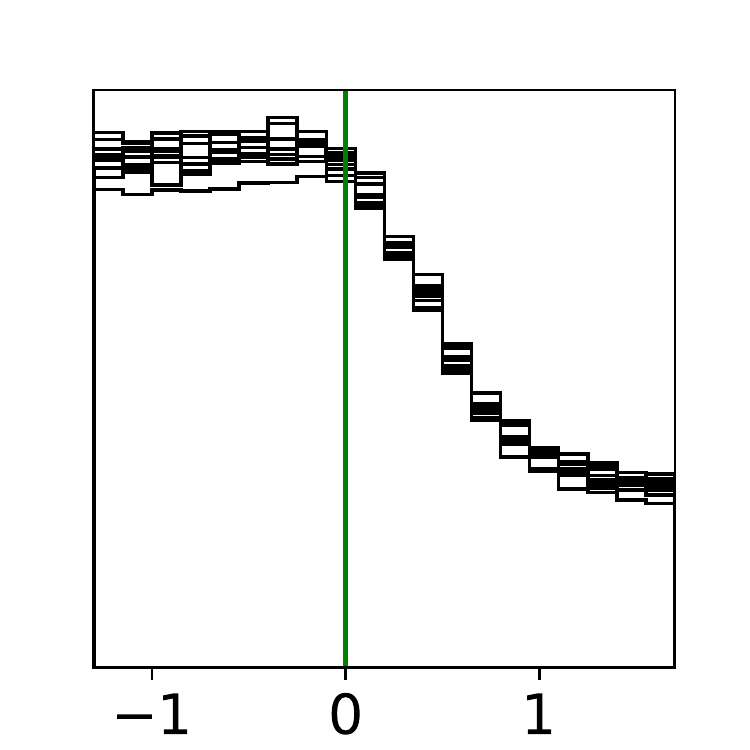}%
    \includegraphics[width=2.5cm]{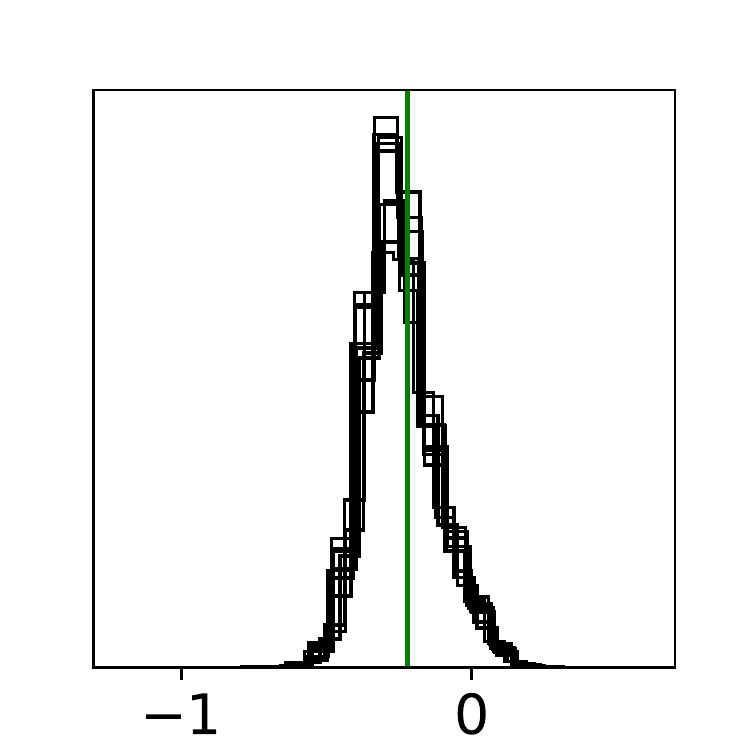}%
    \\
    \includegraphics[width=2.5cm]{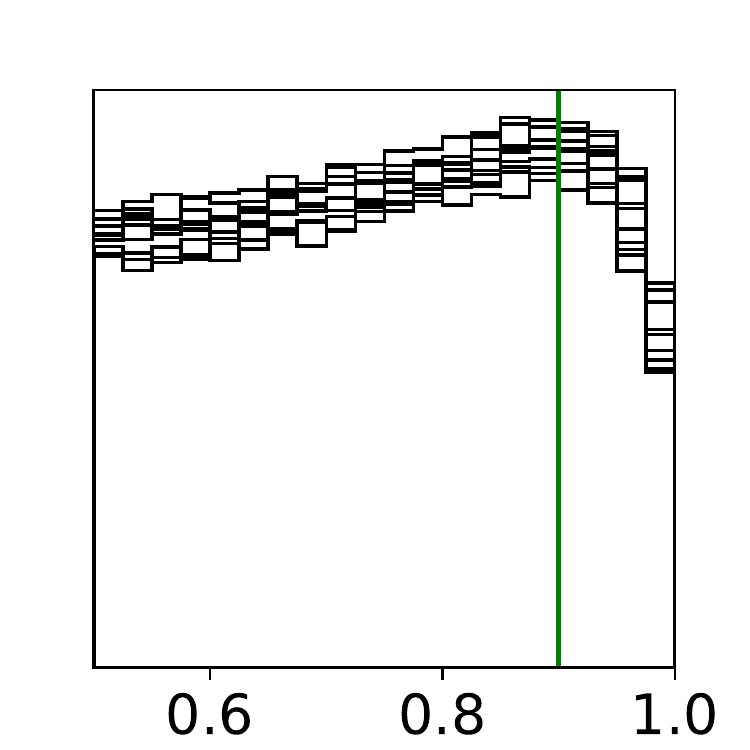}%
    \includegraphics[width=2.5cm]{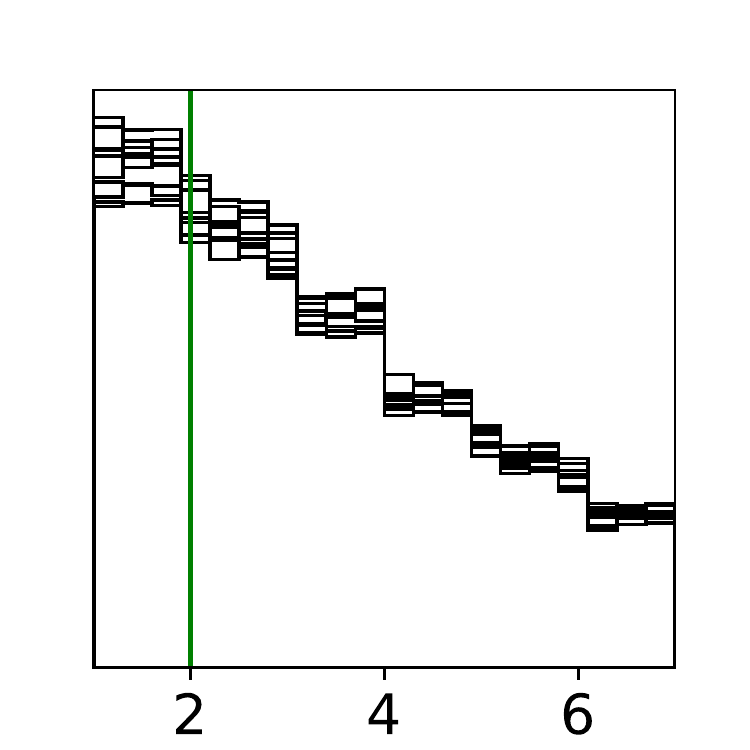}%
    \includegraphics[width=2.5cm]{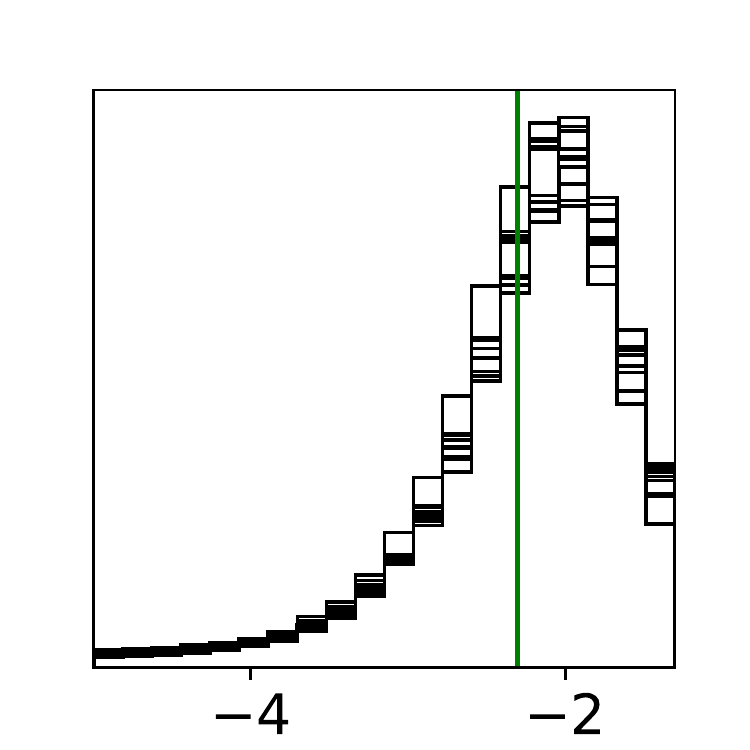}%
    \includegraphics[width=2.5cm]{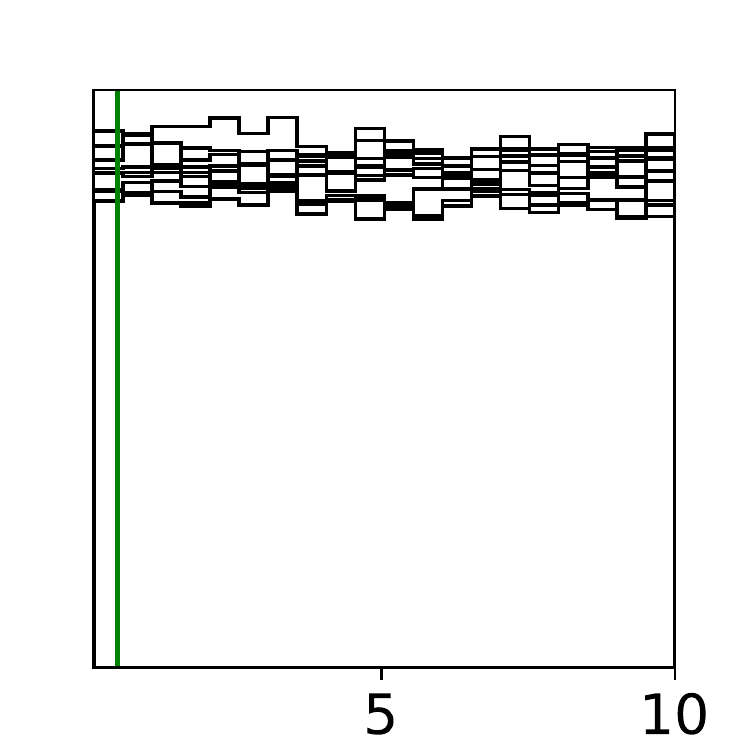}%
    \includegraphics[width=2.5cm]{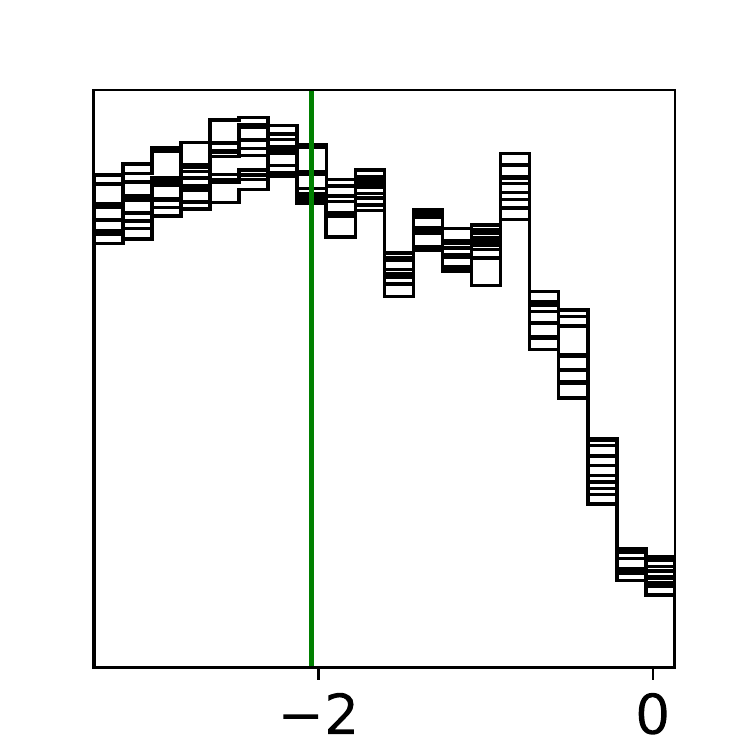}%
    \includegraphics[width=2.5cm]{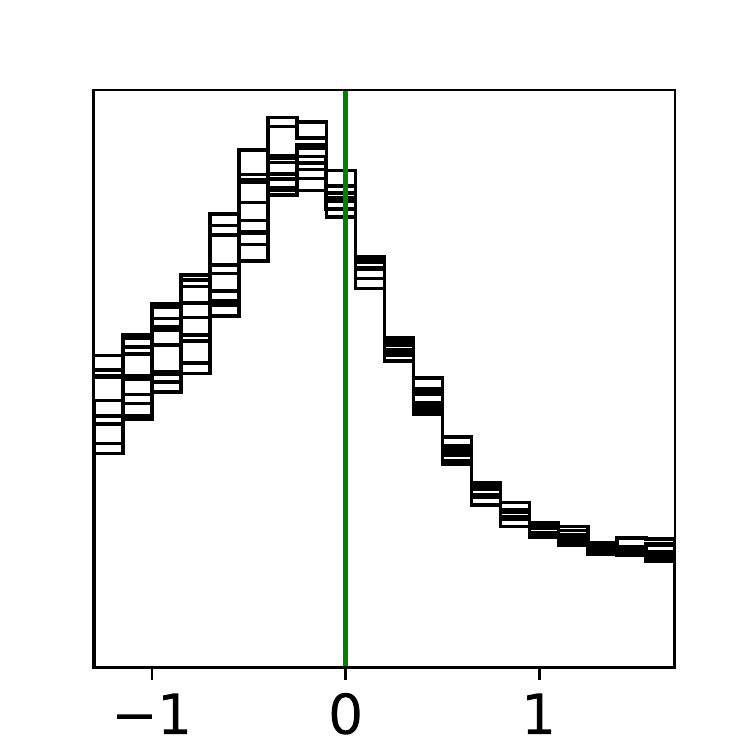}%
    \vspace{0.5cm}
    \\
    \includegraphics[width=2.5cm]{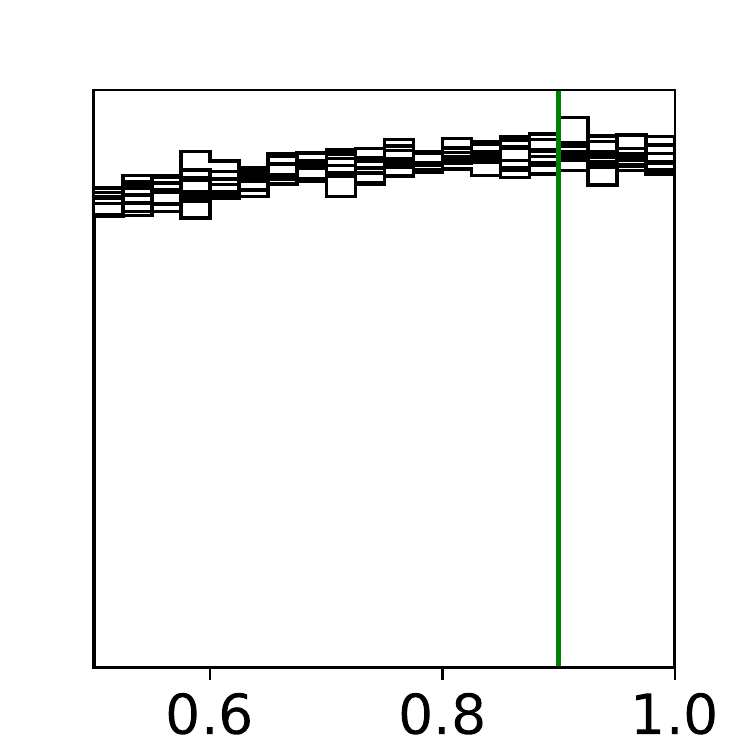}%
    \includegraphics[width=2.5cm]{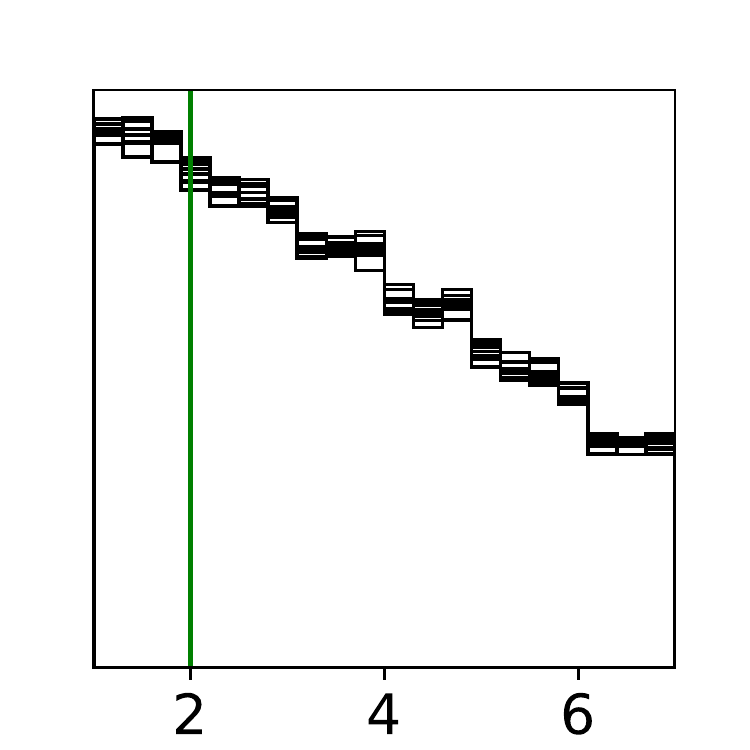}%
    \includegraphics[width=2.5cm]{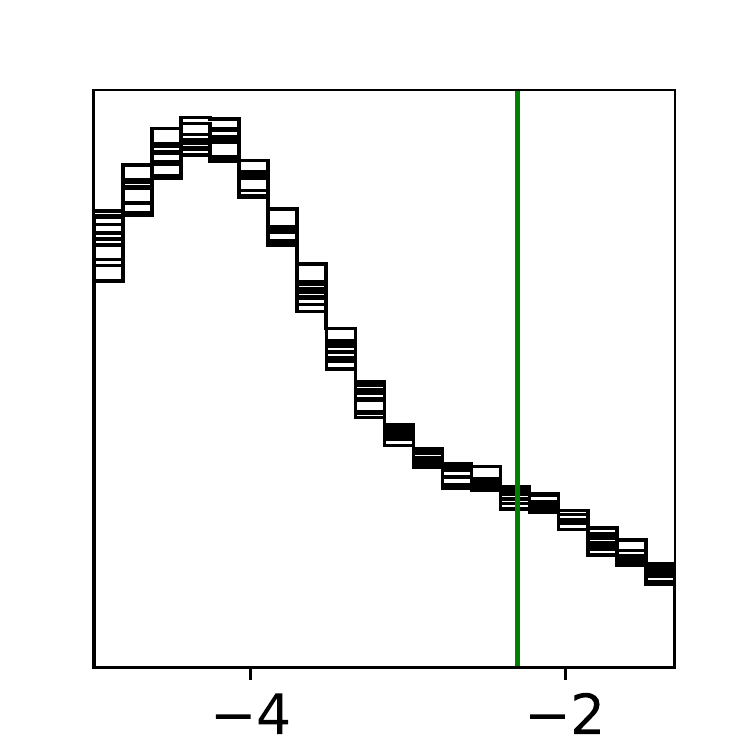}%
    \includegraphics[width=2.5cm]{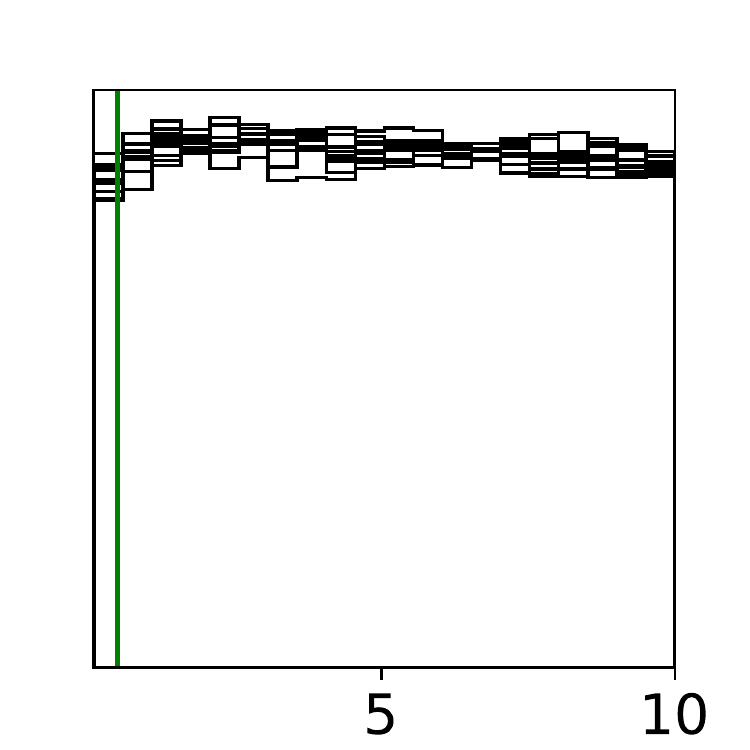}%
    \includegraphics[width=2.5cm]{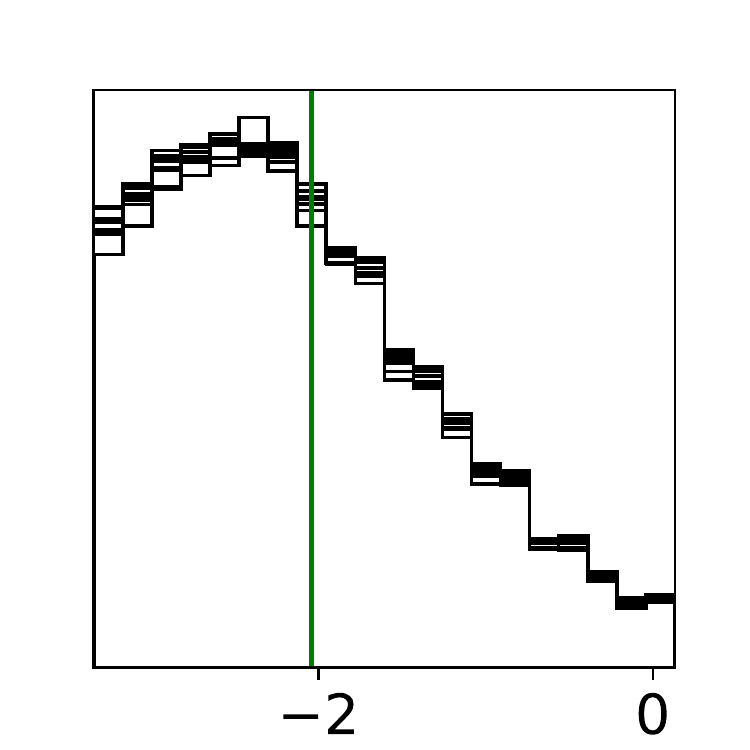}%
    \includegraphics[width=2.5cm]{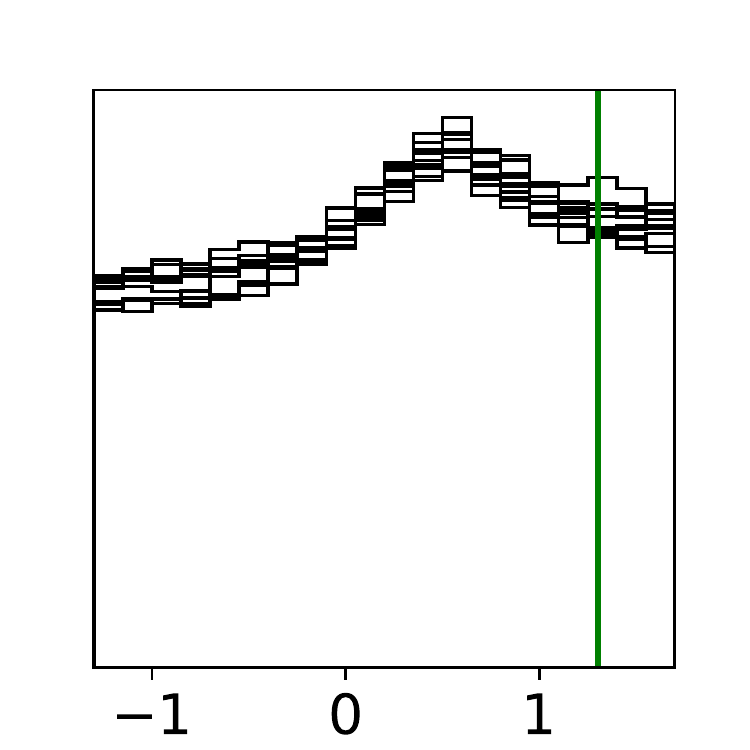}%
    \includegraphics[width=2.5cm]{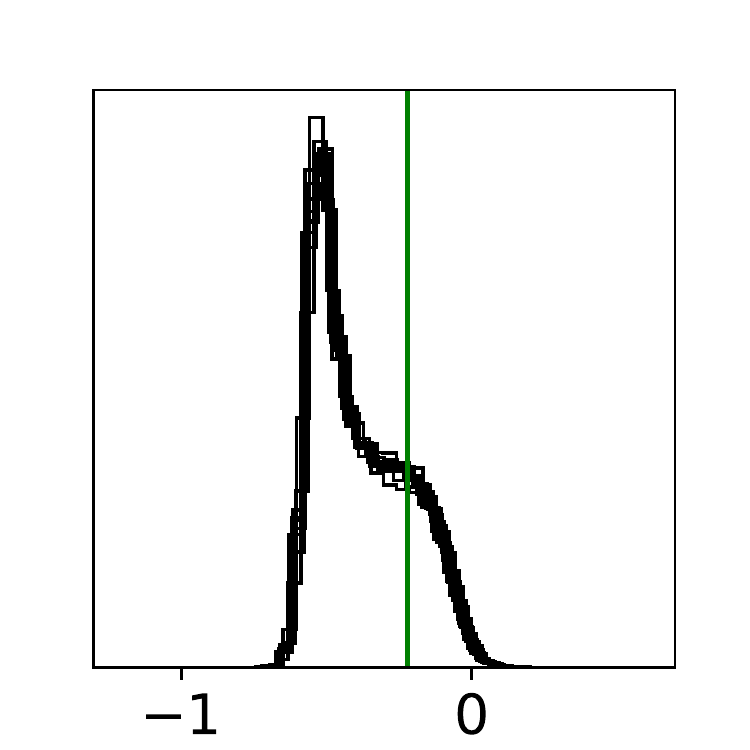}%
    \\
    \includegraphics[width=2.5cm]{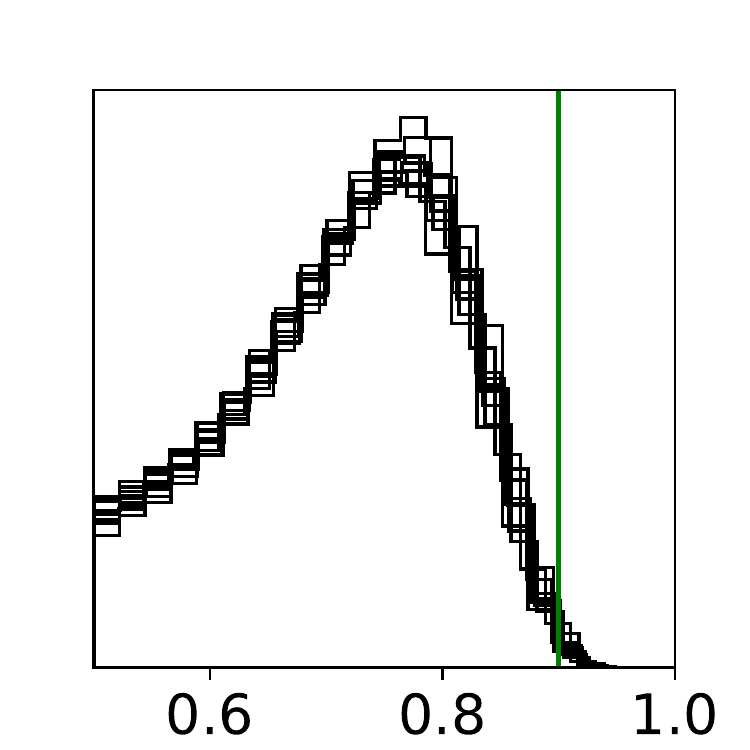}%
    \includegraphics[width=2.5cm]{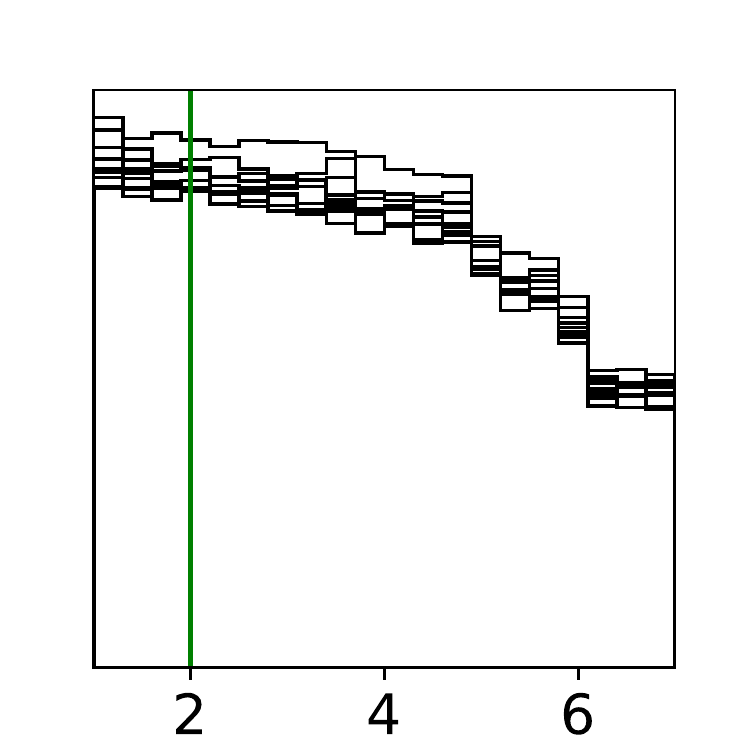}%
    \includegraphics[width=2.5cm]{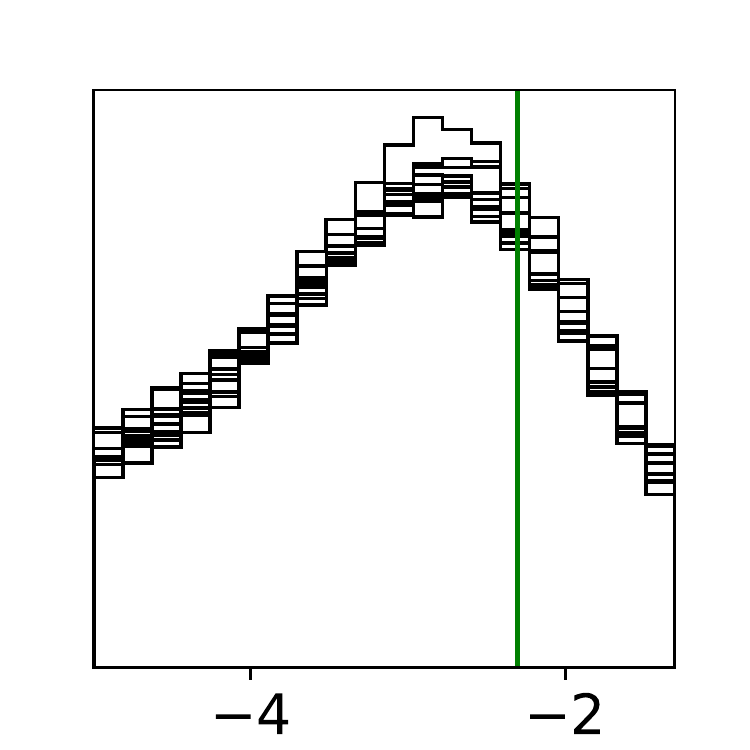}%
    \includegraphics[width=2.5cm]{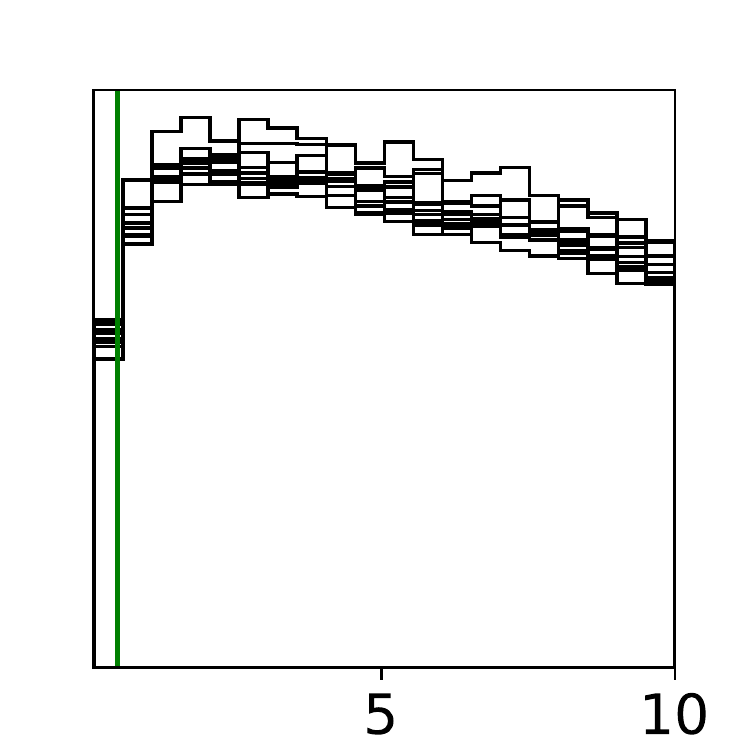}%
    \includegraphics[width=2.5cm]{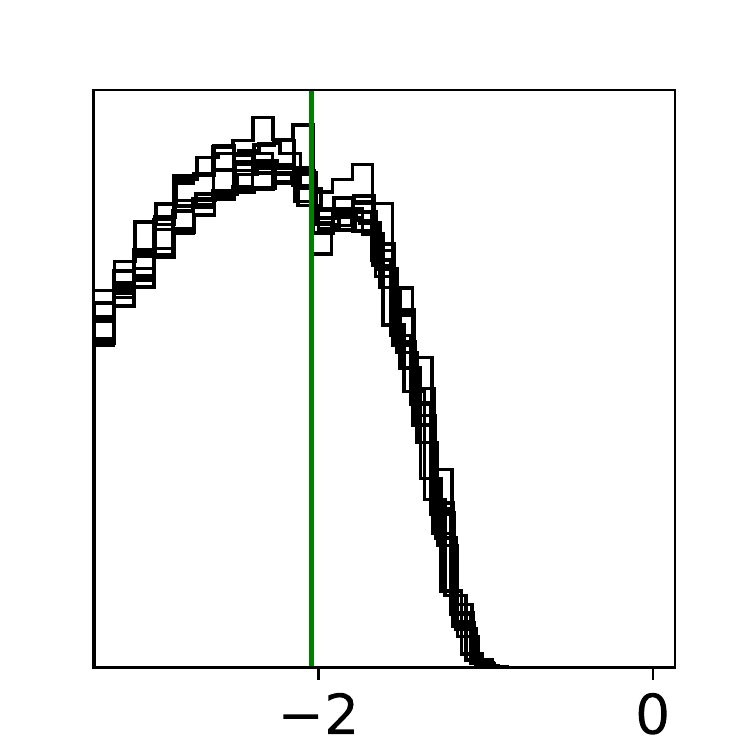}%
    \includegraphics[width=2.5cm]{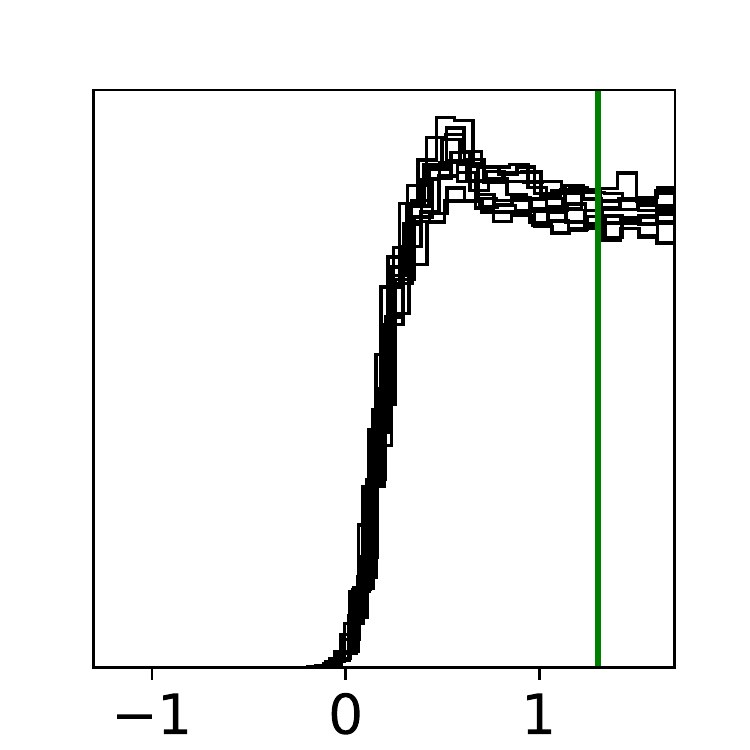}%
    
    \caption{Solid black lines: Marginalized probability distributions for each of the model parameters, given in each of the columns, after performing the retrievals 10 times for as many different noise realizations of the \textit{measured} spectra.
    Top rows: \textit{no-cloud} scenario, both for $R_p$ unknown and known.
    Middle rows:  \textit{thin-cloud} scenario, both for $R_p$ unknown and known.
    Bottom rows:  \textit{thick-cloud} scenario, both for $R_p$ unknown and known.
    Vertical green lines mark the \textit{true} values of the model parameters (see Table \ref{table:truth}) for each scenario.
    }%
    \label{fig:retrievals_several_noises}%
    \end{figure}

\end{appendix}

\end{document}